\documentclass[manuscript]{acmart}

\usepackage[utf8]{inputenc}
\usepackage{url}
\usepackage{colortbl}
\usepackage{balance}
\usepackage{xspace}
\usepackage{graphicx}
\usepackage{array} 
\usepackage{verbatim}
\usepackage{rotating}
\usepackage{bold-extra}
\usepackage{multirow}
\usepackage{listings}
\usepackage{boxedminipage}
\usepackage{soul}
\usepackage{enumitem}
\usepackage[normalem]{ulem}
\setlist{nolistsep,leftmargin=.5cm}
\usepackage{booktabs}
\usepackage{tabularx}
\usepackage{svg}
\usepackage{calc}
\usepackage{float}
\usepackage{multirow}
\usepackage[normalem]{ulem}
\useunder{\uline}{\ul}{}
\usepackage[most]{tcolorbox}
\definecolor{MidnightBlue}{HTML}{006895}
\definecolor{BoxesBlue}{HTML}{DEECFF}
\definecolor{BoxesYellow}{HTML}{FFF2CC}
\definecolor{StateGreen}{HTML}{91C788}
\definecolor{StateRed}{HTML}{FF8080}
\definecolor{ArrowGreen}{HTML}{61B15A}
\definecolor{ArrowViolet}{HTML}{BA94D1}
\usepackage{caption}
\usepackage[compact]{titlesec}
\usepackage{microtype} 
\usepackage{amsmath,amsfonts}
\usepackage{algorithmic}
\usepackage{textcomp}
\usepackage{xcolor}
\usepackage{hyperref}
\usepackage{ifthen}
\usepackage{wrapfig}
\usepackage{subcaption}
\usepackage{cleveref}

\usepackage{threeparttable}

\newboolean{showcomments}
\setboolean{showcomments}{true}

\ifthenelse{\boolean{showcomments}}
{\newcommand{\nb}[2]{
		\fbox{\bfseries\sffamily\scriptsize#1}
		{\sf\small$\blacktriangleright$\textit{#2}$\blacktriangleleft$}
	}
}
{\newcommand{\nb}[2]{}
}

\newcommand{\ie}{\textit{i.e.},\xspace}
\newcommand{\eg}{\textit{e.g.},\xspace}

\newcommand{\etal}{\textit{et al.}\xspace}

\usepackage{tikz}

\definecolor{bug_red}{rgb}{.84,.23,.29}

\definecolor{info-needed-color}{rgb}{1,.8,.12}

\definecolor{lightblue}{rgb}{ .753,  .902,  .961}
\DeclareRobustCommand{\hlblue}[1]{{\sethlcolor{lightblue}\hl{#1}}}

\definecolor{lightgreen}{rgb}{.596, 1, .596}
\DeclareRobustCommand{\hlgreen}[1]{{\sethlcolor{lightgreen}\hl{#1}}}

\definecolor{lightyellow}{rgb}{1,0.94902,0.8}
\DeclareRobustCommand{\hlyellow}[1]{{\sethlcolor{lightyellow}\hl{#1}}}

\newcommand{\ABUFIF}{{ABU}\textsubscript{50\%}\xspace}
\newcommand{\ABU}{{ABU}\xspace}
\newcommand{\BD}{{BD}\xspace}
\newcommand{\BDFIF}{{BD\textsubscript{50\%}}\xspace}
\newcommand{\PBU}{{PBU}\xspace}
\newcommand{\AU}{{AU}\xspace}
\newcommand{\RL}{{RL}\xspace}
\newcommand{\MB}{MB\textsubscript{0}\xspace}
\newcommand{\MBT}{MB\textsubscript{2}\xspace}

\newcommand{\RIAC}{{RI\textsubscript{AC}}\xspace}
\newcommand{\RIRC}{{RI\textsubscript{RC}}\xspace}
\newcommand{\RIDIF}{\boldmath{$\Delta_{RI}$}\xspace}

\newcommand{\dsthree}{DS2\xspace}
\newcommand{\dssix}{DS1\xspace}

\def\BibTeX{{\rm B\kern-.05em{\sc i\kern-.025em b}\kern-.08em
    T\kern-.1667em\lower.7ex\hbox{E}\kern-.125emX}}

\begin{document}

\title{From Absolute to Relative Code Comprehensibility Prediction}

\author{Nadeeshan De Silva}
\affiliation{%
  \institution{William \& Mary}
  \city{Williamsburg, VA}
  \country{USA}}
\email{kgdesilva@wm.edu}

\author{Martin Kellogg}
\affiliation{%
  \institution{New Jersey Institute of Technology}
  \city{Newark, NJ}
  \country{USA}}
\email{martin.kellogg@njit.edu}

\author{Oscar Chaparro}
\affiliation{%
  \institution{William \& Mary}
  \city{Williamsburg, VA}
  \country{USA}}
\email{oscarch@wm.edu}

\begin{abstract}

  Automatically predicting code comprehensibility could support tasks such as refactoring and code review. Existing metrics correlate poorly with human comprehension, motivating ML models that predict comprehensibility directly from code and developer features. Prior models predict absolute comprehensibility (AC), a comprehensibility value for an isolated snippet, but perform poorly since AC is a subjective proxy for a complex cognitive process. 
  We propose relative comprehensibility (RC) as an alternative task: given two snippets, predict which is easier to understand or whether they are comparable. We hypothesize RC is easier to learn, since it only requires identifying distinguishing features between snippets rather than estimating absolute values.
  Using 150 Java snippets and 12,540 human comprehensibility measurements from two prior studies, we compare AC and RC prediction across classical ML models, a CNN, and two LLMs, evaluating both snippet-wise (aggregate) and developer-wise (individual-judgment) predictions. AC models rarely beat simple baselines (at most 33.4\% average relative improvement), while snippet-wise RC models outperform baselines in 96.8\% of configurations, with gains up to 159.8\% consistent across architectures, though developer-wise results are more variable. 
  We surveyed 38 practitioners and found both AC and RC useful, with a stronger preference for RC in comparison-oriented tasks like refactoring and review.
  \looseness=-1
\end{abstract}

\maketitle

\section{Introduction}
\label{sec:intro}

Understanding source code is one of the most frequent and critical activities in software engineering~\cite{Minelli:ICPC15, Xia:TSE18, Maalej:TOSEM14, Borstler:2023developers}. Developers must build a mental model of how the code works to accomplish tasks such as designing new features, refactoring code, reviewing code, and correcting defects. However, code comprehension can be difficult and time-consuming for developers, especially when the code is complex,  poorly designed and written, or lacks adequate documentation~\cite{Tao:FSE12,Peitek:ICSE21,Ammerlaan:SANER15,Maalej:TOSEM14}. Indeed, studies have shown that developers spend between 58\% and 70\% of their time trying to understand code~\cite{Xia:TSE18,Minelli:ICPC15}.
\looseness=-1

To help developers control the \textit{difficulty} to understand code (\ie
``\textit{code comprehensibility}''), researchers have developed various metrics to measure attributes related to how easy code is to understand (\eg McCabe's~\cite{McCabe:TSE76} and Halstead's~\cite{Halstead1977} complexity metrics)~\cite{Nunez-Varela:JSS17,Curtis:TSE79,Zuse:IWCP'93,Sneed:JSMRP95,Ajami:EMSE19,Jbara:EMSE17,Chidamber:TSE94,Henderson-Sellers1995,Halstead1977,McCabe:TSE76,Beyer:ICPC10}. However, understanding code is a complex activity for humans, and many of these metrics correlate poorly with actual human comprehension, as shown by prior studies~\cite{Scalabrino:TSE19,Peitek:ICSE21,Feigenspan:ESEM11}. This has prompted researchers to study the use of machine learning to predict comprehensibility proxies~\cite{Scalabrino:TSE19,Raymond:TSE10,Trockman:MSR18,Lavazza:ESE23} based on various code- and developer-related features. %
\looseness=-1

A recent study~\cite{Scalabrino:TSE19}
evaluated six traditional machine learning models (\eg
Random Forests) to predict six
comprehensibility metrics or proxies (\eg  binary understandability and answers about code correctness) collected from 50 students and 13 professional
developers, who engaged in understanding 50 Java methods from 10 open-source projects. Unfortunately, the trained models performed poorly, achieving F1-scores between 0.59 and 0.77 when predicting the majority class (low understandability), and very poorly on the minority classes (high understandability), with F1-scores of 0.48 and 0.37.
We interpret these results to mean
that these models have low discriminatory power and might perform no better than basic baseline models such as random classifiers, and we validate this interpretation in our study.

Since determining whether a human has understood a piece of code is inherently difficult, the human measurements these models aim to predict (\eg understandability or readability ratings, comprehension time, or answers about code correctness) are proxies for a complex cognitive process. We
define the task of predicting such a proxy for an individual code snippet in
isolation as \emph{absolute comprehensibility} (AC) prediction.

We argue that these proxies are difficult prediction targets for learning-based models: their values can vary across participants because developers differ in their
backgrounds, experience, and interpretations of the code. Moreover, our recent
study~\cite{Arvan:ASE26} shows that the reliability of common comprehension proxies 
varies: some align better than others with expert judgment.
We therefore hypothesize that predicting precise
absolute proxy values makes it difficult for models to learn stable
relationships between code properties and human comprehension measurements.

Instead, we propose predicting the \emph{relative comprehensibility} (RC)
between two snippets rather than the absolute comprehensibility of each snippet
in isolation. At a conceptual level, AC and RC represent different prediction
problems. AC requires a model to estimate where a snippet falls on an absolute
proxy scale. RC requires the model to identify the direction of the difference
between two snippets: which one is more comprehensible, or whether they are
similarly comprehensible.
Pairwise prediction may also reduce the influence of
how different developers interpret and use an absolute rating scale. With this
motivation, we investigate two conjectures. First, predicting RC is a more
learnable and robust task than predicting AC. Second, the advantage of RC, if
present, extends across different model architectures rather than depending on a
particular one.

\begin{figure*}[t]
	\centering
	\includegraphics[width=\textwidth]{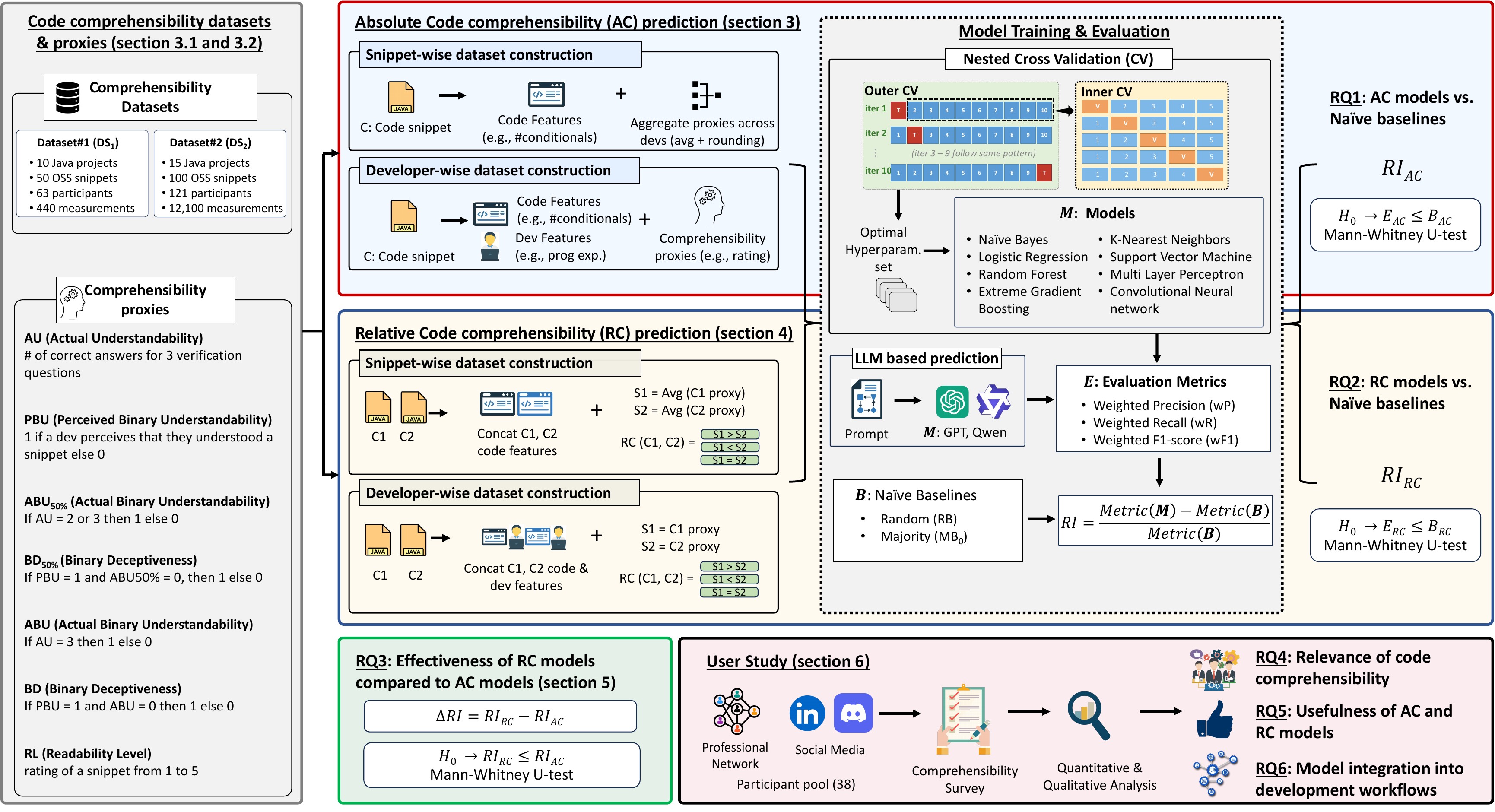}
	\caption{Overview the study.
	}
	\label{fig:study_overview}
\end{figure*}

To validate these conjectures, we conducted a study comprising \textbf{three phases}---see \cref{fig:study_overview} for an overview of the study. 
In the \textbf{first phase}, we replicated and extended the AC prediction study by
Scalabrino \etal~\cite{Scalabrino:TSE19}. We reused their dataset of 50 Java
snippets and 440 human comprehensibility measurements and added a second
dataset containing 100 snippets and 12,100 human readability
measurements~\cite{Raymond:TSE10}. Although readability and comprehensibility
are distinct concepts, readability captures one aspect of how easily code can
be understood; we discuss this distinction in
\Cref{sec:related_work}. Together, the datasets contain 150 Java snippets and
approximately 12,500 human measurements.
We also extended the six classical machine learning models evaluated by
Scalabrino \etal with four modern model types: Extreme Gradient Boosting
(XGBoost), a convolutional neural network (CNN), and two large language models
(LLMs). In total, we evaluated ten model types under two prediction settings.
\emph{Snippet-wise} prediction estimates an aggregate human judgment for an
unseen snippet, whereas \emph{developer-wise} prediction estimates a judgment
associated with an individual developer.

In the \textbf{second phase}, we constructed RC labels from the same human measurements
and evaluated models that predict which of two snippets is more comprehensible
or whether they are similarly comprehensible. We used the same
comprehensibility proxies, model architectures, and prediction settings as in the
AC evaluation. We also examined whether snippet-wise RC results remain robust
when using different thresholds for considering two snippets similarly
comprehensible.

In the \textbf{third phase}, we compared AC and RC prediction. For each task, we measured
model performance against two na\"ive baselines: a random classifier and a
majority-class classifier. Because AC and RC are fundamentally different tasks (\ie they have different output spaces and
class distributions), their weighted F1 scores and other performance metrics are not directly comparable. We
therefore measured each model's relative improvement (RI) over the stronger
baseline for its task and compared the RI achieved by AC and RC models. This allowed us to assess how much the models are learning from the data to accomplish their corresponding task.

Finally, to assess the potential value of RC prediction, we surveyed 38 software practitioners. The survey investigated whether they view code comprehension difficulty as a relevant problem, how useful they perceive AC and RC predictions to be for common software engineering tasks, and how such predictions could be integrated into their development workflows. This allowed us to assess the perceived usefulness of comprehensibility models, with particular attention to RC.
\looseness=-1

Our predictive results show a clear difference between the two tasks. AC
models often fail to outperform the na\"ive baselines and achieve a maximum
average relative improvement of 33.4\%. In contrast, snippet-wise RC models
outperform the baselines in 96.8\% of the evaluated configurations, with
relative improvements of up to 159.8\%. This advantage appears across the
evaluated model architectures, suggesting that it is not specific to a one. Developer-wise RC prediction is more challenging and produces
more variable results, but RC still outperforms AC in most evaluated
model-metric combinations.

The practitioner survey complements these predictive results. Participants
perceived both AC and RC predictions as potentially useful, but showed
stronger support for RC in tasks that naturally involve comparing code
alternatives, particularly refactoring and code review. They also preferred
model outputs that explain the prediction and provide actionable suggestions.

In summary, our work provides evidence that relative prediction is a promising
alternative to absolute comprehensibility prediction. Specifically, we
make the following contributions:

\begin{itemize}
	\item We replicate and extend the study by Scalabrino
	et al.~\cite{Scalabrino:TSE19} with an additional comprehensibility dataset
	and four modern model types, including a CNN and two LLMs. Our results show
	that AC models often fail to outperform simple random and majority-class
	baselines (\Cref{sec:absolute_code_comprehensibility}).
	
	\item We introduce and define relative comprehensibility (RC) between two code
	snippets as an alternative prediction target. We train and evaluate RC models using the
	same comprehensibility proxies, prediction settings, and model architectures
	used for AC prediction. Snippet-wise RC models substantially and
	consistently outperform the na\"ive baselines, while developer-wise
	results are more variable (\Cref{sec:relative_code_comprehensibility}).
	
	\item We compare AC and RC models through their relative improvement over
	task-specific baselines. RC produces larger improvements in nearly all
	snippet-wise comparisons and in most developer-wise model-metric
	combinations. This advantage is observed across the evaluated
	 model architectures rather than being limited to a single
	one (\Cref{sec:comparison}).

	\item We survey 38 software practitioners about the perceived usefulness
	of AC and RC models for common software engineering tasks and how they would prefer to integrate these models into their workflows.
	Participants particularly favored RC for comparison-oriented tasks such
	as refactoring and code review, and preferred explanations and actionable
	suggestions over raw prediction outputs
	(\Cref{sec:user_study}).
	
	\item We discuss the implications of RC prediction for future research and
	practice, highlighting its potential application to SE tasks
	that naturally compare code alternatives, including refactoring and code
	review (\Cref{sec:discussion}).
	
\end{itemize}

All data and materials required to validate and reproduce our study, including the
processed datasets, generated snippet pairs, extracted features, model
configurations, prompts, predictions, statistical-analysis scripts, and
survey materials, are available in our replication package~\cite{repl_pack}.

\section{Background and Related Work}
\label{sec:related_work}

\textbf{Code Comprehensibility},
as defined by Scalabrino \etal~\cite{Scalabrino:TSE19}, ``is a non-trivial mental process that requires building high-level abstractions from code statements or visualizations/models.'' 
Code comprehensibility is a fundamental determinant of software quality~\cite{Aggarwal:2002}, and extensive prior research has established that code comprehension represents one of the most time-consuming aspects of maintenance workflows~\cite{Deimel:1985, Raymond:1991, Rugaber:2000}. 
Buse and Weimer~\cite{Raymond:TSE10} define code \textit{readability} as ``a human judgment of how easy a text is to understand.''  Although readability and
comprehensibility are closely related, they are not equivalent. Readability
primarily concerns how easily the textual and syntactic form of the code can be
processed, whereas comprehensibility also involves semantic reasoning and the
construction of a mental model of the code's behavior. Nevertheless,
readability is commonly used as a proxy for how easily code can be understood.
We therefore use \emph{code comprehensibility} as an umbrella term for the
human-centered measures studied in this paper, including readability and
understandability, while analyzing each proxy separately. Prior studies have adopted this conceptualization, often using both terms to refer to the same concept~\cite{Stapleton:ICPC2020, Wyrich:ACS23,Siegmund:ICSE14, Borstler:TSE16, Jbara:EMSE17, Peitek:TSE18, Ajami:EMSE19}.

Prior work has identified several factors that influence comprehensibility.
\textit{Syntactic factors} include code lexicon and format properties~\cite{Mi:SPE23, Ribeiro:ESE23}, code structure~\cite{Ajami:EMSE19, Johnson:ICSME19}, and complexity metric usage~\cite{Esposito:2023Arxiv, Wyrich:ICSE21};
\textit{Developer factors} include code writing patterns~\cite{Langhout:ICPC21} and code reading behavior~\cite{Borstler:TSE16, Abid:ICSE19, Peitek:ICPC20, Siegmund:SANER16};
and \textit{Factors related to SE tools and practices} include comprehension tool usage~\cite{Storey:SCP00} and code review feedback~\cite{Oliveira:TSE24}.

\textbf{Empirical Validation of Code Complexity Metrics.}
Prior user studies introduced various metrics to measure code comprehensibility from humans, including time to read or understand the code, accuracy metrics based on answered verification questions about the code~\cite{Scalabrino:TSE19}, subject comprehensibility ratings~\cite{Raymond:TSE10}, fMRI scanner measurements~\cite{Peitek:ICSE21, Peitek:TSE18, Siegmund:ICSE14}, biometric sensory data~\cite{Fritz:ICSE14, Fucci:ICPC19, Yeh:FIE17}, and eye-tracking measurements~\cite{Karas:TSE2024, Abbad-Andaloussi:ICPC22, Binkley:EMSE13, Fritz:ICSE14, Peitek:ICPC20, Park:ESE24}. 
\looseness=-1
Researchers have studied whether these metrics correlate with traditional code complexity metrics~\cite{Ajami:EMSE19, Feigenspan:ESEM11, Jbara:EMSE17, Kaner2004, Scalabrino:TSE19}. Scalabrino \etal~\cite{Scalabrino:TSE19} 
found small correlations between the human comprehensibility proxies and  code/developer-related metrics.
Complexity metrics like McCabe's~\cite{McCabe:TSE76} have also been found ineffective for measuring code understandability~\cite{Peitek:ICSE21}. 
In a recent study~\cite{Feldman:FSE23}, we found a small correlation between automated code verifiability (based on program semantics data from verification tools) and human comprehensibility proxies.

\textbf{Predictive Models of Understandability}.
Scalabrino \etal~\cite{Scalabrino:TSE19} developed machine learning models to predict human understandability, outperforming previous models~\cite{Trockman:MSR18}. However, due to low prediction accuracy, they concluded that these models are not practical. We replicated and extended these results by directly comparing model performance to na\"ive baselines, showing that in many cases, the models underperformed. Lavazza \etal~\cite{Lavazza:ESE23} employed regression models with code metrics (\eg Cyclomatic complexity) as features and comprehension time as the target, but their models showed a significant average prediction error of $\approx$30\%.

Prior studies~\cite{Raymond:TSE10, Posnett:SEN21, dorn2012general, Scalabrino:ICPC16, Scalabrino:JSEP18, Mi:JSS22} used code features in binary classifiers to predict code readability. Classifier inputs in these studies, encompass structural features (loops, operators, blank lines)~\cite{Raymond:TSE10, Posnett:SEN21}, aggregated features including visual matrix representations of code tokens and alignment-based metrics\cite{dorn2012general}, and lexical features such as comment readability and textual coherence introduced in recent work~\cite{Scalabrino:ICPC16, Scalabrino:JSEP18}. 
With the emergence of LLMs for code generation, recent studies have focused on examining the readability of AI-generated code~\cite{Sergeyuk:ICPC24, Patel:24BigData, Dantas:2023, Takerngsaksiri:ICSME25, Ye:arxiv26}. These studies emphasize the importance of code readability, and hence we integrated Buse's ``Readability Level'' metric into our analysis. 
\looseness=-1

Prior work only studied how to predict absolute comprehensibility. In contrast, our study investigated how effectively learning-based models (\ie machine and deep learning models, including large language models) can predict relative comprehensibility compared to basic baselines. 
To the best of our knowledge, this is the first study to introduce and empirically validate relative code comprehensibility (RC) prediction and explore its practical implications for software engineering tasks.

\section{Evaluating Predictive Models of Absolute Code Comprehensibility}
\label{sec:absolute_code_comprehensibility}

Predicting absolute code comprehensibility (AC) involves building a model to directly predict measurements of comprehensibility obtained from humans.  Consider a group of software developers (\eg from a user study) that provide Likert-scale understandability ratings about a set of code snippets (\ie methods). Two types of AC models can be trained (see \Cref{tab:ac_task_summary}):
\begin{itemize}
  \item a \textbf{\emph{snippet-wise}} model that predicts the average Likert rating of all participants for a given code snippet based on
  code features such as number of conditionals, loops, and code blocks; or
\item a \textbf{\emph{developer-wise}} model that predicts the Likert rating given by a specific developer to a code snippet,
  considering the same set of code features as the snippet-wise model and additional features about
  the developer (\eg years of programming experience).
\end{itemize}

These AC prediction tasks are classification tasks: the model learns to choose among a set of predefined options (\eg the Likert scale options in
the above scenario) for a given code snippet. 
With this in mind, our first goal is to evaluate the ability of learning-based models at predicting absolute measures of comprehensibility, by  answering the following 
research question (RQ):
\looseness=-1

\vspace{0.1cm}
\begin{enumerate}[label=\textbf{RQ$_\arabic*$:}, ref=\textbf{RQ$_\arabic*$}, itemindent=0.5cm,leftmargin=0.5cm]
	\item \label{rq:predict_ac}{\textit{How effective are learning-based models at predicting absolute comprehensibility,
			compared to na\"ive baselines?\looseness=-1}}
\end{enumerate}
\vspace{0.1cm}

To answer this RQ, we replicate and extend the study by Scalabrino \etal~\cite{Scalabrino:TSE19}, which built and evaluated AC models for developer-wise prediction. 
That is, we developed our own infrastructure to carry out the experiments and answer \ref{rq:predict_ac}; however, while we started from prior work's data~\cite{Scalabrino:TSE19} and overall
experimental design, we made some corrections (noted throughout)
and augmented it in four important ways:
\begin{itemize}
\item we extended the evaluation dataset with the readability data from another prior user study~\cite{Raymond:TSE10}; 
\item we extended the evaluation by adding four new state-of-the-art models, allowing us to investigate how different model architectures perform at AC prediction;   
\item we compared the resulting models not by absolute performance (\ie by reporting
  only F1-scores as prior work does~\cite{Scalabrino:TSE19}), but instead by relative improvement over appropriate ``best'' baselines per task, allowing us to evaluate how well the models learn from the data; and
  \item we evaluated AC models for snippet-wise prediction.
\end{itemize}
We first detail our experimental design before we discuss the results in \Cref{sec:rq1-results}.

\begin{table}[t]
	\centering
	\caption{Overview of the absolute comprehensibility (AC) prediction task and settings.}
	\label{tab:ac_task_summary}
	\resizebox{0.9\columnwidth}{!}{
	\begin{tabular}{m{1.5cm}|c|m{8cm}|m{2cm}}
		\toprule
			\textbf{Setting} & \multicolumn{1}{c|}{\textbf{Model input}} & \multicolumn{1}{c|}{\textbf{Model output}}                                                                                                             & \multicolumn{1}{c}{\textbf{Features}}       \\ 
			\hline
			Snippet-wise     & A code snippet                            & Aggregated comprehensibility proxy across developers (e.g., aggregated Likert rating values given by various developers). Aggregation method: average + rounding & Code features                               \\ 
			\hline
			Developer-wise   & A code snippet                            & Comprehensibility proxy value (e.g., Likert rating of 1) given by a specific developer                                                                           & Code and developer features (concatenated)  \\
		\bottomrule
	\end{tabular}
}
\end{table}

\subsection{Comprehensibility Data Sources}
\label{subsec:ac_data_sources}

We reused two state-of-the-art Java code comprehensibility datasets from the comprehensibility studies~\cite{Scalabrino:TSE19,Raymond:TSE10}, both of which trained ML models on hand-crafted features to predict absolute comprehensibility:
\begin{itemize}
	\item \textbf{Dataset \#1 (\dssix)} from Scalabrino
	\etal~\cite{Scalabrino:TSE19} includes 50 OSS Java methods, which were evaluated by
	50 computer science (CS) students and 13 professional developers.  Participants first read and
	understood the methods, rated whether snippets were
	understandable (or not), and then answered three questions about code behavior
	and output to assess actual comprehension. Task
	completion time was also collected.
	
	\item \textbf{Dataset \#2 (\dsthree)} from Buse and Weimer
	~\cite{Raymond:TSE10} includes partial snippets from 100 OSS Java methods; 121 CS students rated the readability of each snippet on a 5-point Likert scale after reading and understanding the code. It should be noted that these snippets are intentionally simplified by Buse and Weimer to eliminate contextual dependencies and algorithmic complexity, focusing primarily on ``low-level'' readability characteristics~\cite{Raymond:TSE10}.
	
\end{itemize}

These datasets are the largest available collections of human comprehensibility measurements for Java code snippets derived from production-level open-source projects. The average NCNB LoC (Non-Comment, Non-Blank Lines of Code) for \dssix snippets is 37.86 and 10.23 for \dsthree snippets. For context, in Yu \etal's study~\cite{Yu:ICSE24}, production Java function-level datasets (not for code comprehensibility tasks) were snippets of 10.2 average NCNB LoC. Hence, we consider these snippets to be reasonably-sized.

Most evaluators in both datasets were CS students with ``intermediate to high
programming experience''~\cite{Raymond:TSE10,Scalabrino:TSE19}; only \dssix includes professional developers. All comprehensibility measurements were collected individually, with \dssix featuring 6 snippets evaluated by eight and 44 snippets by nine participants, while \dsthree had 121 participants per snippet.
\looseness=-1

\begin{table}[t]
    \caption{Code understandability metrics for absolute comprehensibility. Majority class (*).}
      \label{tab:absolute_proxies}
      \resizebox{\columnwidth}{!}{
      \begin{threeparttable}[b]  
        \begin{tabular}{l|l|l|llll|lcrr} 
          \toprule
          \multicolumn{1}{c|}{\multirow{2}{*}{\textbf{DS}}} & \multicolumn{1}{c|}{\multirow{2}{*}{\textbf{Metric}}} & \multicolumn{1}{c|}{\multirow{2}{*}{\textbf{Definition}}}                                                                                                & \multicolumn{8}{c}{\textbf{Class Distribution}}                                                                                                                                                                                                                                                                                                                                                                                                                                                                                                                          \\
          \multicolumn{1}{c|}{}                             & \multicolumn{1}{c|}{}                                 & \multicolumn{1}{c|}{}                                                                                                                                    & \multicolumn{4}{c|}{\textbf{Snippet-wise}}                                                                                                                                                                                                                                             & \multicolumn{4}{c}{\textbf{Developer-wise}}                                                                                                                                                                                                                                                                \\ 
          \hline
          1                                                 & Actual Understandability (\textbf{\AU})                                                    & \begin{tabular}[c]{@{}l@{}}The number of correct answers\\ for three verification questions\\ about the code. Possible values\\ are 0,1,2,3\end{tabular} & \begin{tabular}[c]{@{}l@{}}0*\\1\end{tabular}    & \begin{tabular}[c]{@{}l@{}}-\\-\end{tabular}    & \begin{tabular}[c]{@{}l@{}}37\\13\end{tabular}     & \begin{tabular}[c]{@{}l@{}}(74.0\%)\\(26.0\%)\end{tabular}                          & \begin{tabular}[c]{@{}l@{}}0*\\ 1\\ 2*\\ 3\end{tabular}      & \begin{tabular}[c]{@{}c@{}}-\\ -\\ -\\ -\end{tabular}     & \begin{tabular}[c]{@{}r@{}}153\\ 72\\ 138\\ 77\end{tabular}             & \begin{tabular}[c]{@{}r@{}}(34.7\%)\\ (16.4\%)\\ (31.4\%)\\ (17.5\%)\end{tabular}            \\ 
          \hline
          1                                                 & Perceived Binary Understandability (\textbf{\PBU})\tnote{1}                                                   & \begin{tabular}[c]{@{}l@{}}1 if a developer perceives that\\ they understood a given snip-\\ pet; 0 otherwise.\end{tabular}                              & \begin{tabular}[c]{@{}l@{}}0*\\1\end{tabular}    & \begin{tabular}[c]{@{}l@{}}-\\-\end{tabular}    & \begin{tabular}[c]{@{}l@{}}41\\9\end{tabular}      & \begin{tabular}[c]{@{}l@{}}(82.0\%)\\(18.0\%)\end{tabular}                          & \begin{tabular}[c]{@{}l@{}}0 \\ 1*\end{tabular}              & \begin{tabular}[c]{@{}c@{}}-\\ -\end{tabular}             & \begin{tabular}[c]{@{}r@{}}136\\ 304\end{tabular}                       & \begin{tabular}[c]{@{}r@{}}(30.9\%)\\ (69.1\%)\end{tabular}                                  \\ 
          \hline
          1                                                 & Actual Binary Understandability (\textbf{\ABU})\tnote{1}                                                  & If \AU = 3 then 1 else 0                                                                                                                                   & \begin{tabular}[c]{@{}l@{}}0*\\1\end{tabular}    & \begin{tabular}[c]{@{}l@{}}-\\-\end{tabular}    & \begin{tabular}[c]{@{}l@{}}48\\2\end{tabular}      & \begin{tabular}[c]{@{}l@{}}(96.0\%)\\(4.0\%)\end{tabular}                           & \begin{tabular}[c]{@{}l@{}}0*\\ 1\end{tabular}               & \begin{tabular}[c]{@{}c@{}}-\\ -\end{tabular}             & \begin{tabular}[c]{@{}r@{}}363\\ 77\end{tabular}                        & \begin{tabular}[c]{@{}r@{}}(82.5\%)\\ (17.5\%)\end{tabular}                                  \\ 
          \hline
          1                                                 & Actual Binary Understandability 50\% (\textbf{\ABUFIF})                                                 & If \AU $=$ 2 or 3 then 1 else 0                                                                                                                           & \begin{tabular}[c]{@{}l@{}}0*\\1\end{tabular}    & \begin{tabular}[c]{@{}l@{}}-\\-\end{tabular}    & \begin{tabular}[c]{@{}l@{}}29\\21\end{tabular}     & \begin{tabular}[c]{@{}l@{}}(58.0\%)\\(42.0\%)\end{tabular}                          & \begin{tabular}[c]{@{}l@{}}0*\\ 1\end{tabular}               & \begin{tabular}[c]{@{}c@{}}-\\ -\end{tabular}             & \begin{tabular}[c]{@{}r@{}}225\\ 215\end{tabular}                       & \begin{tabular}[c]{@{}r@{}}(51.1\%)\\ (48.9\%)\end{tabular}                                  \\ 
          \hline
          1                                                 & Binary Deceptiveness (\textbf{\BD})                                                    & \begin{tabular}[c]{@{}l@{}}If \PBU = 1 and \ABU = 0,\\ then 1 else 0\end{tabular}                                                                                 & \begin{tabular}[c]{@{}l@{}}0*\\1\end{tabular}    & \begin{tabular}[c]{@{}l@{}}-\\-\end{tabular}    & \begin{tabular}[c]{@{}l@{}}28\\22\end{tabular}     & \begin{tabular}[c]{@{}l@{}}(56.0\%)\\(44.0\%)\end{tabular}                          & \begin{tabular}[c]{@{}l@{}}0 \\ 1*\end{tabular}              & \begin{tabular}[c]{@{}c@{}}-\\ -\end{tabular}             & \begin{tabular}[c]{@{}r@{}}213\\ 227\end{tabular}                       & \begin{tabular}[c]{@{}r@{}}(48.4\%)\\ (51.6\%)\end{tabular}                                  \\ 
          \hline
          1                                                 & Binary Deceptiveness 50\% (\textbf{\BDFIF}) \tnote{1}                                                & \begin{tabular}[c]{@{}l@{}}If \PBU = 1 and \ABUFIF = 0, \\ then 1 else 0\end{tabular}                                                                                 & \begin{tabular}[c]{@{}l@{}}0*\\1\end{tabular}    & \begin{tabular}[c]{@{}l@{}}-\\-\end{tabular}    & \begin{tabular}[c]{@{}l@{}}45\\5\end{tabular}      & \begin{tabular}[c]{@{}l@{}}(90.0\%)\\(10.0\%)\end{tabular}                          & \begin{tabular}[c]{@{}l@{}}0* \\ 1\end{tabular}              & \begin{tabular}[c]{@{}c@{}}-\\ -\end{tabular}             & \begin{tabular}[c]{@{}r@{}}351\\ 89\end{tabular}                        & \begin{tabular}[c]{@{}r@{}}(79.8\%)\\ (20.2\%)\end{tabular}                                  \\ 
          \hline
          2                                                 & Readability Level (\textbf{\RL})                                                    & \begin{tabular}[c]{@{}l@{}}Readability rating of a snippet\\ from 1 to 5 (higher value im-\\ plies higher readability)\end{tabular}                      & \begin{tabular}[c]{@{}l@{}}2\\3*\\4*\end{tabular} & \begin{tabular}[c]{@{}l@{}}-\\-\\-\end{tabular} & \begin{tabular}[c]{@{}l@{}}13\\44\\43\end{tabular} & \begin{tabular}[c]{@{}l@{}}(13.0\%)\\(44.0\%)\\(43.0\%) \end{tabular} & \begin{tabular}[c]{@{}l@{}}1\\ 2*\\ 3*\\ 4*\\ 5\end{tabular} & \begin{tabular}[c]{@{}c@{}}-\\ -\\ -\\ -\\ -\end{tabular} & \begin{tabular}[c]{@{}r@{}}889\\ 2481\\ 3240\\ 3290\\ 2200\end{tabular} & \begin{tabular}[c]{@{}r@{}}(7.3\%)\\ (20.5\%)\\ (26.8\%)\\ (27.2\%)\\ (18.2\%)\end{tabular}  \\
          \bottomrule
          \end{tabular}

          \begin{tablenotes}
          	\centering
            \item[1]metric excluded for snippet-wise experiments
          \end{tablenotes}

      \end{threeparttable}
        }
    \vspace{-0.2cm}
      
\end{table}

\subsection{Code Comprehensibility Metrics}
\label{subsec:ac_proxies}

The comprehensibility measures from the prior studies~\cite{Scalabrino:TSE19,Raymond:TSE10} fall into three categories: subjective \textit{ratings} of human understanding of the code; \textit{correctness} of human interpretations of program output or behavior; and \textit{time} taken to read and understand a snippet.

We focus on the \textit{rating} and \textit{correctness}
metrics because \textit{time} measurements are highly influenced by factors like the developer’s cognitive speed, skills, and task complexity. This leads to significant variability, making time measurements less reliable for modeling purposes. For example, the time metric in \dssix (TNPU~\cite{Scalabrino:TSE19}) ranges from 3 to 1,649 seconds, the widest among all collected metrics. 
As a result, we expect that \textit{time} is the
\emph{hardest} metric for training a model; if we cannot
build good models for rating and correctness metrics, there is little value in predicting time metrics.
Excluding time leaves five metrics from prior work as predictive targets for the predictive models:
\begin{itemize}
	\item Actual Understandability (\textbf{\AU}) is the number of questions about the code correctly answered by the subject, out of three.         
        \item Perceived Binary Understandability (\textbf{\PBU}) is 1 if the subject claims to understand the code, and 0 otherwise.
	\item Actual Binary Understandability 50\% (\textbf{\ABUFIF}) is derived from \AU: it is 1 if at least 50\% (\ie 2 or 3) of the questions were correctly answered, and 0 otherwise. 
	\item Binary Deceptiveness 50\% (\textbf{\BDFIF}) is 1 if the subject claimed to understand the code (\PBU~=~1), but failed to answer 50\% of the questions correctly (\ABUFIF~=~0); otherwise, this metric is~0. The interpretation of this metric is the opposite of other metrics: 0 indicates understandable code.
        \item Readability Level (\textbf{\RL}) is a 5-point Likert score where larger values represent higher perceived code readability.
        
\end{itemize}

\noindent We defined two more metrics to augment the ones above:
\begin{itemize}
	\item Actual Binary Understandability (\textbf{\ABU}) is 1 if \AU is 3 (\ie the subject answered all questions correctly), and 0 otherwise.
	\item Binary Deceptiveness (\textbf{\BD}) is 1 if the subject claimed to understand the code (\PBU = 1) but failed to answer all the three questions correctly (\ABU = 0), and 0 otherwise. 
\end{itemize}
\ABU distinguishes \emph{fully}-understood code from \emph{partially}-understood code, and %
\BD distinguishes cases where developers \emph{think} they understood the code, but actually they lacked a complete grasp.

All the metrics provide discrete comprehensibility scores. In total, \dsthree includes 12,100 RL measurements, while \dssix includes 440 measurements for each of the other six metrics. We use these measurements for \textbf{developer-wise} prediction.

For \textbf{snippet-wise} prediction, we averaged the comprehensibility measurements collected from all subjects for each snippet and rounded the result to create discrete metrics for classification. However, for three binary metrics (PBU, ABU, and \BDFIF), the class distribution was highly imbalanced
(see \Cref{tab:absolute_proxies}). This made it impossible to evaluate models using the cross-validation approach described in \Cref{subsec:model_training_evaluation}. As a result, we excluded these metrics and kept the remaining four. 
For the multi-class metric AU, aggregation resulted in classes with too few data points, so we merged them with the ``closest'' class (\ie class 1 with 0 (as class 0) and class 2 with 3 (as class 1)), making AU a binary metric.
This class merge is sound both statistically and conceptually. 
From a class distribution perspective, this will resolve the class imbalance, enabling the models to focus on a broader and more meaningful distinction. 
Conceptually, answering one question is arguably closer to answering none in terms of code understandability, just as answering two is closer to answering all three.
\looseness=-1

\subsection{Features}
\label{subsec:ac_code_features}

We used two kinds of features to train the predictive models. \emph{Code} features measure the code's complexity, size, lexicon, format, or documentation. \emph{Developer} features measure some aspects of the developer's background. For snippet-wise prediction, we use only code features, and for developer-wise prediction, we use both.

\subsubsection{Code Features.}
We began with the 115 code features defined by  Scalabrino \etal~\cite{Scalabrino:TSE19}. After reviewing their data, definitions, and implementations~\cite{Scalabrino:TSE19}, we found that some features: (1) were ambiguously defined (\eg the term ``word'' in ``\# of words'' is unclear), (2)  applied to code constructs do not present in the snippets (\eg ``\# of aligned blocks'' only applies to constructors, but no snippets are constructors), 
and (3) could not be computed for many snippets (with \texttt{NaN} values for 30\%+ of snippets). We excluded 38 features with one of these properties, leaving  77.
We added seven complementary features to ensure consistency, as some attributes (\eg \# of commas)
only had normalized counts instead of totals---for other features, both total and normalized counts were already present, so we just standardized their definitions.
In total, we used 84 code features (see \Cref{tab:code_features}). The complete feature set is in our replication package~\cite{repl_pack}. 
\looseness=-1

\begin{table}[t]
  \centering
  \caption{Code features categorized by type}
  \label{tab:code_features}
  \resizebox{0.95\columnwidth}{!}{%
  \begin{tabular}{l|l|c}
    \toprule
  \textbf{Category} & \textbf{Code Features}                                    & \multicolumn{1}{l}{\textbf{\# Features}} \\ \hline
  Complexity        & Cyclomatic complexity, Nested blocks, Num of loops, Num of comparisons, \ldots    & 10                                              \\
  Size              & LOC, Num of parameters in method, Num of statements, Num of literals, \ldots   & 17                                              \\
  Lexicon           & Num of Identifiers, Num of keywords, Identifier length, Num of operators, \ldots & 27                                              \\
  Format            & Num of blank lines, Num of spaces, Num of parenthesis, Num of commas, \ldots  & 18                                              \\
  Documentation     & Num of comments, Comments readability, Comments and identifiers consistency, \ldots                 & 12                                              \\ \hline
  \multicolumn{2}{c|}{\textbf{Total}}                                                     & \textbf{84}                                            \\                                           
    \bottomrule
\end{tabular}
  }
\end{table}

\textit{Complexity} features include traditional metrics like loop count and cyclomatic complexity, while \textit{size} features account for parameters, statements, and lines of code. \textit{Format} features capture stylistic elements such as blank lines and parentheses, whereas \textit{lexicon} features capture vocabulary, including identifiers and keywords. Finally, \textit{documentation} features account for comments and their readability (\eg via the Flesch reading-ease test~\cite{flesch1979write}).

\subsubsection{Developer Features.}
We reused all developer features from the prior work~\cite{Scalabrino:TSE19,Raymond:TSE10}, since they
are specific to study participants. These features capture aspects of a programmer’s
background and experience. \dsthree only includes a single
developer feature: university class year or academic level: first year, second year, third/fourth year, or graduate level---the original data does not distinguish third- and fourth-year participants. \dssix includes three features: years of general and Java programming experience (values from 1 to 10, representing increasing
levels of experience) and educational/professional position (bachelor student, master
student, Ph.D. student, and professional developer).

\subsection{Models}
\label{subsec:models}
While Scalabrino~\etal~\cite{Scalabrino:TSE19} evaluated only 
six classical ML models, a key goal of our study is to validate 
whether our findings hold regardless of model architecture. To 
this end, we extended the original six classical ML models with 
four additional more-modern architectures spanning a diverse 
range of paradigms, making our evaluation broader 
and more representative of the current state of the art.

\subsubsection{Machine Learning Models.}
\label{subsec:ml_models}

We replicated the six classical ML models from prior work~\cite{Scalabrino:TSE19} (first 6 items in the list below) and extended the evaluation by adding another modern ML model, Extreme Gradient Boosting (XGBoost)~\cite{Chen:KDD16}, that is more representative of the models that might be deployed in practice.
\begin{itemize}
	\item {Naïve Bayes (\textbf{NB})} \cite{naive_bayes} is a probabilistic model based on Bayes’ theorem, which assumes feature independence and assigns a label based on the class with the highest probability.
	\item {K-Nearest Neighbors (\textbf{KNN})} \cite{k_nearest_neighbor} is a non-parametric model that compares a snippet to its \emph{k}-closest instances in the feature space and takes
	a majority vote of those as the output class.
	\item {Logistic Regression (\textbf{LR})} \cite{lr} is a linear model that assigns weights to code features to predict class probabilities, classifying based on a threshold (\eg 0.5).
	\item {Multilayer Perceptron (\textbf{MLP})} \cite{mlp} is a feedforward neural network that learns the relationship between features and the target metric through layers of interconnected neurons.
	\item {Random Forest (\textbf{RF})} \cite{random_forest} is an ensemble model that combines decision trees through bagging, with the final classification based on a majority vote among the trees.
	\item {Support Vector Machines (\textbf{SVM})} \cite{svc} finds a hyperplane to separate classes, maximizing the margin between them; it supports linear and non-linear classification using various kernels.
  \item {Extreme Gradient Boosting (\textbf{XGBoost})}~\cite{Chen:KDD16} is an  ensemble model that builds decision trees sequentially. Each new
  tree focuses on correcting errors made by the previous trees, and the final
  prediction combines the outputs of all trees. XGBoost extends standard
  gradient boosting with regularization and computational optimizations that can
  improve both predictive performance and training efficiency.
\end{itemize}

\subsubsection{Deep Learning Models.}
\label{subsec:dl_models}
The aforementioned ML models rely solely on handcrafted syntactic and developer features, and their underlying architectures may not be well-suited to capture the complex, non-linear patterns inherent in code comprehensibility data. To investigate the effect of model architecture and input representation on prediction performance, we extend our evaluation to include modern learning-based models. While several deep learning architectures (\eg Recurrent Neural Networks, Long Short-Term Memory,  Graph Neural Networks, and Transformers) have demonstrated superior performance over classical ML models on software engineering tasks~\cite{Jha:IEEEAccess19, Yang:22CSUR, Pandey:ISSE25}, we selected the Convolutional Neural Network (CNN) and Transformers (more specifically, Transformer-based large language models) as representative architectures to explore the potential of deep learning for code comprehensibility prediction, rather than exhaustively benchmarking all variants.
\looseness=-1

{Convolutional Neural Network (\textbf{CNN})}~\cite{Lecun:98CNN} is a deep learning architecture originally designed for computer vision tasks, which has since been adapted for diverse software engineering tasks~\cite{ Mi:IST18, Thaller:SANER19, Ren:TOSEM19, Zampetti:SANER20, He:ICPC20, He:SANER20}. We adapted the CNN architecture of Liu \etal~\cite{Liu:18ASE} to our code comprehensibility prediction task, as it natively supports tabular numerical features as an input while also accepting raw code snippets as a complementary input. This architecture comprises two parallel branches. The first branch accepts raw code snippets as text input, converts them into embeddings using a word2vec pretrained model~\cite{Mikolov:13arxiv} and passes them through three convolutional layers. The second branch takes the handcrafted code/developer features and processes them through three analogous convolutional layers. Both branches share a symmetrical convolutional architecture. The outputs of the two branches are subsequently flattened via a max pooling layer. They are concatenated, then passed through a SoftMax layer for final class probability estimation.
The total number of parameters of this model is 346,114.

As for Transformers, we experimented with two specific large language models (LLMs):  \textbf{GPT-5.4}~\cite{OpenAI:GPT5} and \textbf{Qwen2.5-Coder-32B-Instruct}~\cite{hui2024qwen2}. GPT-5.4 is one of the most capable and efficient frontier commercial reasoning models available from OpenAI~\cite{open_ai_documentation}, and it was the latest GPT when we executed this part of the study in April 2026. 
Qwen 2.5-Coder-32B-Instruct is one of the most capable and efficient open-source, code-specific large language models and was  ranked as the top model on the Big Code Models Leaderboard by Hugging Face~\cite{big_code_models_leaderboard} when we executed the study in April 2026.
Models in the GPT family generally use a decoder-only Transformer architecture that processes text autoregressively, predicting each token based on the preceding context through stacked self-attention and feed-forward layers. GPT-5.4 is a proprietary model designed for language understanding, generation, and multi-step reasoning; however, OpenAI has not publicly disclosed details such as its parameter count, number of layers, attention configuration, or whether it uses a dense or mixture-of-experts architecture. 
Qwen2.5-Coder-32B-Instruct is an instruction-tuned, decoder-only Transformer with approximately 32 billion parameters. It follows the Qwen2.5 architecture~\cite{qwen2025qwen25technicalreport}, which uses grouped-query attention, rotary positional embeddings, RMS normalization, and gated feed-forward layers, and is further pretrained on a large code-focused corpus to support code generation, reasoning, completion, and repair tasks.

\subsection{Data Normalization and Feature Selection}
\label{subsec:data_normalization_balancing}

To prepare the data for ML and DL model training and evaluation (as explained below, all models except the LLMs were trained), we applied three standard data normalization and balancing strategies.
First, to prevent overfitting, we removed duplicate data instances on the training data only.
Second, since the ML models may be sensitive to the magnitudes of the code features, we applied standard normalization~\cite{Abdi:ERD10} (on training, validation, and test sets) by transforming their values ($x$) into  $z=\frac{x-\mu}{\sigma}$ values. 
Finally, as \Cref{tab:absolute_proxies} shows, the output classes are imbalanced for most of the comprehensibility metrics. Since class imbalance can negatively affect the prediction capabilities of the models, we applied the Synthetic Minority Over-sampling Technique (SMOTE)~\cite{chawla2002smote} to generate synthetic samples for the minority classes to balance the data.

Since the number of features exceeds the number of data instances, particularly in the AC datasets, we applied feature selection as a dimensionality reduction step prior to oversampling. This approach helps mitigate overfitting and improves the effectiveness of subsequent data balancing techniques.
\looseness=-1

We applied the same correlation-based feature selection approach from
the prior work~\cite{Scalabrino:TSE19}, which ranks the
code features that most correlate with a comprehensibility metric and
selects the top-$k$ features for model training and evaluation. The
correlation is calculated based on Kendall's
$\tau$~\cite{kendall1938new}.
We tested the top 10\%, 20\%, ..., 100\%
most correlated code features for each model. Due to the small number of developer features, we used all of them.

\subsection{Model Training and Evaluation}
\label{subsec:model_training_evaluation}

\subsubsection{Machine and Deep Learning Models.}
\label{subsec:ml_dl_model_training_evaluation}

All ML and DL models, except the LLMs, were trained using the code and developer features described in \cref{subsec:ac_code_features}. We describe how this training was performed in this section while we explain how the LLMs were guided for predicting AC in \cref{subsub:llm_prompting}.

To mitigate model overfitting and potential biases introduced from having a relatively small dataset, we used nested cross-validation~\cite{cross_validation} for machine learning model training and evaluation.
Unlike Scalabrino \etal\cite{Scalabrino:TSE19}, who set aside 10\% of the data for hyperparameter tuning and excluded it from experiments, we adopted nested cross-validation for better generalization. A fixed tuning set risks overfitting to specific instances, whereas nested cross-validation iteratively tunes and validates across different folds, reducing this risk of overfitting~\cite{Cawley:JMLR2010}.

Nested cross-validation (CV) consists of two levels: outer CV and inner CV.
The \textbf{outer CV} randomly splits the data into 10 folds, ensuring each fold roughly maintains the class distribution of the full dataset.  Each fold serves as a test set once  to measure unbiased model performance, while the remaining nine folds are used for training and hyperparameter tuning. 
The \textbf{inner CV} further divides each outer training set into five folds to optimize hyperparameters. 

Each fold was used once as the validation set, with the other four as the training set.  
We tested various hyperparameter values for each model, detailed in our replication package~\cite{repl_pack}. For the ML models, we started with the hyperparameters from Scalabrino  \etal~\cite{Scalabrino:TSE19} and expanded them following %
expert guidance~\cite{Breiman:Springer2001,Amy:Medium22}.
For XGBoost, we selected the hyperparameters to tune based on the 
official XGBoost documentation~\cite{XGBoost_official_documentation} and following prior research~\cite{Santos:20BSSQ}. 
As experiments were executed; we adjusted the hyperparameter sets to accommodate our experiments to the computational resources of our lab.

For the CNN model, we selected the same hyperparameters as Liu~\etal~\cite{Liu:18ASE}.
The model was trained for 5 epochs with a batch size of 32, using the Adam 
optimizer~\cite{Kingma:14arxiv} with a learning rate of 0.001. Moreover, we used early stopping based on 10\% of the validation data for each fold to prevent overfitting, and the model was trained on a single NVIDIA A40 GPU with 40GB of memory.
We omitted hyperparameter tuning for the CNN model due to the high 
computational cost associated with training deep learning models. Also, our goal was to include a
representative deep-learning architecture in the AC evaluation, rather than to
identify the best possible CNN configuration.

We systematically trained and evaluated the model across all possible parameter combinations. The optimal values were selected based on the highest weighted F1-score (defined in \Cref{subsec:metrics}) across validation sets.

\begin{figure*}[t]
    \centering %
    
    \begin{minipage}[c]{0.44\textwidth}
        \begin{subfigure}{\textwidth}
            \centering
            \includegraphics[width=\textwidth, height=4.5cm, keepaspectratio]{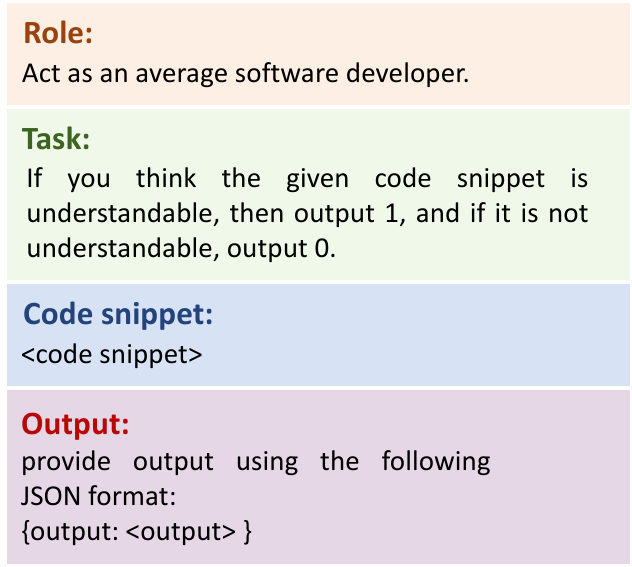}
            \caption{\PBU Snippet-wise Prompt Template}
            \label{fig:top_left}
        \end{subfigure}
        
        \vspace{1em} 
        
        \begin{subfigure}{\textwidth}
            \centering
            \includegraphics[width=\textwidth, height=4.5cm, keepaspectratio]{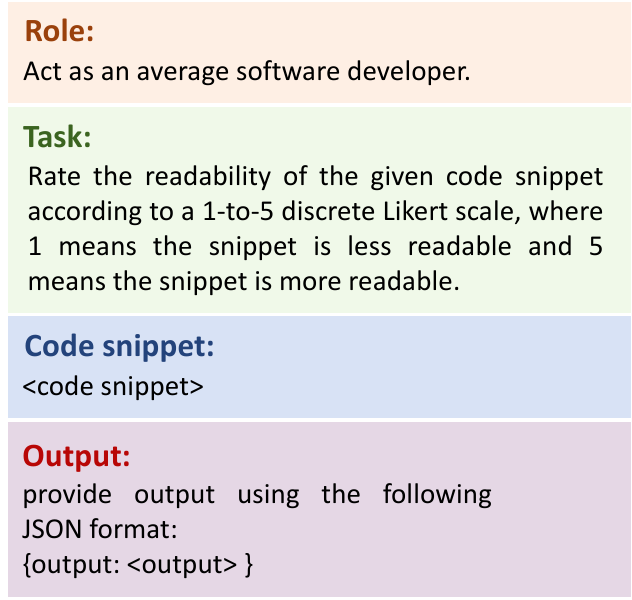}
            \caption{\RL Snippet-wise prompt}
            \label{fig:bottom_left}
        \end{subfigure}
    \end{minipage}%
    \hspace{0.04\textwidth} %
    \begin{minipage}[c]{0.5\textwidth}
        \begin{subfigure}{\textwidth}
            \centering
            \includegraphics[width=\textwidth, height=10cm, keepaspectratio]{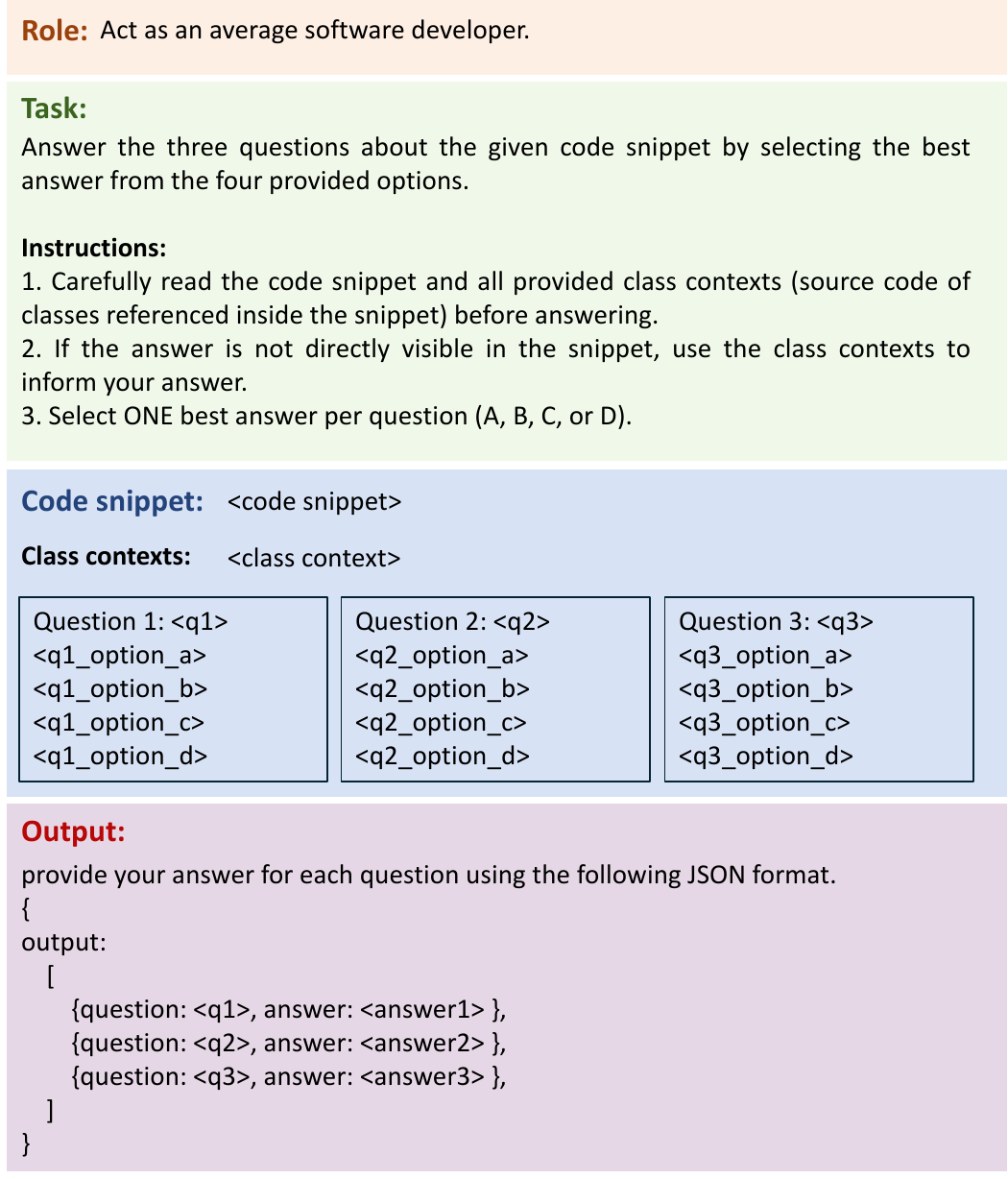} 
            \caption{\AU Snippet-wise prompt }
            \label{fig:au_snippet_prompt_template}
        \end{subfigure}
    \end{minipage}

    \caption{Prompt templates for snippet-wise AC prediction. Content inside the <brackets> are placeholders.}
    \label{fig:ac_prompt_templated_snippet}
\end{figure*}

To balance accuracy and training time, we chose 10 outer folds and 5 inner folds. More folds reduce test set size, making results less reliable, while fewer folds risk overfitting. After determining the best hyperparameter sets across all 10 outer training sets, we trained the models using each optimal set and evaluated them on every test fold. Unlike prior work~\cite{Scalabrino:TSE19}, which selects a single best hyperparameter set, our approach reduces bias by testing multiple optimal configurations, leading to a more reliable performance estimate.
\looseness=-1

\subsubsection{Large Language Models.}
\label{subsub:llm_prompting}

Unlike the previous models, the LLMs were not trained or fine-tuned on the
study datasets. Instead, we evaluated them through zero-shot prompting using
the templates shown in \Cref{fig:ac_prompt_templated_snippet}.

Our goal was to measure whether general-purpose and code-specialized LLMs can
perform AC prediction without task-specific training or calibration. We
therefore used a simple and consistent prompting strategy across all
comprehensibility proxies and prediction settings. Few-shot prompting, prompt
optimization, and fine-tuning could improve performance, but evaluating these
strategies would require additional design choices and validation experiments
that are outside the scope of this study. Accordingly, we interpret the LLM
results as zero-shot baselines rather than estimates of the models' best
achievable performance.

\textbf{Prompt Design.}
For AC prediction, we defined three prompt templates corresponding to the comprehensibility proxies/metrics \AU, \PBU, and \RL.  Results for the remaining dependent \dssix metrics (\ABU/\ABUFIF depend on \AU and \BD/\BDFIF depend on both \AU and \PBU---see metric definitions in  \Cref{tab:absolute_proxies}) were derived using the \AU and \PBU results. To calculate \AU, we compared the predicted answers to each question and the ground truth answers provided in the replication packages of \dssix and \dsthree~\cite{Scalabrino:TSE19,Raymond:TSE10}.

Each template included four parts: (1) the \textit{role} asked the LLM to act 
as an average developer (in the snippet-wise setting) or a particular person 
with given developer features (in the developer-wise setting), (2) the \textit{task} instructed the LLM to 
evaluate the code snippet according to the metric's definition, (3) the 
\textit{snippet} itself, where for \AU we additionally provided surrounding 
code context (source code of the classes referenced inside the snippet) to replicate the experimental conditions of Scalabrino~\etal~\cite{Scalabrino:TSE19}, 
who exposed participants to the same context in their user study, and  (4) the \textit{output format} asked for the response in JSON format. \Cref{fig:ac_prompt_templated_snippet} shows the prompt templates we used in our experiments.
Using these templates, we created 200 prompts for the snippet-wise setting (50 snippets $\times$ 2 comprehensibility proxies for \dssix, plus 100 snippets for \dsthree's RL) and 1,166 prompts for the developer-wise setting (50 snippets $\times$ 2 proxies $\times$ all unique combinations of the three developer features(number of developers per each snippet is vary because each snippet is evaluated either 8 or 9 participants and there are participants with same developer background.) for \dssix, plus 100 unique snippets $\times$ four levels of college-level experience for \dsthree).
\looseness=-1

\textbf{Prompt Execution.} For GPT-5.4, we executed the experiments via the OpenAI 
API~\cite{open_ai_api_reference}. For Qwen-2.5-Coder-32B-Instruct, 
an open-weights model available on HuggingFace~\cite{huggingface-Qwen}, we locally hosted it 
in our lab on a machine with 8 NVIDIA A30 GPUs, each with 24GB of memory. We used vLLM~\cite{Kwon:23SOOSP} as the inference engine for Qwen, because it is 
compatible with the OpenAI API and enables efficient inference — 
allowing us to reuse the same experimental infrastructure for both 
the commercial and open-source LLMs. To account for non-deterministic 
model outputs, we set the temperature to 0, used reasoning-effort as none for GPT-5.4 (for Qwen there is not a reasoning effort setting because it is a standard causal model), and executed each prompt 
three times per LLM. We then computed the evaluation metrics (defined 
in \Cref{subsec:metrics}) for each run and averaged them to obtain 
the final performance scores. For the snippet-wise setting, the total input tokens were 984,926 and total output tokens were 5,552. For the developer-wise setting, the total input tokens were 7,254,741 and total output tokens were 39,328.
\looseness=-1

\subsection{Evaluation Metrics}
\label{subsec:metrics}

We evaluated model performance using standard classification metrics~\cite{hossin:IJDKP2015, Naidu:Springer23} \textit{precision (P)}, \textit{recall (R)} and \textit{F1-score (F1)}. These were computed per target class ($P_i$, $R_i$, $F1_i$) and aggregated into \textbf{weighted} \textbf{precision} ($wP$), \textbf{recall}~($wR$) and  \textbf{F1-score} ($wF1$), accounting for the class distribution in the data. This approach ensures a fair evaluation across both majority and minority classes. 
Since we used cross-validation for traditional ML and DL models, overall precision, recall, and F1 were obtained by summing true positives, false positives, and false negatives across folds before computing the metrics. 
This approach provides a more reliable estimate than averaging per-fold scores, which can be skewed by outlier folds~\cite{Forman:SigKDD2010}.
\looseness=-1

We also compute two correlation-based metrics: Matthews Correlation Coefficient ($\pmb{MCC}$) and Cohen’s Kappa ($\pmb{Cohen}$). MCC incorporates all four entries of the confusion matrix (TP, TN, FP, FN), making it robust for imbalanced datasets. Cohen’s Kappa quantifies agreement between predicted and true labels, adjusting for chance-level agreement.
We use standard interpretation guidelines for effect size ($r$) thresholds~\cite{Wattanakriengkrai:TSE20}: Large ($r > 0.5$), Medium ($0.3 < r \leq 0.5$), Small ($0.1 < r \leq 0.3$), and Negligible ($r < 0.1$).

\subsection{Comparison with Baseline Models}
\label{subsec:baseline_models}
To compare models, we computed the \textbf{relative improvement} (\textbf{RI}) of a model $M$ over a baseline $B$, defined as $RI = \frac{Metric(M) - Metric(B)}{Metric(B)}$, where $Metric(X)$ is the performance of model $X$ under a given metric (\eg weighted F1). RI provides a normalized measure of improvement relative to the baseline.

We compared all trained models against two na\"ive baselines: 
(1) a ``lazy'' model ($LB_i$) that always predicts a single class $i$ regardless of the snippet (if $i$ is a majority class, we label the model as $MB_i$), and (2) a ``random'' model ($RB$) that randomly predicts a class for a snippet based on the class distribution for a metric.
This comparison allowed us to determine if the models are indeed learning from the data and can outperform basic classifiers. We measured how much the models learn by computing the relative improvement (RI) against the best baseline for each metric.
Baseline performance was measured by simple calculation considering the class distribution of each metric.

\subsection{\ref{rq:predict_ac}: Absolute Comprehensibility Results}
\label{sec:rq1-results}

In total, we trained 863 %
and 1,660 
classifiers that predict snippet-wise and developer-wise absolute comprehensibility (AC), respectively. These classifiers were trained under different configurations: eight machine and deep learning models, ten code feature sets, seven comprehensibility metrics, and different sets of best hyperparameters found during cross-validation for each metric (see \Cref{subsec:model_training_evaluation}). Further, we evaluated the two state-of-the-art large language models with 200 prompts (100 for \dssix and 100 for \dsthree) in the snippet-wise setting and 1,166 prompts in the developer-wise setting---each prompt executed three times to account for LLM non-determinism.

\Cref{tab:ac_results_dev,tab:ac_results_snippet} show the model performances aggregated by metric and model, compared to the best baseline model obtained for each metric. We focus our analysis on $wF1$, but the prediction results based on all the metrics we computed, described in \cref{subsec:metrics}, are found in our replication package~\cite{repl_pack}.  
The performance is averaged across trained models in terms of $wF1$ and across the three LLM predictions. The performance is compared against the best baseline via the relative improvement (RI) metric.

\begin{table}[t]
    \caption{Absolute comprehensibility (AC) results. Each column reports [$\pmb{wF1}$, (RI)]. $\pmb{wF1}$: average weighted F1 across all trained models and three LLM executions. RI: average relative improvement over the best baseline, either a: {$^\ddagger$}Random ({RB}) or {$^\dagger$}Majority Class (\MB) classifier.  \hlgreen{Green}: positive RI.}
    \label{tab:ac_results}
    \setlength{\tabcolsep}{5pt}
    \centering

\begin{subtable}[c]{\textwidth}
    \centering
    \caption{Snippet-wise prediction}
    \label{tab:ac_results_snippet}
    \resizebox{0.6\columnwidth}{!}{%
        \begin{tabular}{l|c|c|c|c}
            \toprule
            \textbf{Model} & \textbf{\AU} & \textbf{\ABUFIF} & \textbf{\BD} & \textbf{\RL} \\
            \hline

            \textbf{Baseline} & 
            \makebox[0.8cm][l]{0.629{$^\dagger$}} & 
            \makebox[0.8cm][l]{0.513{$^\ddagger$}} & 
            \makebox[0.8cm][l]{0.507{$^\ddagger$}} & 
            \makebox[0.8cm][l]{0.395{$^\ddagger$}}\\
            \hline

            \textbf{NB} & 
            \makebox[0.8cm][l]{0.572} \makebox[0.9cm][r]{(-9.1\%)} & 
            \makebox[0.8cm][l]{0.429} \makebox[0.9cm][r]{(-16.3\%)} & 
            \cellcolor{lightgreen}\makebox[0.8cm][l]{0.572} \makebox[0.9cm][r]{(12.8\%)} & 
            \makebox[0.8cm][l]{0.232} \makebox[0.9cm][r]{(-41.3\%)} \\

            \textbf{KNN} & 
            \makebox[0.8cm][l]{0.618} \makebox[0.9cm][r]{(-1.8\%)} & 
            \makebox[0.8cm][l]{0.438} \makebox[0.9cm][r]{(-14.6\%)} & 
            \cellcolor{lightgreen}\makebox[0.8cm][l]{0.592} \makebox[0.9cm][r]{(16.8\%)} & 
            \makebox[0.8cm][l]{0.341} \makebox[0.9cm][r]{(-13.8\%)} \\

            \textbf{LR} & 
            \makebox[0.8cm][l]{0.599} \makebox[0.9cm][r]{(-4.9\%)} & 
            \cellcolor{lightgreen}\makebox[0.8cm][l]{0.524} \makebox[0.9cm][r]{(2.2\%)} & 
            \cellcolor{lightgreen}\makebox[0.8cm][l]{0.646} \makebox[0.9cm][r]{(27.4\%)} & 
            \cellcolor{lightgreen}\makebox[0.8cm][l]{0.422} \makebox[0.9cm][r]{(6.8\%)} \\

            \textbf{MLP} & 
            \makebox[0.8cm][l]{0.565} \makebox[0.9cm][r]{(-10.2\%)} & 
            \makebox[0.8cm][l]{0.470} \makebox[0.9cm][r]{(-8.3\%)} & 
            \makebox[0.8cm][l]{0.456} \makebox[0.9cm][r]{(-10.0\%)} & 
            \makebox[0.8cm][l]{0.338} \makebox[0.9cm][r]{(-14.6\%)} \\

            \textbf{RF} & 
            \cellcolor{lightgreen}\makebox[0.8cm][l]{0.645} \makebox[0.9cm][r]{(2.4\%)} & 
            \makebox[0.8cm][l]{0.420} \makebox[0.9cm][r]{(-18.1\%)} & 
            \cellcolor{lightgreen}\makebox[0.8cm][l]{0.677} \makebox[0.9cm][r]{(33.4\%)} & 
            \makebox[0.8cm][l]{0.391} \makebox[0.9cm][r]{(-1.0\%)} \\

            \textbf{SVM} & 
            \cellcolor{lightgreen}\makebox[0.8cm][l]{0.661} \makebox[0.9cm][r]{(5.0\%)} & 
            \cellcolor{lightgreen}\makebox[0.8cm][l]{0.530} \makebox[0.9cm][r]{(3.4\%)} & 
            \cellcolor{lightgreen}\makebox[0.8cm][l]{0.668} \makebox[0.9cm][r]{(31.7\%)} & 
            \makebox[0.8cm][l]{0.385} \makebox[0.9cm][r]{(-2.5\%)} \\
            \textbf{XGBoost} & 
            \makebox[0.8cm][l]{0.626} \makebox[0.9cm][r]{(-0.6\%)} & 
            \makebox[0.8cm][l]{0.448} \makebox[0.9cm][r]{(-12.6\%)} & 
            \cellcolor{lightgreen}\makebox[0.8cm][l]{0.510} \makebox[0.9cm][r]{(0.5\%)} & 
            \cellcolor{lightgreen}\makebox[0.8cm][l]{0.420} \makebox[0.9cm][r]{(6.3\%)} \\

            \hline

            \textbf{CNN} & 
            \makebox[0.8cm][l]{0.562} \makebox[0.9cm][r]{(-10.6\%)}& 
            \cellcolor{lightgreen}\makebox[0.8cm][l]{0.530} \makebox[0.9cm][r]{(3.4\%)}& 
            \cellcolor{lightgreen}\makebox[0.8cm][l]{0.579} \makebox[0.9cm][r]{(14.1\%)}& 
            \makebox[0.8cm][l]{0.348} \makebox[0.9cm][r]{(-12.0\%)} \\

            \textbf{GPT-5.4} & 
            \makebox[0.8cm][l]{0.134} \makebox[0.9cm][r]{(-78.7\%)} & 
            \makebox[0.8cm][l]{0.237} \makebox[0.9cm][r]{(-53.8\%)} & 
            \cellcolor{lightgreen}\makebox[0.8cm][l]{0.597} \makebox[0.9cm][r]{(17.8\%)} & 
            \makebox[0.8cm][l]{0.371} \makebox[0.9cm][r]{(-6.3\%)} \\

            \textbf{Qwen-2.5} & 
            \makebox[0.8cm][l]{0.140} \makebox[0.9cm][r]{(-77.7\%)} & 
            \makebox[0.8cm][l]{0.271} \makebox[0.9cm][r]{(-47.2\%)} & 
            \cellcolor{lightgreen}\makebox[0.8cm][l]{0.557} \makebox[0.9cm][r]{(9.8\%)} & 
            \makebox[0.8cm][l]{0.229} \makebox[0.9cm][r]{(-42.0\%)} \\

            \bottomrule
        \end{tabular}
    }
\end{subtable}

   \vspace{0.1cm}

    \begin{subtable}[c]{\textwidth}
    \centering
    \caption{Developer-wise prediction}
    \label{tab:ac_results_dev}
    \resizebox{\columnwidth}{!}{%
        \begin{tabular}{l|c|c|c|c|c|c|c}
            \toprule
            \textbf{Model} &
            \textbf{\AU} &
            \textbf{\PBU} &
            \textbf{\ABU} &
            \textbf{\ABUFIF} &
            \textbf{\BD} &
            \textbf{\BDFIF} &
            \textbf{\RL} \\
            \hline

            \textbf{Baseline} &
            \makebox[0.8cm][l]{0.277{$^\ddagger$}} &
            \makebox[0.8cm][l]{0.573{$^\ddagger$}} &
            \makebox[0.8cm][l]{0.746{$^\dagger$}} &
            \makebox[0.8cm][l]{0.500{$^\ddagger$}} &
            \makebox[0.8cm][l]{0.501{$^\ddagger$}} &
            \makebox[0.8cm][l]{0.708{$^\dagger$}} &
            \makebox[0.8cm][l]{0.226{$^\ddagger$}} \\
            \hline

            \textbf{NB} &
            \makebox[0.8cm][l]{0.273} \makebox[0.9cm][r]{(-1.3\%)} &
            \makebox[0.8cm][l]{0.556} \makebox[0.9cm][r]{(-2.9\%)} &
            \makebox[0.8cm][l]{0.639} \makebox[0.9cm][r]{(-14.4\%)} &
            \makebox[0.8cm][l]{0.600} \makebox[0.9cm][r]{(\cellcolor{lightgreen}19.9\%)} &
            \makebox[0.8cm][l]{0.515} \makebox[0.9cm][r]{(\cellcolor{lightgreen}2.8\%)} &
            \makebox[0.8cm][l]{0.579} \makebox[0.9cm][r]{(-18.2\%)} &
            \makebox[0.8cm][l]{0.179} \makebox[0.9cm][r]{(-20.6\%)} \\

            \textbf{KNN} &
            \makebox[0.8cm][l]{0.278} \makebox[0.9cm][r]{(\cellcolor{lightgreen}0.3\%)} &
            \makebox[0.8cm][l]{0.537} \makebox[0.9cm][r]{(-6.3\%)} &
            \makebox[0.8cm][l]{0.631} \makebox[0.9cm][r]{(-15.4\%)} &
            \makebox[0.8cm][l]{0.559} \makebox[0.9cm][r]{(\cellcolor{lightgreen}11.7\%)} &
            \makebox[0.8cm][l]{0.502} \makebox[0.9cm][r]{(\cellcolor{lightgreen}0.4\%)} &
            \makebox[0.8cm][l]{0.567} \makebox[0.9cm][r]{(-19.9\%)} &
            \makebox[0.8cm][l]{0.199} \makebox[0.9cm][r]{(-12.2\%)} \\

            \textbf{LR} &
            \makebox[0.8cm][l]{0.324} \makebox[0.9cm][r]{(\cellcolor{lightgreen}17.0\%)} &
            \makebox[0.8cm][l]{0.593} \makebox[0.9cm][r]{(\cellcolor{lightgreen}3.5\%)} &
            \makebox[0.8cm][l]{0.710} \makebox[0.9cm][r]{(-4.8\%)} &
            \makebox[0.8cm][l]{0.614} \makebox[0.9cm][r]{(\cellcolor{lightgreen}22.8\%)} &
            \makebox[0.8cm][l]{0.540} \makebox[0.9cm][r]{(\cellcolor{lightgreen}7.8\%)} &
            \makebox[0.8cm][l]{0.572} \makebox[0.9cm][r]{(-19.2\%)} &
            \makebox[0.8cm][l]{0.178} \makebox[0.9cm][r]{(-21.2\%)} \\

            \textbf{MLP} &
            \makebox[0.8cm][l]{0.255} \makebox[0.9cm][r]{(-7.7\%)} &
            \makebox[0.8cm][l]{0.516} \makebox[0.9cm][r]{(-10.0\%)} &
            \makebox[0.8cm][l]{0.626} \makebox[0.9cm][r]{(-16.1\%)} &
            \makebox[0.8cm][l]{0.489} \makebox[0.9cm][r]{(-2.2\%)} &
            \makebox[0.8cm][l]{0.471} \makebox[0.9cm][r]{(-5.9\%)} &
            \makebox[0.8cm][l]{0.573} \makebox[0.9cm][r]{(-19.1\%)} &
            \makebox[0.8cm][l]{0.202} \makebox[0.9cm][r]{(-10.7\%)} \\

            \textbf{RF} &
            \makebox[0.8cm][l]{0.340} \makebox[0.9cm][r]{(\cellcolor{lightgreen}22.8\%)} &
            \makebox[0.8cm][l]{0.666} \makebox[0.9cm][r]{(\cellcolor{lightgreen}16.2\%)} &
            \makebox[0.8cm][l]{0.749} \makebox[0.9cm][r]{(\cellcolor{lightgreen}0.4\%)} &
            \makebox[0.8cm][l]{0.611} \makebox[0.9cm][r]{(\cellcolor{lightgreen}22.2\%)} &
            \makebox[0.8cm][l]{0.569} \makebox[0.9cm][r]{(\cellcolor{lightgreen}13.7\%)} &
            \makebox[0.8cm][l]{0.702} \makebox[0.9cm][r]{(-0.8\%)} &
            \makebox[0.8cm][l]{0.162} \makebox[0.9cm][r]{(-28.4\%)} \\

            \textbf{SVM} &
            \makebox[0.8cm][l]{0.304} \makebox[0.9cm][r]{(\cellcolor{lightgreen}10.0\%)} &
            \makebox[0.8cm][l]{0.584} \makebox[0.9cm][r]{(\cellcolor{lightgreen}2.0\%)} &
            \makebox[0.8cm][l]{0.710} \makebox[0.9cm][r]{(-4.8\%)} &
            \makebox[0.8cm][l]{0.589} \makebox[0.9cm][r]{(\cellcolor{lightgreen}17.8\%)} &
            \makebox[0.8cm][l]{0.528} \makebox[0.9cm][r]{(\cellcolor{lightgreen}5.5\%)} &
            \makebox[0.8cm][l]{0.565} \makebox[0.9cm][r]{(-20.1\%)} &
            \makebox[0.8cm][l]{0.165} \makebox[0.9cm][r]{(-27.2\%)} \\

            \textbf{XGBoost} &
            \makebox[0.8cm][l]{0.334} \makebox[0.9cm][r]{(\cellcolor{lightgreen}20.7\%)} &
            \makebox[0.8cm][l]{0.631} \makebox[0.9cm][r]{(\cellcolor{lightgreen}10.1\%)} &
            \makebox[0.8cm][l]{0.746} \makebox[0.9cm][r]{(\cellcolor{lightgreen}0.0\%)} &
            \makebox[0.8cm][l]{0.600} \makebox[0.9cm][r]{(\cellcolor{lightgreen}20.0\%)} &
            \makebox[0.8cm][l]{0.537} \makebox[0.9cm][r]{(\cellcolor{lightgreen}7.4\%)} &
            \makebox[0.8cm][l]{0.664} \makebox[0.9cm][r]{(-6.1\%)} &
            \makebox[0.8cm][l]{0.226} \makebox[0.9cm][r]{(\cellcolor{lightgreen}0.1\%)} \\
            
            \hline

            \textbf{CNN} &
             \makebox[0.8cm][l]{0.182} \makebox[0.9cm][r]{(-5.9\%)} &
             \makebox[0.8cm][l]{0.448} \makebox[0.9cm][r]{(-21.7\%)} &
             \makebox[0.8cm][l]{0.464} \makebox[0.9cm][r]{(-37.9\%)} &
             \makebox[0.8cm][l]{0.445} \makebox[0.9cm][r]{(-11.0\%)} &
             \makebox[0.8cm][l]{0.454} \makebox[0.9cm][r]{(-9.2\%)} &
             \makebox[0.8cm][l]{0.542} \makebox[0.9cm][r]{(-23.4\%)} &
             \makebox[0.8cm][l]{0.182} \makebox[0.9cm][r]{(-19.7\%)} \\

            \textbf{GPT-5.4} &
            \makebox[0.8cm][l]{0.159} \makebox[0.9cm][r]{(-42.6\%)} &
            \makebox[0.8cm][l]{0.563} \makebox[0.9cm][r]{(-1.7\%)} &
            \makebox[0.8cm][l]{0.430} \makebox[0.9cm][r]{(-42.3\%)} &
            \makebox[0.8cm][l]{0.342} \makebox[0.9cm][r]{(-31.7\%)} &
            \makebox[0.8cm][l]{0.511} \makebox[0.9cm][r]{(\cellcolor{lightgreen}2.1\%)} &
            \makebox[0.8cm][l]{0.713} \makebox[0.9cm][r]{(\cellcolor{lightgreen}0.7\%)} &
            \makebox[0.8cm][l]{0.220} \makebox[0.9cm][r]{(-2.6\%)} \\

            \textbf{Qwen-2.5} &
            \makebox[0.8cm][l]{0.153} \makebox[0.9cm][r]{(-44.8\%)} &
            \makebox[0.8cm][l]{0.616} \makebox[0.9cm][r]{(\cellcolor{lightgreen}7.5\%)} &
            \makebox[0.8cm][l]{0.496} \makebox[0.9cm][r]{(-33.5\%)} &
            \makebox[0.8cm][l]{0.343} \makebox[0.9cm][r]{(-31.5\%)} &
            \makebox[0.8cm][l]{0.515} \makebox[0.9cm][r]{(\cellcolor{lightgreen}2.9\%)} &
            \makebox[0.8cm][l]{0.711} \makebox[0.9cm][r]{(\cellcolor{lightgreen}0.5\%)} &
            \makebox[0.8cm][l]{0.177} \makebox[0.9cm][r]{(-21.7\%)} \\

            \bottomrule
        \end{tabular}
        
    }
    
    \vspace{0.3cm}
    \begin{minipage}{\columnwidth}
    	\footnotesize
    	\textit{Comprehensibility Metrics:} \textbf{\AU}, Actual Understandability;
    	\textbf{\PBU}, Perceived Binary Understandability;
    	\textbf{\ABU}, Actual Binary Understandability;
    	\textbf{\ABUFIF}, Actual Binary Understandability 50\%;
    	\textbf{\BD}, Binary Deceptiveness;
    	\textbf{\BDFIF}, Binary Deceptiveness 50\%; and
    	\textbf{\RL}, Readability Level.
    	
    	\smallskip
    	\smallskip
    	
    	\textit{Models:} \textbf{NB}, Naïve Bayes;
    	\textbf{KNN}, K-Nearest Neighbors;
    	\textbf{LR}, Logistic Regression;
    	\textbf{MLP}, Multilayer Perceptron;
    	\textbf{RF}, Random Forest;
    	\textbf{SVM}, Support Vector Machine;
    	\textbf{XGBoost}, Extreme Gradient Boosting;
    	\textbf{CNN}, Convolutional Neural Network; and
    	\textbf{Qwen-2.5}, Qwen2.5-Coder-32B-Instruct.
    	
    \end{minipage}
\end{subtable}

\end{table}

\subsubsection{Snippet-wise results.} 
\Cref{tab:ac_results_snippet} shows the model performance for the four comprehensibility metrics for which we were able to train and evaluate the models. As explained in \cref{subsec:ac_data_sources}, aggregating the binary metrics \ABU, \PBU, and \BDFIF snippet-wise resulted in extremely imbalanced data that prevented us from evaluating the models using cross validation; hence, they were excluded.

None of the metrics show consistent improvement across all models. \AU and \RL are the most challenging metrics to predict, with RI values ranging from -78.7\% to 6.8\%; only two models achieve a positive RI (RF, SVM for AU and LR, XGBoost for RL). In contrast, \BD is the easiest metric, with nine out of ten models outperforming the baseline (RI of 0.5\% to 33.4\%). \ABUFIF is moderately challenging, with three out of ten models showing improvement (RI of 2.2\% to 3.4\%).

LR and SVM demonstrate RI improvements of 2.2\% to 31.7\% for three out of four metrics, while RF, XGBoost and CNN show improvement for two metrics. The remaining models improve over the baseline for one metric.
\looseness=-1
All positive RI values correspond to small to medium effect sizes in both Cohen’s Kappa and MCC, except for LR on \RL and \ABUFIF;  XGBoost on \BD and \RL; RF on \AU; SVM, CNN on \ABUFIF and Qwen-2.5 on \BD. All these exceptions correspond to negligible effect sizes.

To assess the consistency of LLM predictions across iterations,  we computed Fleiss' kappa~\cite{Fleiss:1971measuring} over three independent runs. The agreement varied substantially across metrics: \BD achieved almost perfect agreement ($\kappa = 1.00$), \RL and \ABUFIF reached substantial agreement ($\kappa = 0.7978$ 
and $\kappa = 0.7432$, respectively), while \AU yielded poor agreement ($\kappa = -0.0135$), indicating near-chance consistency. These results suggest that LLM predictions are stable for \BD, \RL, and \ABUFIF, but highly inconsistent for \AU, which undermines the reliability of any observed performance differences on this metric.

\looseness=-1

\subsubsection{Developer-wise results.}

Table \ref{tab:ac_results_dev} shows the developer-wise AC prediction results.
\ABUFIF and \BD are the easiest metrics to predict, with the positive RI values ranging from 0.4\% to 22.8\%; six and eight out of ten models outperform the baselines for \ABUFIF and \BD metrics, respectively. In contrast, \ABU, \BDFIF, and \RL are the most challenging metrics, with almost all the models (8 for \ABU and \BDFIF and 9 for \RL out of 10 models) showing performance degradation compared to the baselines (RI from -0.8\% to -28.4\%). For the remaining metrics, only five models achieve better performance than the baseline.

The XGBoost model consistently demonstrates positive RI across all metrics except \BDFIF (RI from 0.01\% to 20.7\%) while other models achieve improvement on five or fewer metrics.
\looseness=-1
All positive RI values correspond to small effect sizes under both 
Cohen's Kappa and MCC, with the following exceptions, which 
correspond to negligible effect sizes: all models on \AU; all models 
except RF on \BD; Qwen-2.5 on \PBU (Cohen's Kappa only); 
XGBoost on \RL; and both LLMs on \BDFIF.

\subsubsection{Results analysis.}

RI distribution analysis via box plots (see our replication package~\cite{repl_pack}) suggests outliers have no impact on model performance averages; rather, the performance stems from the models’ ability (or inability) to learn from the data.
The performance trends across models and metrics align with the proportion of classifiers that outperform the baselines. In snippet-wise and developer-wise prediction, only 48\% and 50\% of classifiers outperform the baselines, respectively.
In order to assess the statistical significance of the RI values, we conducted Mann-whitney U test~\cite{mann-witney} with a significance level of 0.05.
The positive RI values for snippet-wise AC models are statistically significant across all metrics and models, except for \AU in RF, \ABUFIF in CNN, and \BD in XGBoost. Similarly, for dev-wise AC models, statistical significance holds across all metrics and models, except for \AU and \BD in KNN, and \ABU in RF and XGBoost. The full statistical results, including p-values, are available in our replication package~\cite{repl_pack}.

Comparing the RI of developer-wise \textit{vs.} snippet-wise AC models, we observe mixed trends. \BD shows consistent improvement in both snippet- and dev-wise settings across the board for every model except for two cases; both developer- and snippet-wise for MLP and developer-wise for CNN. 
No trend is observed between code understandability metrics and the readability metric (\RL) in both settings.

\looseness=-1

\begin{tcolorbox}[boxsep=5pt, bottom=1pt]
	\vspace{0.5em}
\textbf{\ref{rq:predict_ac} Findings:} Predictive models of different architectures (including class machine learning and deep learning models) do not reliably predict snippet- and developer-wise absolute comprehensibility values. Many of the models are unable to learn from the data and underperform na\"ive baselines. Our results are in line with Scalabrino \etal~\cite{Scalabrino:TSE19}, who concluded that absolute-prediction models are far from being practical.
\vspace{0.5em}
\end{tcolorbox}
\looseness=-1

\section{Evaluating Predictive Models of Relative Code Comprehensibility}
\label{sec:relative_code_comprehensibility}

As discussed in \Cref{sec:intro}, predicting an absolute
comprehensibility proxy requires a model to estimate a precise value for an
individual code snippet in isolation. This task may be difficult because human
comprehensibility measurements vary across participants and capture different
aspects of a complex cognitive process. The results of \ref{rq:predict_ac} are consistent with
this concern: AC models often fail to outperform the na\"ive baselines across
the evaluated proxies and model types.

We investigate relative comprehensibility (RC) as an alternative prediction
task. Instead of estimating an absolute proxy value for each snippet, an RC
model compares two snippets and predicts which one is more comprehensible or
whether they are similarly comprehensible. We hypothesize that this task is
more learnable because the model must identify the direction of the difference
between the snippets rather than estimate a precise value for each one
independently. 
To validate this conjecture, we answer the following RQ:
\vspace{0.1cm}
\begin{enumerate}[label=\textbf{RQ$_\arabic*$:}, ref=\textbf{RQ$_\arabic*$}, itemindent=0.5cm,leftmargin=0.5cm,resume]
  \item \label{rq:predict_rc}{\textit{How effective are learning-based models at predicting the relative comprehensibility between two snippets,
  		compared to na\"ive baselines?}}
\end{enumerate}
\vspace{0.1cm}

We built and evaluated models that predict RC  in both snippet-wise and developer-wise settings.  Given two code snippets, an RC model
is a ternary classifier: it can predict that one snippet is more comprehensible than the other, or
that the two snippets are equally comprehensible.
To answer \ref{rq:predict_rc}, we used the same data sources, features, comprehensibility metrics, models, evaluation metrics, baselines, and model training/evaluation approach used for answering \ref{rq:predict_ac} in \cref{sec:absolute_code_comprehensibility}. 
The key difference is that the dataset here consists of snippet pairs with relative comprehensibility measures.
Table~\ref{tab:rc_task_summary} summarizes the RC prediction task.
\looseness=-1

\begin{table}[t]
	\centering
	\caption{Overview of the relative comprehensibility (RC) prediction task and settings.}
	\label{tab:rc_task_summary}
	\resizebox{0.95\columnwidth}{!}{
	\begin{tabular}{m{1.5cm}|m{4cm}|m{7cm}|m{2.5cm}}
		\toprule
			\textbf{Setting} & \multicolumn{1}{c|}{\textbf{Model input}}                  & \multicolumn{1}{c|}{\textbf{Model output}}                                                                                                & \multicolumn{1}{c}{\textbf{Features}}                                 \\ 
			\hline
			Snippet-wise     & A pair of two code snippets                                & RC value (0, 1, or 2) based on aggregating a comprehensibility proxy (e.g., PBU) across developers. Aggregation method: average           & Code features for each snippet (concatenated)                         \\ 
			\hline
			Developer-wise   & A pair of two code snippets (judged by the same developer) & RC value (0, 1, or 2) based on the comprehensibility proxy values (e.g., PBU values) obtained from a specific developer for both snippets & Code features for each snippet and developer features (concatenated)  \\
			\bottomrule
	\end{tabular}
}
\end{table}

\subsection{Dataset Construction and RC Definition}
\label{subsec:rc_datasets}

We created ordered snippet pairs from the 50 and 100 snippets in \dssix and \dsthree, respectively, resulting in 2,500 pairs for \dssix and 10,000 pairs for \dsthree.
We concatenate the code features of each snippet of a pair $(c_1, c_2)$ into a single vector of features as the input to a model.
This dataset was used for \textbf{snippet-wise} comprehensibility prediction.

For \textbf{developer-wise} prediction, we generated ordered triplets $(c_1, c_2, \text{\textit{p}})$, where each snippet pair was understood by the same participant $p$ in the original studies of \dssix and \dsthree. A triplet is represented by concatenating the code features of 
$c_1$ and $c_2$ followed by $p$'s developer features. There are 3,323 triplets for \dssix and 1.21 million for \dsthree.
\looseness=-1

To prevent data leakage across training and test sets, we performed
cross-validation at the snippet level rather than at the pair level. For each
outer fold, we first partitioned the code snippets into disjoint
training and test sets. We then constructed the training pairs exclusively
from snippets in the training partition and the test pairs exclusively from
snippets in the test partition. Consequently, no snippet appearing in a test
pair appeared in any training pair. The same snippet-level partitioning procedure was applied within the inner
cross-validation folds used for hyperparameter selection. All data-dependent preprocessing steps,
including feature selection, normalization, SMOTE, and hyperparameter tuning,
were performed using only the training partition within each outer fold. 

RC is a categorical metric with three possible values.
A snippet pair's RC is derived from the metrics from \cref{subsec:ac_proxies}, based on the human evaluations.
For developer-wise prediction, RC is defined using participant-specific measurements.
For snippet-wise prediction, we aggregated individual measurements by averaging them.
For instance, in Scalabrino \etal's study~\cite{Scalabrino:TSE19}, each \dssix snippet received eight or nine binary understandability (\PBU) ratings, which we averaged to obtain a single, aggregated comprehensibility score.

\begin{table}[t]
	\centering
	\caption{Class distribution for relative comprehensibility.}
	\label{tab:rc_class_distribution}
	\resizebox{0.6\columnwidth}{!}{
		\begin{tabular}{l|rrrr|rrrr}
			\toprule
			\multicolumn{1}{c|}{\multirow{2}{*}{\textbf{Metric}}} & \multicolumn{4}{c}{\textbf{Snippet-wise}}                                               & \multicolumn{4}{c}{\textbf{Developer-wise}}                                            \\
			\multicolumn{1}{c|}{}                        & \multicolumn{1}{c}{\textbf{0}} & \multicolumn{1}{c}{\textbf{1}} & \multicolumn{1}{c}{\textbf{2}} & \textbf{Total}  & \multicolumn{1}{c}{\textbf{0}} & \multicolumn{1}{c}{\textbf{1}} & \multicolumn{1}{c}{\textbf{2}} & \textbf{Total} \\
			\hline
			\AU                                         & 46.7\%                & 46.7\%                & 6.6\%                 & 2,500  & 29.4\%                & 29.4\%                & 41.2\%                & 3,323 \\
			\PBU                                        & 43.6\%                & 43.6\%                & 12.8\%                & 2,500  & 16.9\%                & 16.9\%                & 66.2\%                & 3,323 \\
			\ABU                                        & 39.9\%                & 39.9\%                & 20.2\%                & 2,500  & 12.0\%                & 12.0\%                & 76.0\%                & 3,323 \\
			\ABUFIF                                     & 44.3\%                & 44.3\%                & 11.4\%                & 2,500  & 19.2\%                & 19.2\%                & 61.6\%                & 3,323 \\
			\BD                                         & 42.3\%                & 42.3\%                & 15.4\%                & 2,500  & 20.3\%                & 20.3\%                & 59.3\%                & 3,323 \\
			\BDFIF                                      & 41.4\%                & 41.4\%                & 17.1\%                & 2,500  & 12.8\%                & 12.8\%                & 74.4\%                & 3,323 \\
			\hline
			\RL                          				& 49.3\%                & 49.3\%                & 1.4\%                 & 10,000 & 36.4\%                & 36.4\%                & 27.1\%                & 1.21M \\
			\bottomrule
	\end{tabular}
	}
\end{table}

More precisely, let $S_1$ and $S_2$ be the individual or aggregated comprehensibility of snippets $c_1$ and $c_2$, respectively. $RC(c_1,c_2)$ is defined as follows: 
\begin{equation*}
	RC(c_1,c_2) = 
	\begin{cases} 
		0 & \text{if } S_1 > S_2 \quad (c_1 \text{ is more understandable}) \\
		1 & \text{if } S_2 > S_1 \quad (c_2 \text{ is more understandable})\\
		2 & \text{if } S_1 = S_2 \quad (\text{both are equally  understandable})
	\end{cases}
\end{equation*}

For the \BD and \BDFIF metrics, where lower values indicate higher comprehensibility, we invert the comparison signs ($>$ to $<$).
For each comprehensibility metric, we computed the relative comprehensibility (RC) for all snippet pairs from the respective data source. \Cref{tab:rc_class_distribution} shows the class distribution for each metric.

The RC dataset was normalized, pre-processed, and split for model training and evaluation %
following the same methodology used for \ref{rq:predict_ac} (see \Cref{subsec:data_normalization_balancing,subsec:model_training_evaluation,subsec:metrics}).
\looseness=-1

\begin{figure*}[t]
    \centering %
    
    \begin{minipage}[c]{0.31\textwidth}
        \begin{subfigure}{\textwidth}
            \centering
            \includegraphics[width=\textwidth, height=5cm, keepaspectratio]{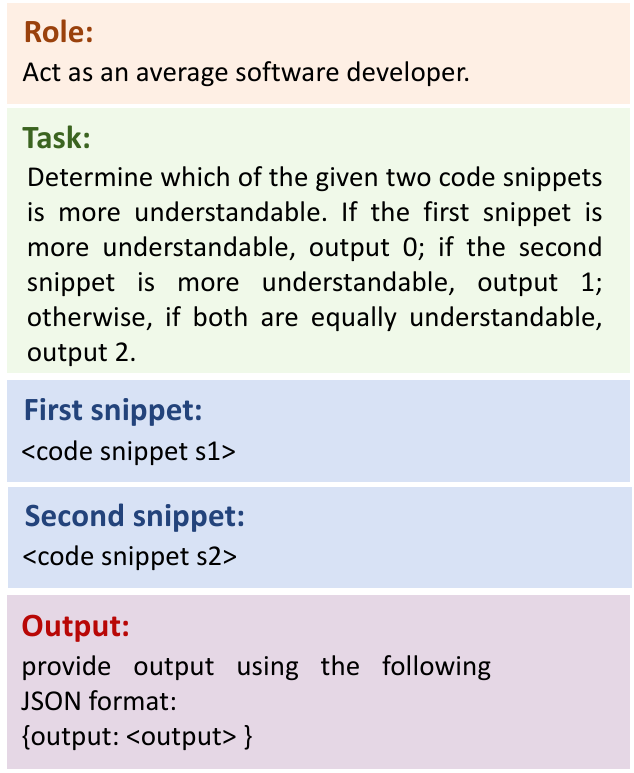}
            \caption{RC snippet-wise prompt}
            \label{fig:side_left}
        \end{subfigure}
    \end{minipage}%
    \hspace{0.02\textwidth} 
    \begin{minipage}[c]{0.31\textwidth}
        \begin{subfigure}{\textwidth}
            \centering
            \includegraphics[width=\textwidth, height=5cm, keepaspectratio]{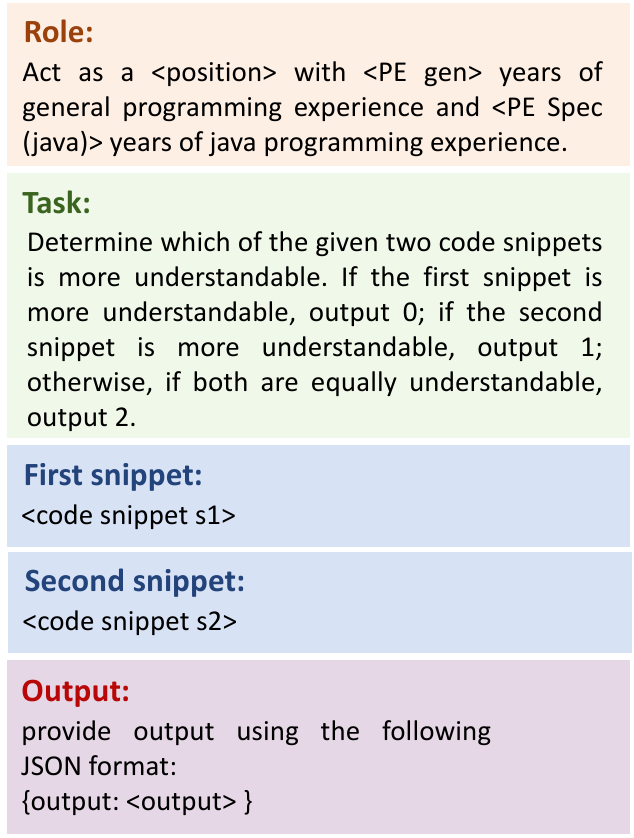}
            \caption{RC developer-wise prompt for \dssix}
            \label{fig:side_middle}
        \end{subfigure}
    \end{minipage}%
    \hspace{0.02\textwidth} 
    \begin{minipage}[c]{0.31\textwidth}
        \begin{subfigure}{\textwidth}
            \centering
            \includegraphics[width=\textwidth, height=5cm, keepaspectratio]{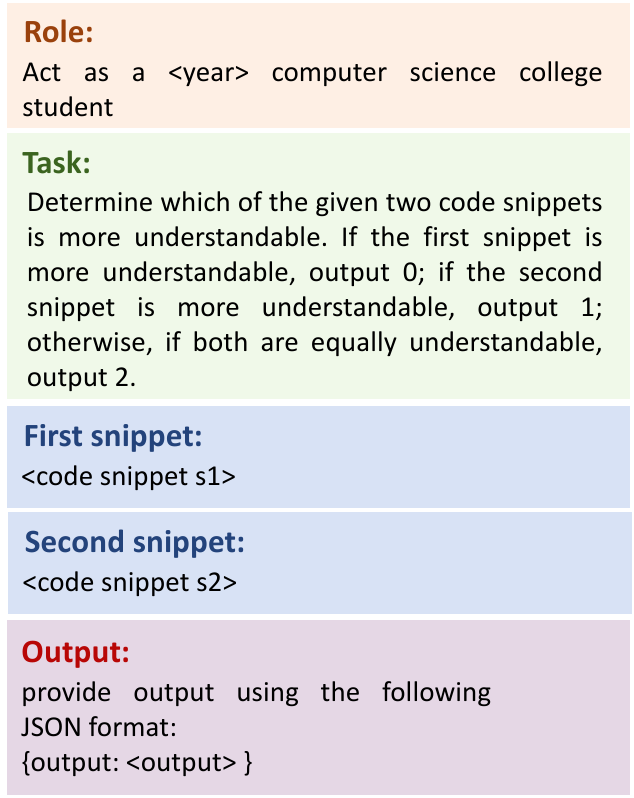} 
            \caption{RC developer-wise prompt for \dsthree}
            \label{fig:side_right}
        \end{subfigure}
    \end{minipage}

    \caption{Prompt templates for snippet- and developer-wise RC prediction. Content inside the <brackets> are placeholders.}
    \label{fig:rc_prompts}
\end{figure*}

\textbf{Prompt Design and Execution.}
We created and executed 12,500 prompts (2,500 for \dssix and 10,000 for \dsthree) for snippet-wise RC prediction, and 43,097 prompts (3,097 for \dssix and 40,000 for \dsthree) for developer-wise RC prediction. \Cref{fig:rc_prompts} presents the prompt templates used for each setting. In the snippet-wise templates, the LLM was instructed to  adopt the perspective of an average software engineer, whereas in the developer-wise templates, it was instructed to assume the role of a specific software engineer with a given background. In both 
settings, the LLM is expected to output one of three values 0, 1, or 2, corresponding to the RC definitions described above. As in AC prediction, we executed each prompt three times and averaged the performance metrics across the three executions to measure overall model performance.

\subsection{\ref{rq:predict_rc}: Relative Comprehensibility Results}
We trained 1,100 
and 1,204
classifiers that predict snippet- and developer-wise relative comprehensibility, respectively. 
These classifiers were developed under different configurations: seven ML model types, three DL model types including two state-of-the-art LLMs, ten feature sets, RC metrics defined for the seven individual comprehensibility metrics, and different sets of best hyperparameters found during cross-validation for each metric. 

\begin{table*}[t]
    \caption{Relative comprehensibility (RC) results. Each column reports [$\pmb{wF1}$, (RI)]. $\pmb{wF1}$: average weighted F1 across all trained models and three executions of the LLMs. RI: average relative improvement over the best baseline, either a {$^\dagger$}Random ({RB}) or a Majority Class ({$^\ddagger$}\MB or {$^\star$}\MBT) classifier. \hlgreen{Green}: positive RI.}
    \label{tab:rc_results}
    \centering
    \setlength{\tabcolsep}{5pt}

    \begin{subtable}{\textwidth}
    \centering
    \caption{Snippet-wise prediction}
    \label{tab:rc_results_snippet}
    \resizebox{\columnwidth}{!}{%
        \begin{tabular}{l|c|c|c|c|c|c|c}
            \toprule
            \textbf{Model} &
            \textbf{\AU} &
            \textbf{\PBU} &
            \textbf{\ABU} &
            \textbf{\ABUFIF} &
            \textbf{\BD} &
            \textbf{\BDFIF} &
            \textbf{\RL} \\
            \hline

            \textbf{Baseline} &
            \makebox[0.8cm][l]{0.440{$^\ddagger$}}  &
            \makebox[0.8cm][l]{0.396{$^\ddagger$}}  &
            \makebox[0.8cm][l]{0.359{$^\dagger$}}  &
            \makebox[0.8cm][l]{0.406{$^\ddagger$}}  &
            \makebox[0.8cm][l]{0.382{$^\ddagger$}}  &
            \makebox[0.8cm][l]{0.373{$^\dagger$}}  &
            \makebox[0.8cm][l]{0.487{$^\ddagger$}}  \\
            \hline
            
            \textbf{NB} &
            \makebox[0.8cm][l]{0.501} \makebox[0.9cm][r]{\cellcolor{lightgreen}(13.7\%)} &
            \makebox[0.8cm][l]{0.585} \makebox[0.9cm][r]{\cellcolor{lightgreen}(47.6\%)} &
            \makebox[0.8cm][l]{0.553} \makebox[0.9cm][r]{\cellcolor{lightgreen}(53.9\%)} &
            \makebox[0.8cm][l]{0.541} \makebox[0.9cm][r]{\cellcolor{lightgreen}(33.4\%)} &
            \makebox[0.8cm][l]{0.580} \makebox[0.9cm][r]{\cellcolor{lightgreen}(51.9\%)} &
            \makebox[0.8cm][l]{0.558} \makebox[0.9cm][r]{\cellcolor{lightgreen}(49.6\%)} &
            \makebox[0.8cm][l]{0.453} \makebox[0.9cm][r]{-7.0\%} \\

            \textbf{KNN} &
            \makebox[0.8cm][l]{0.655} \makebox[0.9cm][r]{\cellcolor{lightgreen}(48.9\%)} &
            \makebox[0.8cm][l]{0.582} \makebox[0.9cm][r]{\cellcolor{lightgreen}(47.0\%)} &
            \makebox[0.8cm][l]{0.584} \makebox[0.9cm][r]{\cellcolor{lightgreen}(62.5\%)} &
            \makebox[0.8cm][l]{0.557} \makebox[0.9cm][r]{\cellcolor{lightgreen}(37.3\%)} &
            \makebox[0.8cm][l]{0.591} \makebox[0.9cm][r]{\cellcolor{lightgreen}(54.8\%)} &
            \makebox[0.8cm][l]{0.501} \makebox[0.9cm][r]{\cellcolor{lightgreen}(34.3\%)} &
            \makebox[0.8cm][l]{0.693} \makebox[0.9cm][r]{\cellcolor{lightgreen}(42.4\%)} \\

            \textbf{LR} &
            \makebox[0.8cm][l]{0.696} \makebox[0.9cm][r]{\cellcolor{lightgreen}(58.0\%)} &
            \makebox[0.8cm][l]{0.712} \makebox[0.9cm][r]{\cellcolor{lightgreen}(79.7\%)} &
            \makebox[0.8cm][l]{0.670} \makebox[0.9cm][r]{\cellcolor{lightgreen}(86.5\%)} &
            \makebox[0.8cm][l]{0.690} \makebox[0.9cm][r]{\cellcolor{lightgreen}(70.1\%)} &
            \makebox[0.8cm][l]{0.701} \makebox[0.9cm][r]{\cellcolor{lightgreen}(83.5\%)} &
            \makebox[0.8cm][l]{0.653} \makebox[0.9cm][r]{\cellcolor{lightgreen}(75.3\%)} &
            \makebox[0.8cm][l]{0.666} \makebox[0.9cm][r]{\cellcolor{lightgreen}(36.8\%)} \\

            \textbf{MLP} &
            \makebox[0.8cm][l]{0.729} \makebox[0.9cm][r]{\cellcolor{lightgreen}(65.7\%)} &
            \makebox[0.8cm][l]{0.698} \makebox[0.9cm][r]{\cellcolor{lightgreen}(76.2\%)} &
            \makebox[0.8cm][l]{0.636} \makebox[0.9cm][r]{\cellcolor{lightgreen}(77.0\%)} &
            \makebox[0.8cm][l]{0.713} \makebox[0.9cm][r]{\cellcolor{lightgreen}(75.6\%)} &
            \makebox[0.8cm][l]{0.718} \makebox[0.9cm][r]{\cellcolor{lightgreen}(88.0\%)} &
            \makebox[0.8cm][l]{0.622} \makebox[0.9cm][r]{\cellcolor{lightgreen}(66.8\%)} &
            \makebox[0.8cm][l]{0.831} \makebox[0.9cm][r]{\cellcolor{lightgreen}(70.7\%)} \\

            \textbf{RF} &
            \makebox[0.8cm][l]{0.841} \makebox[0.9cm][r]{\cellcolor{lightgreen}(91.2\%)} &
            \makebox[0.8cm][l]{0.858} \makebox[0.9cm][r]{\cellcolor{lightgreen}(116.7\%)} &
            \makebox[0.8cm][l]{0.848} \makebox[0.9cm][r]{\cellcolor{lightgreen}(135.9\%)} &
            \makebox[0.8cm][l]{0.867} \makebox[0.9cm][r]{\cellcolor{lightgreen}(113.7\%)} &
            \makebox[0.8cm][l]{0.908} \makebox[0.9cm][r]{\cellcolor{lightgreen}(137.8\%)} &
            \makebox[0.8cm][l]{0.863} \makebox[0.9cm][r]{\cellcolor{lightgreen}(131.5\%)} &
            \makebox[0.8cm][l]{0.804} \makebox[0.9cm][r]{\cellcolor{lightgreen}(65.2\%)} \\

            \textbf{SVM} &
            \makebox[0.8cm][l]{0.774} \makebox[0.9cm][r]{\cellcolor{lightgreen}(75.9\%)} &
            \makebox[0.8cm][l]{0.769} \makebox[0.9cm][r]{\cellcolor{lightgreen}(94.2\%)} &
            \makebox[0.8cm][l]{0.766} \makebox[0.9cm][r]{\cellcolor{lightgreen}(113.2\%)} &
            \makebox[0.8cm][l]{0.792} \makebox[0.9cm][r]{\cellcolor{lightgreen}(95.2\%)} &
            \makebox[0.8cm][l]{0.775} \makebox[0.9cm][r]{\cellcolor{lightgreen}(102.9\%)} &
            \makebox[0.8cm][l]{0.758} \makebox[0.9cm][r]{\cellcolor{lightgreen}(103.3\%)} &
            \makebox[0.8cm][l]{0.690} \makebox[0.9cm][r]{\cellcolor{lightgreen}(41.8\%)} \\

            \textbf{XGBoost} &
            \makebox[0.8cm][l]{0.886} \makebox[0.9cm][r]{\cellcolor{lightgreen}(101.3\%)} &
            \makebox[0.8cm][l]{0.859} \makebox[0.9cm][r]{\cellcolor{lightgreen}(116.8\%)} &
            \makebox[0.8cm][l]{0.811} \makebox[0.9cm][r]{\cellcolor{lightgreen}(125.7\%)} &
            \makebox[0.8cm][l]{0.876} \makebox[0.9cm][r]{\cellcolor{lightgreen}(115.8\%)} &
            \makebox[0.8cm][l]{0.914} \makebox[0.9cm][r]{\cellcolor{lightgreen}(139.4\%)} &
            \makebox[0.8cm][l]{0.829} \makebox[0.9cm][r]{\cellcolor{lightgreen}(103.3\%)} &
            \makebox[0.8cm][l]{0.897} \makebox[0.9cm][r]{\cellcolor{lightgreen}(84.2\%)} \\

            \hline
            \textbf{CNN} &
            \makebox[0.8cm][l]{0.828} \makebox[0.9cm][r]{\cellcolor{lightgreen}(88.0\%)} &
            \makebox[0.8cm][l]{0.847} \makebox[0.9cm][r]{\cellcolor{lightgreen}(113.9\%)} &
            \makebox[0.8cm][l]{0.934} \makebox[0.9cm][r]{\cellcolor{lightgreen}{(159.8\%)}} &
            \makebox[0.8cm][l]{0.852} \makebox[0.9cm][r]{\cellcolor{lightgreen}(110.1\%)} &
            \makebox[0.8cm][l]{0.880} \makebox[0.9cm][r]{\cellcolor{lightgreen}(130.6\%)} &
            \makebox[0.8cm][l]{0.898} \makebox[0.9cm][r]{\cellcolor{lightgreen}(140.9\%)} &
            \makebox[0.8cm][l]{0.897} \makebox[0.9cm][r]{\cellcolor{lightgreen}(84.2\%)} \\

            \textbf{GPT-5.4} &
            \makebox[0.8cm][l]{0.558} \makebox[0.9cm][r]{\cellcolor{lightgreen}(26.6\%)} &
            \makebox[0.8cm][l]{0.564} \makebox[0.9cm][r]{\cellcolor{lightgreen}(42.4\%)} &
            \makebox[0.8cm][l]{0.380} \makebox[0.9cm][r]{\cellcolor{lightgreen}(5.8\%)} &
            \makebox[0.8cm][l]{0.534} \makebox[0.9cm][r]{\cellcolor{lightgreen}(31.5\%)} &
            \makebox[0.8cm][l]{0.302} \makebox[0.9cm][r]{(-20.9\%)} &
            \makebox[0.8cm][l]{0.429} \makebox[0.9cm][r]{\cellcolor{lightgreen}(15.0\%)} &
            \makebox[0.8cm][l]{0.511} \makebox[0.9cm][r]{\cellcolor{lightgreen}(5.0\%)} \\

            \textbf{Qwen-2.5} &
            \makebox[0.8cm][l]{0.494} \makebox[0.9cm][r]{\cellcolor{lightgreen}(12.3\%)} &
            \makebox[0.8cm][l]{0.489} \makebox[0.9cm][r]{\cellcolor{lightgreen}(23.5\%)} &
            \makebox[0.8cm][l]{0.365} \makebox[0.9cm][r]{\cellcolor{lightgreen}(1.6\%)} &
            \makebox[0.8cm][l]{0.466} \makebox[0.9cm][r]{\cellcolor{lightgreen}(14.8\%)} &
            \makebox[0.8cm][l]{0.293} \makebox[0.9cm][r]{(-23.2\%)} &
            \makebox[0.8cm][l]{0.401} \makebox[0.9cm][r]{\cellcolor{lightgreen}(7.5\%)} &
            \makebox[0.8cm][l]{0.436} \makebox[0.9cm][r]{(-10.4\%)} \\

            \bottomrule
        \end{tabular}
    }
\end{subtable}

    \vspace{0.1cm} %

    \begin{subtable}{\textwidth}
    \centering
    \caption{Developer-wise prediction}
    \label{tab:rc_results_dev}
    \resizebox{\columnwidth}{!}{%
    \begin{threeparttable}[b]
    \begin{tabular}{l|c|c|c|c|c|c|c}
        \toprule
        \textbf{Model} &
        \textbf{\AU} &
        \textbf{\PBU} &
        \textbf{\ABU} &
        \textbf{\ABUFIF} &
        \textbf{\BD} &
        \textbf{\BDFIF} &
        \textbf{\RL} \\
        \hline

        \textbf{Baseline} &
        \makebox[0.8cm][l]{0.343{$^\ddagger$}}  &
        \makebox[0.8cm][l]{0.528{$^\star$}}  &
        \makebox[0.8cm][l]{0.656{$^\star$}}  &
        \makebox[0.8cm][l]{0.407{$^\star$}}  &
        \makebox[0.8cm][l]{0.442{$^\star$}}  &
        \makebox[0.8cm][l]{0.635{$^\star$}}  &
        \makebox[0.8cm][l]{0.339{$^\ddagger$}}  \\
        \hline

        \textbf{NB} &
        \makebox[0.8cm][l]{0.378} \makebox[0.9cm][r]{(\cellcolor{lightgreen}10.4\%)} &
        \makebox[0.8cm][l]{0.328} \makebox[0.9cm][r]{(-37.8\%)} &
        \makebox[0.8cm][l]{0.483} \makebox[0.9cm][r]{(-26.4\%)} &
        \makebox[0.8cm][l]{0.354} \makebox[0.9cm][r]{(-24.6\%)} &
        \makebox[0.8cm][l]{0.337} \makebox[0.9cm][r]{(-23.6\%)} &
        \makebox[0.8cm][l]{0.451} \makebox[0.9cm][r]{(-28.9\%)} &
        \makebox[0.8cm][l]{0.411} \makebox[0.9cm][r]{(\cellcolor{lightgreen}21.1\%)} \\

        \textbf{KNN} &
        \makebox[0.8cm][l]{0.518} \makebox[0.9cm][r]{(\cellcolor{lightgreen}51.2\%)} &
        \makebox[0.8cm][l]{0.573} \makebox[0.9cm][r]{(\cellcolor{lightgreen}8.6\%)} &
        \makebox[0.8cm][l]{0.645} \makebox[0.9cm][r]{(-1.7\%)} &
        \makebox[0.8cm][l]{0.560} \makebox[0.9cm][r]{(\cellcolor{lightgreen}19.3\%)} &
        \makebox[0.8cm][l]{0.493} \makebox[0.9cm][r]{(\cellcolor{lightgreen}11.6\%)} &
        \makebox[0.8cm][l]{0.601} \makebox[0.9cm][r]{(-5.4\%)} &
        \makebox[0.8cm][l]{0.431} \makebox[0.9cm][r]{(\cellcolor{lightgreen}27.1\%)} \\

        \textbf{LR} &
        \makebox[0.8cm][l]{0.432} \makebox[0.9cm][r]{(\cellcolor{lightgreen}26.0\%)} &
        \makebox[0.8cm][l]{0.474} \makebox[0.9cm][r]{(-10.3\%)} &
        \makebox[0.8cm][l]{0.593} \makebox[0.9cm][r]{(-9.7\%)} &
        \makebox[0.8cm][l]{0.426} \makebox[0.9cm][r]{(-9.2\%)} &
        \makebox[0.8cm][l]{0.379} \makebox[0.9cm][r]{(-14.1\%)} &
        \makebox[0.8cm][l]{0.529} \makebox[0.9cm][r]{(-16.7\%)} &
        \makebox[0.8cm][l]{0.450} \makebox[0.9cm][r]{(\cellcolor{lightgreen}32.6\%)} \\

        \textbf{MLP} &
        \makebox[0.8cm][l]{0.478} \makebox[0.9cm][r]{(\cellcolor{lightgreen}39.5\%)} &
        \makebox[0.8cm][l]{0.563} \makebox[0.9cm][r]{(\cellcolor{lightgreen}6.7\%)} &
        \makebox[0.8cm][l]{0.640} \makebox[0.9cm][r]{(-2.4\%)} &
        \makebox[0.8cm][l]{0.524} \makebox[0.9cm][r]{(\cellcolor{lightgreen}11.6\%)} &
        \makebox[0.8cm][l]{0.460} \makebox[0.9cm][r]{(\cellcolor{lightgreen}4.1\%)} &
        \makebox[0.8cm][l]{0.605} \makebox[0.9cm][r]{(-4.8\%)} &
        \makebox[0.8cm][l]{0.545} \makebox[0.9cm][r]{(\cellcolor{lightgreen}60.8\%)} \\

        \textbf{RF} &
        \makebox[0.8cm][l]{0.598} \makebox[0.9cm][r]{(\cellcolor{lightgreen}74.7\%)} &
        \makebox[0.8cm][l]{0.710} \makebox[0.9cm][r]{(\cellcolor{lightgreen}34.5\%)} &
        \makebox[0.8cm][l]{0.747} \makebox[0.9cm][r]{(\cellcolor{lightgreen}13.8\%)} &
        \makebox[0.8cm][l]{0.649} \makebox[0.9cm][r]{(\cellcolor{lightgreen}38.3\%)} &
        \makebox[0.8cm][l]{0.614} \makebox[0.9cm][r]{(\cellcolor{lightgreen}39.1\%)} &
        \makebox[0.8cm][l]{0.737} \makebox[0.9cm][r]{(\cellcolor{lightgreen}16.0\%)} &
        \makebox[0.8cm][l]{0.543} \makebox[0.9cm][r]{(\cellcolor{lightgreen}60.2\%)} \\

        \textbf{SVM} &
        \makebox[0.8cm][l]{0.427} \makebox[0.9cm][r]{(\cellcolor{lightgreen}24.5\%)} &
        \makebox[0.8cm][l]{0.463} \makebox[0.9cm][r]{(-12.3\%)} &
        \makebox[0.8cm][l]{0.587} \makebox[0.9cm][r]{(-10.5\%)} &
        \makebox[0.8cm][l]{0.420} \makebox[0.9cm][r]{(-10.7\%)} &
        \makebox[0.8cm][l]{0.376} \makebox[0.9cm][r]{(-14.9\%)} &
        \makebox[0.8cm][l]{0.530} \makebox[0.9cm][r]{(-16.6\%)} &
        \makebox[0.8cm][l]{0.446} \makebox[0.9cm][r]{(\cellcolor{lightgreen}31.4\%)} \\

        \textbf{XGBoost} &
        \makebox[0.8cm][l]{0.577} \makebox[0.9cm][r]{(\cellcolor{lightgreen}68.5\%)} &
        \makebox[0.8cm][l]{0.671} \makebox[0.9cm][r]{(\cellcolor{lightgreen}27.1\%)} &
        \makebox[0.8cm][l]{0.761} \makebox[0.9cm][r]{(\cellcolor{lightgreen}15.9\%)} &
        \makebox[0.8cm][l]{0.642} \makebox[0.9cm][r]{(\cellcolor{lightgreen}36.6\%)} &
        \makebox[0.8cm][l]{0.581} \makebox[0.9cm][r]{(\cellcolor{lightgreen}31.6\%)} &
        \makebox[0.8cm][l]{0.720} \makebox[0.9cm][r]{(\cellcolor{lightgreen}13.4\%)} &
        \makebox[0.8cm][l]{0.531} \makebox[0.9cm][r]{(\cellcolor{lightgreen}56.7\%)} \\

        \hline
        \textbf{CNN} &
        \makebox[0.8cm][l]{0.430}\makebox[0.9cm][r]{\cellcolor{lightgreen}(25.5\%)} &
        \makebox[0.8cm][l]{0.594} \makebox[0.9cm][r]{\cellcolor{lightgreen}(12.5\%)} &
        \makebox[0.8cm][l]{0.694} \makebox[0.9cm][r]{\cellcolor{lightgreen}(5.8\%)} &
        \makebox[0.8cm][l]{0.587} \makebox[0.9cm][r]{\cellcolor{lightgreen}(25.0\%)} &
        \makebox[0.8cm][l]{0.523} \makebox[0.9cm][r]{\cellcolor{lightgreen}(18.4\%)} &
        \makebox[0.8cm][l]{0.668} \makebox[0.9cm][r]{\cellcolor{lightgreen}(5.1\%)} &
        \makebox[0.8cm][l]{0.532} \makebox[0.9cm][r]{\cellcolor{lightgreen}(56.9\%)} \\

        \textbf{GPT-5.4} &
        \makebox[0.8cm][l]{0.465} \makebox[0.9cm][r]{\cellcolor{lightgreen}(35.7\%)} &
        \makebox[0.8cm][l]{0.334} \makebox[0.9cm][r]{(-36.7\%)} &
        \makebox[0.8cm][l]{0.276} \makebox[0.9cm][r]{(-57.9\%)} &
        \makebox[0.8cm][l]{0.359} \makebox[0.9cm][r]{(-23.6\%)} &
        \makebox[0.8cm][l]{0.321} \makebox[0.9cm][r]{(-27.3\%)} &
        \makebox[0.8cm][l]{0.282} \makebox[0.9cm][r]{(-55.6\%)} &
        \makebox[0.8cm][l]{0.329} \makebox[0.9cm][r]{(-3.0\%)} \\

        \textbf{Qwen-2.5} &
        \makebox[0.8cm][l]{0.454} (\cellcolor{lightgreen}32.6\%) &
        \makebox[0.8cm][l]{0.374} \makebox[0.9cm][r]{(-29.2\%)} &
        \makebox[0.8cm][l]{0.334} \makebox[0.9cm][r]{(-49.1\%)} &
        \makebox[0.8cm][l]{0.385} \makebox[0.9cm][r]{(-17.9\%)} &
        \makebox[0.8cm][l]{0.347} \makebox[0.9cm][r]{(-21.3\%)} &
        \makebox[0.8cm][l]{0.326} \makebox[0.9cm][r]{(-48.7\%)} &
        \makebox[0.8cm][l]{0.337} \makebox[0.9cm][r]{(-0.8\%)} \\

        \bottomrule
    \end{tabular}
    \end{threeparttable}
    }

    \vspace{0.3cm}
    \begin{minipage}{\columnwidth}
    	\footnotesize
    	\textit{Comprehensibility Metrics:} \textbf{\AU}, Actual Understandability;
    	\textbf{\PBU}, Perceived Binary Understandability;
    	\textbf{\ABU}, Actual Binary Understandability;
    	\textbf{\ABUFIF}, Actual Binary Understandability 50\%;
    	\textbf{\BD}, Binary Deceptiveness;
    	\textbf{\BDFIF}, Binary Deceptiveness 50\%; and
    	\textbf{\RL}, Readability Level.
    	
    	\smallskip
    	\smallskip
    	
    	\textit{Models:} \textbf{NB}, Naïve Bayes;
    	\textbf{KNN}, K-Nearest Neighbors;
    	\textbf{LR}, Logistic Regression;
    	\textbf{MLP}, Multilayer Perceptron;
    	\textbf{RF}, Random Forest;
    	\textbf{SVM}, Support Vector Machine;
    	\textbf{XGBoost}, Extreme Gradient Boosting;
    	\textbf{CNN}, Convolutional Neural Network; and
    	\textbf{Qwen-2.5}, Qwen2.5-Coder-32B-Instruct.
    	
    \end{minipage}
\end{subtable}

\end{table*}

\subsubsection{Snippet-wise results.} \Cref{tab:rc_results_snippet} presents the model performance for snippet-wise prediction, revealing a clear trend: virtually all the models outperform the baselines across all metrics, 
with improvements RI ranging from 1.6\% to 159.8\%.
One exception we see in the results is the Na\"ive Bayes (NB) classifier for the RL metric, which applies Bayes’ theorem assuming each feature is statistically independent; given that the \dsthree snippets are quite short (5 - 13 NCLOC), the model may lack enough features to statistically learn patterns between them and RL RC values.

Another notable exception is that both LLMs underperform the strongest baseline for \BD. When prompted to determine which snippet is more understandable, the LLMs may rely primarily on surface-level syntax and structural clarity, which more closely resemble perceived understandability (\PBU). In contrast, \BD identifies deceptive snippets: code that participants initially perceive as understandable (\PBU = 1) but subsequently fail to understand correctly (\ABU = 0).
We hypothesized that the LLMs rely on the same cues that drive perceived understandability rather than those required to identify deceptiveness. If this is the case, they should perform better on the relaxed deceptiveness metric (\BDFIF), which labels fewer snippets as deceptive.
To investigate this hypothesis, we analyzed all snippet pairs for which the expected RC labels based on \PBU and \BD/\BDFIF disagree. There are two types of conflicting pairs:
\begin{enumerate}
	
	\item participants perceive Snippet~1 as more understandable than Snippet 2 ($RC_{\PBU}=0$), but Snippet~1 is actually deceptive, making Snippet~2 the correct choice according to \BD ($RC_{\BD}=1$); and
	
	\item  the opposite case, where $RC_{\PBU}=1$ but $RC_{\BD}=0$: Snippet 2 was perceived as more understandable, but it was actually deceptive in favor of Snippet 1.
\end{enumerate}
For \BD, there are 731 pairs of each type (1,462 conflicting pairs in total). Under the relaxed deceptiveness metric \BDFIF, the number of conflicting pairs decreases to 488 pairs of each type (976 conflicting pairs in total).
For the first type of conflict, the LLMs produced incorrect RC predictions for 480--488 of the 731 pairs (65.8\%--66.8\%) across the three LLM executions. For the second type, they mispredicted 557--563 pairs (76.2\%--77.0\%). In contrast, when using \BDFIF, the number of mispredictions dropped to 274 of 488 pairs (56.1\%) for the first type and 331--340 pairs (68.0\%--70.0\%) for the second. These results help explain why the LLMs perform better on \BDFIF than on \BD.
Overall, these findings suggest that the information provided in the prompts is sufficient for judging perceived understandability but is less effective at identifying deceptive snippets. Richer context or additional reasoning about program behavior may therefore be needed to improve prediction of deceptiveness.

Both MCC and Cohen’s Kappa indicate positive correlations between predicted and actual labels, with effect sizes ranging from small to large depending on the model and metric for all positive RI cases. Exceptions are GPT-5.4 and Qwen-2.5 on \ABU, \BDFIF, and GPT-5.4 on \RL, which are negligible.
The Fleiss kappa agreement for all the metrics for the two LLMs are respectively for GPT-5.4 and Qwen-2.5 are 0.898 and 0.98, indicating high consistency across prompt executions and that the observed improvements are likely not due to non-deterministic LLM behavior.
\looseness=-1

\subsubsection{Developer-wise results.} \Cref{tab:rc_results_dev} shows different  developer-wise  prediction performance across models and metrics with no consistent overall trend in terms of weighted F1. However, there is a clear consistent trend in \AU, where all the models outperform the baselines with RI ranging from 10.4\% to 74.7\%. 
A similar trend is observed with \RL, where all the classical ML models and the CNN  improve over the best baseline.

To better understand the cases in which models do not outperform the
baselines, we examined several characteristics of the developer-wise
predictions and datasets:
\begin{itemize}
	
	 \item We investigated \emph{conflicting groups}: sets of instances with the
	same code and developer features but different comprehensibility labels.
	These groups represent cases in which developers with the same recorded
	profile give different judgments for the same snippets.
	Since \ABU and \BDFIF are the most challenging metrics to predict, we focused on these two for this analysis.
Across the developer-wise datasets for these two metrics, we found 3,097 unique groups, of which only 187 (6\%) and 157 (5.1\%) were conflicting for \ABU and \BDFIF, respectively. This suggests that conflicting groups
	alone do not explain the generally weak performance.
	
	    \item We observed that linear models, including LR, SVM, and NB, rarely
	improve over the baselines for the understandability metrics. In contrast,
	non-linear models such as Random Forest and XGBoost achieve positive RI for
	some of these metrics. One possible explanation for this is that the relationships
	between the code and developer features and the labels are non-linear or
	depend on interactions between features, which non-linear models can handle.
	
	 \item 
	The LLMs underperform the strongest baseline across the developer-wise metrics. To investigate whether the developer information included in the prompt contributed to this result, we repeated the developer-wise experiments after removing all developer-specific information from the prompts. (Effectively, this reduces the developer-wise setting to snippet-wise prediction.) However, we observed the same trend: the LLMs continued to underperform the strongest baseline---see the full results in our replication package~\cite{repl_pack}.
	This suggests that the LLMs rely primarily on information contained in the code snippets rather than on the developer information when making predictions. This interpretation is supported by the high agreement between developer-wise predictions with and without developer information: Cohen's $\kappa = 0.8601$--$0.8668$ across the three executions.
	Overall, this analysis is inconclusive with respect to explaining the LLMs' underperformance. While developer information appears to have little influence on the predictions, it is possible that the information provided to the LLM was insufficient to capture meaningful developer-specific characteristics that could improve prediction accuracy.

\end{itemize}

Both MCC and Cohen’s Kappa indicate positive correlations between predicted and actual labels, with effect sizes ranging from small to medium depending on the model and metric for all positive RI cases.

\subsubsection{Results analysis.}
Results distribution analysis showed that only two model-metric combination were impacted by outliers: RF for both snippet- and dev-wise \ABUFIF and RF for dev-wise \RL,
where outliers drag the mean down slightly; for details, see the replication package~\cite{repl_pack}.  
The positive RI for both snippet- and dev-wise RC models is statistically significant across all metrics and model types.

Snippet-wise RC models outperform the baselines across all 
comprehensibility proxies and model architectures, with only 
four exceptions. This strongly suggests that RC is more effective 
than AC for snippet-wise prediction, further confirming the 
model-agnostic nature of RC. 
\looseness=-1

\begin{tcolorbox}[boxsep=2pt, bottom=0pt]
	\vspace{0.5em}
\textbf{\ref{rq:predict_rc} Findings:} Snippet-wise RC models consistently outperform the na\"ive baselines and, in
the best cases, achieve high absolute performance
(\eg 0.934 $wF1$ for the CNN when predicting \ABU). This indicates that the
models capture useful patterns in the data for predicting RC. 
Developer-wise performance is more variable, although several models and
metrics still show clear improvements over the baselines.
\vspace{0.5em}
\end{tcolorbox}

\subsection{Relaxing the Definition of RC}
\label{sec:additional_experiments}

\ref{rq:predict_rc} 
shows that models outperform na\"ive baselines in predicting snippet-level relative comprehensibility (RC). We investigate the robustness of this finding by relaxing the strict RC definition from \cref{subsec:rc_datasets} and varying RC's class distribution, noting that the na\"ive baselines perform better with less balanced class distributions (\ie less class   entropy~\cite{Lin:1991TIT}).
\looseness=-1 

\subsubsection{Methodology.}

We relax the definition of RC by 
adding a margin of error $\epsilon (\ge 0)$ when comparing the aggregated similarity scores $S_1$ and $S_2$ of  snippets $c_1$ and $c_2$ in RC definition:
\begin{equation*}
	RC(c_1,c_2) = 
	\begin{cases} 
		0 & \text{if } S_1 -  S_2 > \epsilon  \quad\;\; (c_1 \text{ is more understandable}) \\
		1 & \text{if } S_2  -  S_1 > \epsilon \quad\; \ (c_2 \text{ is more understandable})\\
		2 & \text{if } |S_1 - S_2| \le  \epsilon \quad (\text{similarly understandable})
	\end{cases}
\end{equation*}
A larger $\epsilon$ increases the number of class-2 pairs while reducing class-0 and class-1 pairs.

We report experiments with two $\epsilon$ values: $0.11$ and $0.22$.
The first $\epsilon$ value, $0.11 = 1/9$, is based on the
smallest possible change in aggregated comprehensibility if one additional participant judged a snippet
in Scalabrino \etal~\cite{Scalabrino:TSE19}'s study, which had 8 or 9 participants per snippet.\footnote{We also
considered the same experiment with the Buse and Weimer study~\cite{Raymond:TSE10}, but its 121 participants' results
in a tiny $\epsilon$ with no impact on class distribution. Same for the developer-wise setting for both datasets.}
We then selected $\epsilon=0.22$ (twice the first $\epsilon$) to get a sense of model performance with a large $\epsilon$.
Using these $\epsilon$ values, we followed the same methodology as in \ref{rq:predict_rc}: training models, evaluating performance, and comparing them to the best baselines. We then compared the results to those from \ref{rq:predict_rc} (effectively $\epsilon=0$).

\begin{table}[t]
	\centering
	\caption{Snippet-wise class distribution for different $\epsilon$ values used by the relaxed definition of RC.}
	\label{tab:epsilon_class_distribution}
	\setlength{\tabcolsep}{3pt}
	\resizebox{0.6\columnwidth}{!}{
	\begin{tabular}{l|rrr|rrr|rrr}
		\toprule
		\multicolumn{1}{c|}{\multirow{2}{*}{\textbf{Metric}}} & \multicolumn{3}{c|}{\boldmath{$\epsilon=0$}}                                         & \multicolumn{3}{c|}{\boldmath{$\epsilon=0.11$}}                                      & \multicolumn{3}{c}{\boldmath{$\epsilon=0.22$}}                                     \\ 
		\multicolumn{1}{c|}{}                        & \multicolumn{1}{c}{\textbf{0}} & \multicolumn{1}{c}{\textbf{1}} & \multicolumn{1}{c|}{\textbf{2}} & \multicolumn{1}{c}{\textbf{0}} & \multicolumn{1}{c}{\textbf{1}} & \multicolumn{1}{c|}{\textbf{2}} & \multicolumn{1}{c}{\textbf{0}} & \multicolumn{1}{c}{\textbf{1}} & \multicolumn{1}{c}{\textbf{2}} \\
		\hline
		AU                                          & 46.7\%               & 46.7\%               & 6.6\%                & 46.8\%               & 41.1\%               & 12.1\%               & 37.9\%               & 35.3\%               & 26.8\%               \\
		PBU                                         & 43.6\%               & 43.6\%               & 12.9\%               & 37.0\%               & 33.2\%               & 29.8\%               & 17.6\%               & 17.6\%               & 64.7\%               \\
		ABU                                         & 39.9\%               & 39.9\%               & 20.2\%               & 23.0\%               & 23.0\%               & 54.0\%               & 11.5\%               & 11.5\%               & 77.0\%               \\
		\ABUFIF                                       & 44.3\%               & 44.3\%               & 11.4\%               & 36.0\%               & 35.0\%               & 29.0\%               & 25.3\%               & 25.3\%               & 49.4\%               \\
		BD                                          & 42.3\%               & 42.3\%               & 15.4\%               & 33.0\%               & 33.0\%               & 34.1\%               & 16.8\%               & 16.8\%               & 66.4\%               \\
		\BDFIF                                        & 41.4\%               & 41.4\%               & 17.1\%               & 24.9\%               & 24.9\%               & 50.2\%               & 13.8\%               & 13.8\%               & 72.5\%               \\
		\hline
		\RL                                          & 49.3\%               & 49.3\%               & 1.4\%                & 44.9\%               & 44.9\%               & 10.3\%               & 40.1\%               & 40.1\%               & 19.9\%        \\      
		\bottomrule  	
	\end{tabular}
	}
\end{table}

\begin{table}[t]
	\centering
	\caption{Relative improvement of snippet-wise RC models for different $\epsilon$ values based on the relaxed RC definition. Best baseline models: Random ({RB}) and Majority Class (\MBT).}
	\label{tab:epsilon_results}
	\setlength{\tabcolsep}{3pt}
	\resizebox{0.9\columnwidth}{!}{%
		\begin{tabular}{l|c|lr|ccccccc|ccc}
		\toprule
		\textbf{Metric}    & \boldmath{$\epsilon$} & \multicolumn{2}{c|}{\textbf{Baseline \boldmath{$wF1$}}} & \textbf{NB}                                           	& \textbf{KNN}                                          & \textbf{LR}                                            	& \textbf{MLP}                                           	& \textbf{RF}                                            	& \textbf{SVM}                                      & \textbf{XGBoost}  & \textbf{CNN} & \textbf{GPT-5.4} &\textbf{Qwen-2.5} \\

		\hline
		\multirow{3}{*}{\centering AU}              & 0                & (RB)                & 0.440               & \cellcolor{lightgreen}{ 13.7\%} 						& \cellcolor{lightgreen}{ 48.9\%} 							& \cellcolor{lightgreen}{ 58.0\%}  								& \cellcolor{lightgreen}{ 65.7\%}  							& \cellcolor{lightgreen}{ 91.2\%}  							& \cellcolor{lightgreen}{ 75.9\%} 		& \cellcolor{lightgreen}101.3\%	 &	\cellcolor{lightgreen}88.0\% & \cellcolor{lightgreen}26.6\%	&	\cellcolor{lightgreen}12.3\%	\\
													& 0.11             & (RB)                & 0.403               & \cellcolor{lightgreen}{ 42.4\%} 						& \cellcolor{lightgreen}{ 55.2\%} 							& \cellcolor{lightgreen}{ 71.0\%}  								& \cellcolor{lightgreen}{ 74.8\%}  							& \cellcolor{lightgreen}{ 117.0\%} 							& \cellcolor{lightgreen}{ 81.8\%} 		& \cellcolor{lightgreen}97.2\%	 &	\cellcolor{lightgreen}85.9\% & \cellcolor{lightgreen}17.1\%	&	\cellcolor{lightgreen}9.7\%		\\
													& 0.22             & (RB)                & 0.340               & \cellcolor{lightgreen}{ 67.9\%} 						& \cellcolor{lightgreen}{ 82.0\%} 							& \cellcolor{lightgreen}{ 91.8\%}  								& \cellcolor{lightgreen}{ 92.3\%}  							& \cellcolor{lightgreen}{ 161.8\%} 							& \cellcolor{lightgreen}{ 111.6\%} 		& \cellcolor{lightgreen}89.0\%	 &	\cellcolor{lightgreen}68.0\% & -5.6\%						&	-9.1\%				\\

		\hline
		\multirow{3}{*}{\centering \PBU}            & 0                & (RB)                & 0.396               & \cellcolor{lightgreen}{ 47.6\%} 						& \cellcolor{lightgreen}{ 47.0\%} 							& \cellcolor{lightgreen}{ 79.7\%}  								& \cellcolor{lightgreen}{ 76.2\%}  							& \cellcolor{lightgreen}{ 116.7\%} 							& \cellcolor{lightgreen}{ 94.2\%} 		& \cellcolor{lightgreen}116.8\%  & \cellcolor{lightgreen}113.9\% & \cellcolor{lightgreen}42.4\% & \cellcolor{lightgreen}23.5\%				\\
													& 0.11             & (RB)                & 0.336               & \cellcolor{lightgreen}{ 62.2\%} 						& \cellcolor{lightgreen}{ 70.3\%} 							& \cellcolor{lightgreen}{ 95.2\%}  								& \cellcolor{lightgreen}{ 92.5\%}  							& \cellcolor{lightgreen}{ 170.8\%} 							& \cellcolor{lightgreen}{ 103.0\%} 		& \cellcolor{lightgreen}119.7\%  & \cellcolor{lightgreen}95.2\% & \cellcolor{lightgreen}6.2\% &	\cellcolor{lightgreen}3.4\%				\\
													& 0.22             & (\MBT)              & 0.509               & -15.5\%                                               	& \cellcolor{lightgreen}{ 11.3\%} 							& \cellcolor{lightgreen}{ 38.8\%}  								& \cellcolor{lightgreen}{ 54.0\%}  							& \cellcolor{lightgreen}{ 77.9\%}  							& \cellcolor{lightgreen}{ 44.1\%} 		& \cellcolor{lightgreen}130.8\%  & \cellcolor{lightgreen}126.3\% & -56.2\% 					&	-39.1\%				\\

		\hline
		\multirow{3}{*}{\centering \ABU}            & 0                & (RB)                & 0.359               & \cellcolor{lightgreen}{ 53.9\%} 						& \cellcolor{lightgreen}{ 62.5\%} 							& \cellcolor{lightgreen}{ 86.5\%}  								& \cellcolor{lightgreen}{ 77.0\%}  							& \cellcolor{lightgreen}{ 135.9\%} 							& \cellcolor{lightgreen}{ 113.2\%}      & \cellcolor{lightgreen}125.7\% & \cellcolor{lightgreen}159.8\% & \cellcolor{lightgreen}5.8\% &	\cellcolor{lightgreen}1.6\%	\\
													& 0.11             & (RB)                & 0.397               & \cellcolor{lightgreen}{ 17.6\%} 						& \cellcolor{lightgreen}{ 32.8\%} 							& \cellcolor{lightgreen}{ 76.0\%}  								& \cellcolor{lightgreen}{ 72.1\%}  							& \cellcolor{lightgreen}{ 118.7\%} 							& \cellcolor{lightgreen}83.5\%          & \cellcolor{lightgreen}129.2\% & \cellcolor{lightgreen}155.6\% & -49.1\%  					  &	-34.7\%	\\
													& 0.22             & (\MBT)              & 0.669               & -9.1\%                                                	& \cellcolor{lightgreen}{ 7.0\%}  							& \cellcolor{lightgreen}{ 14.8\%}  								& \cellcolor{lightgreen}{ 22.8\%}  							& \cellcolor{lightgreen}{ 38.2\%}  							& \cellcolor{lightgreen}24.2\%          & \cellcolor{lightgreen}165.9\% & \cellcolor{lightgreen}160.1\% & -76.7\% 					  &	-50.7\%	\\

		\hline
		\multirow{3}{*}{\centering \ABUFIF}         & 0                & (RB)                & 0.406               & \cellcolor{lightgreen}{ 33.4\%} 						& \cellcolor{lightgreen}{ 37.3\%} 							& \cellcolor{lightgreen}{ 70.1\%}  								& \cellcolor{lightgreen}{ 75.6\%}  							& \cellcolor{lightgreen}{ 113.7\%} 							& \cellcolor{lightgreen}{ 95.2\%}      & \cellcolor{lightgreen}115.8\% & \cellcolor{lightgreen}110.1\% & \cellcolor{lightgreen}31.5\%  & \cellcolor{lightgreen}14.8\%		\\
													& 0.11             & (RB)                & 0.336               & \cellcolor{lightgreen}{ 59.7\%} 						& \cellcolor{lightgreen}{ 60.7\%} 							& \cellcolor{lightgreen}{ 107.7\%} 								& \cellcolor{lightgreen}{ 78.7\%}  							& \cellcolor{lightgreen}{ 166.1\%} 							& \cellcolor{lightgreen}{ 116.8\%} 	   & \cellcolor{lightgreen}108.0\% & \cellcolor{lightgreen}87.8\% & -0.5\% 						& -2.8\%		\\
													& 0.22             & (RB)                & 0.372               & \cellcolor{lightgreen}{ 30.5\%} 						& \cellcolor{lightgreen}{ 34.4\%} 							& \cellcolor{lightgreen}{ 83.5\%}  								& \cellcolor{lightgreen}{ 79.9\%}  							& \cellcolor{lightgreen}{ 141.8\%} 							& \cellcolor{lightgreen}{ 87.6\%} 	   & \cellcolor{lightgreen}113.5\% & \cellcolor{lightgreen}112.0\% & -36.5\%					&	-29.6\%	\\

		\hline
		\multirow{3}{*}{\centering \BD}             & 0                & (RB)                & 0.382               & \cellcolor{lightgreen}{ 51.9\%} 						& \cellcolor{lightgreen}{ 54.8\%} 							& \cellcolor{lightgreen}{ 83.5\%}  								& \cellcolor{lightgreen}{ 88.0\%}  							& \cellcolor{lightgreen}{ 137.8\%} 							& \cellcolor{lightgreen}{ 102.9\%} & \cellcolor{lightgreen}139.4\% & \cellcolor{lightgreen}130.6\% & -20.9\% & -23.2\% \\
													& 0.11             & (RB)                & 0.333               & \cellcolor{lightgreen}{ 56.7\%} 						& \cellcolor{lightgreen}{ 68.7\%} 							& \cellcolor{lightgreen}{ 94.7\%}  								& \cellcolor{lightgreen}{ 83.3\%}  							& \cellcolor{lightgreen}{ 174.8\%} 							& \cellcolor{lightgreen}{ 104.5\%} & \cellcolor{lightgreen}129.5\% & \cellcolor{lightgreen}74.0\% & -45.8\% & -40.4\% \\
													& 0.22             & (\MBT)              & 0.530               & -22.9\%                                               	& \cellcolor{lightgreen}{ 4.0\%}  							& \cellcolor{lightgreen}{ 30.5\%}  								& \cellcolor{lightgreen}{ 42.0\%}  							& \cellcolor{lightgreen}{ 77.5\%}  							& \cellcolor{lightgreen}39.1\%     & \cellcolor{lightgreen}142.3\% & \cellcolor{lightgreen}136.0\% & -76.5\% & -60.0\% \\
		
		\hline
		\multirow{3}{*}{\centering \BDFIF}          & 0                & (RB)                & 0.373               & \cellcolor{lightgreen}{ 49.6\%} 						& \cellcolor{lightgreen}{ 34.3\%} 							& \cellcolor{lightgreen}{ 75.3\%}  								& \cellcolor{lightgreen}{ 66.8\%}  							& \cellcolor{lightgreen}{ 131.5\%} 							& \cellcolor{lightgreen}{ 103.3\%} & \cellcolor{lightgreen}122.4\% & \cellcolor{lightgreen}140.9\% & \cellcolor{lightgreen}15.0\% & \cellcolor{lightgreen}7.5\% \\
													& 0.11             & (RB)                & 0.376               & \cellcolor{lightgreen}{ 22.9\%} 						& \cellcolor{lightgreen}{ 15.4\%} 							& \cellcolor{lightgreen}{ 68.8\%}  								& \cellcolor{lightgreen}{ 82.4\%}  							& \cellcolor{lightgreen}{ 145.2\%} 							& \cellcolor{lightgreen}{ 100.8\%} & \cellcolor{lightgreen}137.9\% & \cellcolor{lightgreen}142.8\% & -44.3\% & -37.5\% \\
													& 0.22             & (\MBT)              & 0.609               & -1.7\%                                                	& -19.3\%                         							& \cellcolor{lightgreen}{ 26.2\%}  								& \cellcolor{lightgreen}{ 36.8\%}  							& \cellcolor{lightgreen}{ 55.0\%}  							& \cellcolor{lightgreen}37.0\%     & \cellcolor{lightgreen}156.9\% & \cellcolor{lightgreen}152.0\% & -75.0\% & -59.2\% \\

		\hline
		\multirow{3}{*}{\centering \RL} 			& 0                & (RB)                & 0.487               & -7.0\%                                                	& \cellcolor{lightgreen}{ 42.4\%} 							& \cellcolor{lightgreen}{ 36.8\%}  								& \cellcolor{lightgreen}{ 70.7\%}  							& \cellcolor{lightgreen}{ 65.2\%}  							& \cellcolor{lightgreen}{ 41.8\%} & \cellcolor{lightgreen}84.2\% & \cellcolor{lightgreen}84.2\% & \cellcolor{lightgreen}5.0\% & -10.4\% \\
													& 0.11             & (RB)                & 0.413               & \cellcolor{lightgreen}{ 11.4\%} 						& \cellcolor{lightgreen}{ 39.6\%} 							& \cellcolor{lightgreen}{ 48.2\%}  								& \cellcolor{lightgreen}{ 83.0\%}  							& \cellcolor{lightgreen}{ 102.8\%}  						& \cellcolor{lightgreen}57.0\%    & \cellcolor{lightgreen}66.1\% & \cellcolor{lightgreen}78.4\% & -7.1\% & -19.6\% \\
													& 0.22             & (RB)                & 0.360               & \cellcolor{lightgreen}{ 27.7\%} 						& \cellcolor{lightgreen}{ 50.8\%} 							& \cellcolor{lightgreen}{ 56.8\%}  								& \cellcolor{lightgreen}{ 102.8\%} 							& \cellcolor{lightgreen}{ 129.5\%} 							& \cellcolor{lightgreen}66.0\%    & \cellcolor{lightgreen}58.1\% & \cellcolor{lightgreen}83.6\% & -20.7\% & -27.0\% \\                                    
		\bottomrule
		\end{tabular}
	}

	\vspace{0.3cm}
	\begin{minipage}{\columnwidth}
		\footnotesize
		\textit{Comprehensibility Metrics:} \textbf{\AU}, Actual Understandability;
		\textbf{\PBU}, Perceived Binary Understandability;
		\textbf{\ABU}, Actual Binary Understandability;
		\textbf{\ABUFIF}, Actual Binary Understandability 50\%;
		\textbf{\BD}, Binary Deceptiveness;
		\textbf{\BDFIF}, Binary Deceptiveness 50\%; and
		\textbf{\RL}, Readability Level.
		
		\smallskip
		\smallskip
		
		\textit{Models:} \textbf{NB}, Naïve Bayes;
		\textbf{KNN}, K-Nearest Neighbors;
		\textbf{LR}, Logistic Regression;
		\textbf{MLP}, Multilayer Perceptron;
		\textbf{RF}, Random Forest;
		\textbf{SVM}, Support Vector Machine;
		\textbf{XGBoost}, Extreme Gradient Boosting;
		\textbf{CNN}, Convolutional Neural Network; and
		\textbf{Qwen-2.5}, Qwen2.5-Coder-32B-Instruct.
		
	\end{minipage}
\end{table}

\subsubsection{Results.}

\Cref{tab:epsilon_class_distribution} shows the class distributions for the different $\epsilon$ values. 
As $\epsilon$ increases, class 2 becomes larger while classes 0 and 1 shrink. When $\epsilon = 0$, classes 0 and 1 exhibit identical distributions across all metrics, while class 2 is consistently underrepresented. As $\epsilon$ increases, class 2 becomes dominant across most metrics. For example, in \ABU, its share rises from 20.2\% at $\epsilon = 0$ to 77\% at $\epsilon = 0.22$, while class 0 drops from 39.9\% to 11.5\%. This shift is evident in all metrics except for \AU and \RL, where the distribution remains more balanced.
For more balanced metrics, the baselines perform worse, whereas for less balanced metrics, they perform better, due to their dependence on class balance.

\Cref{tab:epsilon_results} shows that RC models effectively predict nearly all metrics across different 
$\epsilon$ values, with improvements over the baselines of 1.6\% to 174.8\% RI. As 
$\epsilon$ increases, class 2 includes more cases where snippets that humans qualitatively judge as having different comprehensibility, which may explain the learning challenges at $\epsilon=0.22$.
Note that selecting the best $\epsilon$ per metric would consistently achieve the highest RI across all models.
Another notable observation is that the performance degradation of the LLMs as the $\epsilon$ increases can be primarily attributed to a structural limitation of the model prediction under zero-shot prompting. Unlike traditional machine learning models and the CNN, which are explicitly trained and cross-validated on the updated target distributions for each specific $\epsilon$ threshold, the LLMs are evaluated without any parameter-specific fine-tuning. As a result, these models are entirely blind to the hidden $\epsilon$ parameter. While the traditional models successfully learn the relaxed mathematical boundaries that group more snippets into the correct class, the LLMs may persistently differentiate snippet pairs based on syntactic variations. This fundamental disconnect between the LLMs' static evaluation criteria and the artificially relaxed ground truth ultimately penalizes their predictive accuracy, driving their performance down as $\epsilon$ grows.

Overall, these results suggest that snippet-wise machine and deep learning RC models are robust to both strict and relaxed RC definitions, as indicated by small and large 
$\epsilon$ margins, while the LLMs show a performance degradation with increasing $\epsilon$ values, suggesting that zero-shot RC LLM-based prediction is less robust to relaxed RC definitions.
\looseness=-1

\section{Comparing Relative  vs Absolute Comprehensibility Prediction}
\label{sec:comparison}

In this section, we directly compare model performance on AC and RC prediction to validate our conjecture that predicting RC is more effective than predicting AC, and to measure the difference in effectiveness. We do this based on our answers to \textbf{\ref{rq:predict_ac}} and \textbf{\ref{rq:predict_rc}} in the previous sections. We aim to answer the following research question:

\vspace{0.1cm}
\begin{enumerate}[label=\textbf{RQ$_\arabic*$:}, ref=\textbf{RQ$_\arabic*$}, itemindent=0.5cm,leftmargin=0.5cm, resume]
  \item \label{rq:compare_ac_rc}{\textit{How effective are relative comprehensibility models compared to absolute comprehensibility models}?}
\end{enumerate}

\subsection{Methodology}

It is tempting to compare the performance of the models directly
in terms of weighted F1, which makes sense if we are interested
in knowing which model ``performs best'' in an absolute sense.
However, such a direct comparison is not really fair, since the predictive tasks are different in nature: AC prediction estimates an
absolute comprehensibility value for a snippet while RC prediction
estimates a relative comprehensibility relationship between two
snippets. Hence, we \textbf{cannot} and \textbf{must not} compare model prediction performance directly between tasks.
A fairer method is to compare the relative
performance improvement (RI) that the models achieve compared to the
respective baselines for each task ($RI_{RC}$ and $ RI_{AC}$), and
calculate how much RI difference there is between the two tasks
($\Delta_{RI} = RI_{RC} - RI_{AC}$). 
RI is a normalized metric that lets us quantify how much the RC models are learning from the RC data compared to how much the AC models are learning from the AC data.
Effectively, $\Delta_{RI}$ quantifies how much
more RC models learn compared to AC models to answer
\ref{rq:compare_ac_rc}. %

Since we compare the RI of sets of RC and AC models across the ten model types and seven RC metrics (defined for each individual comprehensibility metric), we employed the non-parametric, unpaired Mann-Whitney U test~\cite{mann-witney} to assess statistical significance (evaluated at a confidence level of 95\%: $p < 0.05$.). 
The null hypothesis (H\textsubscript{0}) posits that $RI\textsubscript{RC} \le RI\textsubscript{AC}$ and, hence, the alternative hypothesis (H\textsubscript{a}) posits that  $\RIRC > \RIAC$.

\subsection{\ref{rq:compare_ac_rc}: RC vs. AC Results}

\begin{table*}[t]
	\centering
	\caption{Comparison of the relative improvement (RI) of ML models over the best baselines for predicting absolute (AC) and relative comprehensibility (RC). \hlgreen{Green}: positive \RIDIF $= RI_{RC} - RI_{AC}$. 	\hlyellow{Yellow}: positive $RI_{RC}$. 	\hlblue{Blue}: postive $RI_{AC}$.}
	\label{tab:ac_rc_results}%
	\setlength{\tabcolsep}{3pt}

	\begin{subtable}[c]{\textwidth}
		\centering
		\caption{Snippet-wise prediction}
		\resizebox{0.63\columnwidth}{!}{%
		\begin{tabular}{l|ccc|ccc|ccc|ccc}
		\toprule
		\multirow{2}{*}{\textbf{Model}} &
		\multicolumn{3}{c|}{\textbf{\AU}} &
		\multicolumn{3}{c|}{\textbf{\ABUFIF}} &
		\multicolumn{3}{c|}{\textbf{\BD}} &
		\multicolumn{3}{c}{\textbf{\RL}} \\
		&
		\textbf{\RIAC} & \textbf{\RIRC} & \textbf{\RIDIF} &
		\textbf{\RIAC} & \textbf{\RIRC} & \textbf{\RIDIF} &
		\textbf{\RIAC} & \textbf{\RIRC} & \textbf{\RIDIF} &
		\textbf{\RIAC} & \textbf{\RIRC} & \textbf{\RIDIF} \\
		\hline

		\textbf{NB}
		& -9.1\% & \cellcolor{lightyellow}13.7\% & \cellcolor{lightgreen}22.8\%
		& -16.3\% & \cellcolor{lightyellow}33.4\% & \cellcolor{lightgreen}49.7\%
		& \cellcolor{lightblue}12.8\% & \cellcolor{lightyellow}51.9\% & \cellcolor{lightgreen}39.0\%
		& -41.3\% & -7.0\% & \cellcolor{lightgreen}34.3\% \\

		\textbf{KNN}
		& -1.8\% & \cellcolor{lightyellow}48.9\% & \cellcolor{lightgreen}50.7\%
		& -14.6\% & \cellcolor{lightyellow}37.3\% & \cellcolor{lightgreen}52.0\%
		& \cellcolor{lightblue}16.8\% & \cellcolor{lightyellow}54.8\% & \cellcolor{lightgreen}37.9\%
		& -13.8\% & \cellcolor{lightyellow}42.4\% & \cellcolor{lightgreen}56.2\% \\

		\textbf{LR}
		& -4.9\% & \cellcolor{lightyellow}58.0\% & \cellcolor{lightgreen}62.9\%
		& \cellcolor{lightblue}2.2\% & \cellcolor{lightyellow}70.1\% & \cellcolor{lightgreen}67.9\%
		& \cellcolor{lightblue}27.4\% & \cellcolor{lightyellow}83.5\% & \cellcolor{lightgreen}56.1\%
		& \cellcolor{lightblue}6.8\% & \cellcolor{lightyellow}36.8\% & \cellcolor{lightgreen}30.0\% \\

		\textbf{MLP}
		& -10.2\% & \cellcolor{lightyellow}65.7\% & \cellcolor{lightgreen}75.9\%
		& -8.3\% & \cellcolor{lightyellow}75.6\% & \cellcolor{lightgreen}84.0\%
		& -10.0\% & \cellcolor{lightyellow}88.0\% & \cellcolor{lightgreen}98.1\%
		& -14.6\% & \cellcolor{lightyellow}70.7\% & \cellcolor{lightgreen}85.3\% \\

		\textbf{RF}
		& \cellcolor{lightblue}2.4\% & \cellcolor{lightyellow}91.2\% & \cellcolor{lightgreen}88.7\%
		& -18.1\% & \cellcolor{lightyellow}113.7\% & \cellcolor{lightgreen}131.9\%
		& \cellcolor{lightblue}33.4\% & \cellcolor{lightyellow}137.8\% & \cellcolor{lightgreen}104.4\%
		& -1.0\% & \cellcolor{lightyellow}65.2\% & \cellcolor{lightgreen}66.2\% \\

		\textbf{SVM}
		& \cellcolor{lightblue}5.0\% & \cellcolor{lightyellow}75.9\% & \cellcolor{lightgreen}70.9\%
		& \cellcolor{lightblue}3.4\% & \cellcolor{lightyellow}95.2\% & \cellcolor{lightgreen}91.8\%
		& \cellcolor{lightblue}31.7\% & \cellcolor{lightyellow}102.9\% & \cellcolor{lightgreen}71.2\%
		& -2.5\% & \cellcolor{lightyellow}41.8\% & \cellcolor{lightgreen}44.4\% \\

		\textbf{XGBoost}
		& -0.6\% & \cellcolor{lightyellow}101.3\% & \cellcolor{lightgreen}102.0\%
		& -12.6\% & \cellcolor{lightyellow}115.8\% & \cellcolor{lightgreen}128.4\%
		& \cellcolor{lightblue}0.5\% & \cellcolor{lightyellow}139.4\% & \cellcolor{lightgreen}138.8\%
		& \cellcolor{lightblue}6.3\% & \cellcolor{lightyellow}84.2\% & \cellcolor{lightgreen}77.9\% \\

		\hline

		\textbf{CNN}
		& -10.6\% & \cellcolor{lightyellow}88.0\% & \cellcolor{lightgreen}98.7\%
		& \cellcolor{lightblue}3.4\%  & \cellcolor{lightyellow}110.1\% & \cellcolor{lightgreen}106.7\%
		& \cellcolor{lightblue}14.1\% & \cellcolor{lightyellow}130.6\% & \cellcolor{lightgreen}116.5\%
		& -12.0\% & \cellcolor{lightyellow}84.2\% & \cellcolor{lightgreen}96.2\% \\
		
		\textbf{GPT-5.4}
		& -78.7\%  & \cellcolor{lightyellow}26.6\% & \cellcolor{lightgreen}105.4\% 
		& -53.8\% & \cellcolor{lightyellow}31.5\% & \cellcolor{lightgreen}85.2\%
		& \cellcolor{lightblue}17.8\% & -20.9\% & -38.7\%
		& -6.3\% & \cellcolor{lightyellow}5.0\% &  \cellcolor{lightgreen}11.3\% \\

		\textbf{Qwen-2.5}
		& -77.7\% & \cellcolor{lightyellow}12.3\% & \cellcolor{lightgreen}90.0\% 
		& -47.2\%  &  \cellcolor{lightyellow}14.8\%  &  \cellcolor{lightgreen}62.0\%
		& \cellcolor{lightblue}9.8\% & -23.2\%  & -33.0\% 
		& -42.0\% & -10.4\% & \cellcolor{lightgreen}31.6\% \\

		\bottomrule
		\end{tabular}%
		}
		\label{tab:ac_rc_results_snippet}
	\end{subtable}

	\vspace{0.1cm}

	\begin{subtable}[c]{\textwidth}
	\centering
	\caption{Developer-wise prediction}
	\resizebox{\columnwidth}{!}{%
	\begin{tabular}{l|ccc|ccc|ccc|ccc|ccc|ccc|ccc}
	\toprule
	\multirow{2}{*}{\textbf{Model}}
	& \multicolumn{3}{c|}{\textbf{\AU}}
	& \multicolumn{3}{c|}{\textbf{\PBU}}
	& \multicolumn{3}{c|}{\textbf{\ABU}}
	& \multicolumn{3}{c|}{\textbf{\ABUFIF}}
	& \multicolumn{3}{c|}{\textbf{\BD}}
	& \multicolumn{3}{c|}{\textbf{\BDFIF}}
	& \multicolumn{3}{c}{\textbf{\RL}} \\
	&
	\textbf{\RIAC} & \textbf{\RIRC} & \textbf{\RIDIF}
	& \textbf{\RIAC} & \textbf{\RIRC} & \textbf{\RIDIF}
	& \textbf{\RIAC} & \textbf{\RIRC} & \textbf{\RIDIF}
	& \textbf{\RIAC} & \textbf{\RIRC} & \textbf{\RIDIF}
	& \textbf{\RIAC} & \textbf{\RIRC} & \textbf{\RIDIF}
	& \textbf{\RIAC} & \textbf{\RIRC} & \textbf{\RIDIF}
	& \textbf{\RIAC} & \textbf{\RIRC} & \textbf{\RIDIF} \\
	\hline

	\textbf{NB}
	& -1.3\% & \cellcolor{lightyellow}10.4\% & \cellcolor{lightgreen}11.7\%
	& -2.9\% & -37.8\% & -34.9\%
	& -14.4\% & -26.4\% & -12.0\%
	& \cellcolor{lightblue}19.9\% & -24.6\% & -44.5\%
	& \cellcolor{lightblue}2.8\% & -23.6\% & -26.4\%
	& -18.2\% & -28.9\% & -10.7\%
	& -20.6\% & \cellcolor{lightyellow}21.1\% & \cellcolor{lightgreen}41.7\% \\

	\textbf{KNN}
	& \cellcolor{lightblue}0.3\% & \cellcolor{lightyellow}51.2\% & \cellcolor{lightgreen}50.9\%
	& -6.3\% & \cellcolor{lightyellow}8.6\% & \cellcolor{lightgreen}14.9\%
	& -15.4\% & -1.7\% & \cellcolor{lightgreen}13.7\%
	& \cellcolor{lightblue}11.7\% & \cellcolor{lightyellow}19.3\% & \cellcolor{lightgreen}7.6\%
	& \cellcolor{lightblue}0.4\% & \cellcolor{lightyellow}11.6\% & \cellcolor{lightgreen}11.2\%
	& -19.9\% & -5.4\% & \cellcolor{lightgreen}14.5\%
	& -12.2\% & \cellcolor{lightyellow}27.1\% & \cellcolor{lightgreen}39.3\% \\

	\textbf{LR}
	& \cellcolor{lightblue}17.0\% & \cellcolor{lightyellow}26.0\% & \cellcolor{lightgreen}9.0\%
	& \cellcolor{lightblue}3.5\% & -10.3\% & -13.7\%
	& -4.8\% & -9.7\% & -4.9\%
	& \cellcolor{lightblue}22.8\% & -9.2\% & -32.0\%
	& \cellcolor{lightblue}7.8\% & -14.1\% & -22.0\%
	& -19.2\% & -16.7\% & \cellcolor{lightgreen}2.5\%
	& -21.2\% & \cellcolor{lightyellow}32.6\% & \cellcolor{lightgreen}53.7\% \\

	\textbf{MLP}
	& -7.7\% & \cellcolor{lightyellow}39.5\% & \cellcolor{lightgreen}47.2\%
	& -10.0\% & \cellcolor{lightyellow}6.7\% & \cellcolor{lightgreen}16.7\%
	& -16.1\% & -2.4\% & \cellcolor{lightgreen}13.7\%
	& -2.2\% & \cellcolor{lightyellow}11.6\% & \cellcolor{lightgreen}13.8\%
	& -5.9\% & \cellcolor{lightyellow}4.1\% & \cellcolor{lightgreen}10.0\%
	& -19.1\% & -4.8\% & \cellcolor{lightgreen}14.3\%
	& -10.7\% & \cellcolor{lightyellow}60.8\% & \cellcolor{lightgreen}71.6\% \\

	\textbf{RF}
	& \cellcolor{lightblue}22.8\% & \cellcolor{lightyellow}74.7\% & \cellcolor{lightgreen}51.9\%
	& \cellcolor{lightblue}16.2\% & \cellcolor{lightyellow}34.5\% & \cellcolor{lightgreen}18.2\%
	& \cellcolor{lightblue}0.4\% & \cellcolor{lightyellow}13.8\% & \cellcolor{lightgreen}13.5\%
	& \cellcolor{lightblue}22.2\% & \cellcolor{lightyellow}38.3\% & \cellcolor{lightgreen}16.1\%
	& \cellcolor{lightblue}13.7\% & \cellcolor{lightyellow}39.1\% & \cellcolor{lightgreen}25.4\%
	& -0.8\% & \cellcolor{lightyellow}16.0\% & \cellcolor{lightgreen}16.8\%
	& -28.4\% & \cellcolor{lightyellow}60.2\% & \cellcolor{lightgreen}88.7\% \\

	\textbf{SVM}
	& \cellcolor{lightblue}10.0\% & \cellcolor{lightyellow}24.5\% & \cellcolor{lightgreen}14.5\%
	& \cellcolor{lightblue}2.0\% & -12.3\% & -14.2\%
	& -4.8\% & -10.5\% & -5.7\%
	& \cellcolor{lightblue}17.8\% & -10.7\% & -28.4\%
	& \cellcolor{lightblue}5.5\% & -14.9\% & -20.4\%
	& -20.1\% & -16.6\% & \cellcolor{lightgreen}3.6\%
	& -27.2\% & \cellcolor{lightyellow}31.4\% & \cellcolor{lightgreen}58.6\% \\

	\textbf{XGBoost}
	& \cellcolor{lightblue}20.7\% & \cellcolor{lightyellow}68.5\% & \cellcolor{lightgreen}47.9\%
	& \cellcolor{lightblue}10.1\% & \cellcolor{lightyellow}27.1\% & \cellcolor{lightgreen}17.0\%
	& 0.0\% & \cellcolor{lightyellow}15.9\% & \cellcolor{lightgreen}15.9\%
	& \cellcolor{lightblue}20.0\% & \cellcolor{lightyellow}36.6\% & \cellcolor{lightgreen}16.6\%
	& \cellcolor{lightblue}7.4\% & \cellcolor{lightyellow}31.6\% & \cellcolor{lightgreen}24.3\%
	& -6.1\% & \cellcolor{lightyellow}13.4\% & \cellcolor{lightgreen}19.5\%
	& \cellcolor{lightblue}0.1\% & \cellcolor{lightyellow}56.7\% & \cellcolor{lightgreen}56.6\% \\

	\hline
	\textbf{CNN}
	& -5.9\% & \cellcolor{lightyellow}25.5\% & \cellcolor{lightgreen}31.4\%
	& -21.7\% & \cellcolor{lightyellow}12.5\% & \cellcolor{lightgreen}34.3\%
	& -37.9\% & \cellcolor{lightyellow}5.8\% & \cellcolor{lightgreen}43.7\%
	& -11.0\% & \cellcolor{lightyellow}25.0\% & \cellcolor{lightgreen}36.0\%
	& -9.2\% & \cellcolor{lightyellow}18.4\% & \cellcolor{lightgreen}27.6\%
	& -23.4\% & \cellcolor{lightyellow}5.1\% & \cellcolor{lightgreen}28.5\%
	& -19.7\% & \cellcolor{lightyellow}56.9\% & \cellcolor{lightgreen}76.6\% \\

	\textbf{GPT-5.4}
	& -42.6\% 						& \cellcolor{lightyellow}35.7\% 						& \cellcolor{lightgreen}78.2\%
	& -1.7\% 						& -36.7\% & -35.0\%
	& -42.3\% 						& -57.9\% 						& -15.6\%
	& -31.7\% 						& -23.6\% 						& \cellcolor{lightgreen}8.0\%
	& \cellcolor{lightblue}2.1\% 	& -27.3\% 						& -29.4\%
	& \cellcolor{lightblue}0.7\% 	& -55.6\% 						& -56.3\%
	& -2.6\% 						& -3.0\% 						& -0.4\% \\

	\textbf{Qwen-2.5}
	& -44.8\% 						& \cellcolor{lightyellow}32.6\% & \cellcolor{lightgreen}77.4\%
	& \cellcolor{lightblue}7.5\% 	& -29.2\% 						& -36.6\%
	& -33.5\% 						& -49.1\% 						& -15.6\%
	& -31.5\% 						& -17.9\% 						& \cellcolor{lightgreen}13.6\%
	& \cellcolor{lightblue}2.9\% 	& -21.3\% 						& -24.3\%
	& \cellcolor{lightblue}0.5\% 	& -48.7\% 						& -49.2\%
	& -21.7\% 						& -0.8\% 						& \cellcolor{lightgreen}21.0\% \\

	\bottomrule
	\end{tabular}%
	}
	\label{tab:ac_rc_results_dev}
	\end{subtable}

	\vspace{0.3cm}
	\begin{minipage}{\columnwidth}
		\footnotesize
		\textit{Comprehensibility Metrics:} \textbf{\AU}, Actual Understandability;
		\textbf{\PBU}, Perceived Binary Understandability;
		\textbf{\ABU}, Actual Binary Understandability;
		\textbf{\ABUFIF}, Actual Binary Understandability 50\%;
		\textbf{\BD}, Binary Deceptiveness;
		\textbf{\BDFIF}, Binary Deceptiveness 50\%; and
		\textbf{\RL}, Readability Level.
		
		\smallskip
		\smallskip
		
		\textit{Models:} \textbf{NB}, Naïve Bayes;
		\textbf{KNN}, K-Nearest Neighbors;
		\textbf{LR}, Logistic Regression;
		\textbf{MLP}, Multilayer Perceptron;
		\textbf{RF}, Random Forest;
		\textbf{SVM}, Support Vector Machine;
		\textbf{XGBoost}, Extreme Gradient Boosting;
		\textbf{CNN}, Convolutional Neural Network; and
		\textbf{Qwen-2.5}, Qwen2.5-Coder-32B-Instruct.
		
	\end{minipage}

\end{table*}%

\Cref{tab:ac_rc_results_snippet,tab:ac_rc_results_dev} compare the models' relative improvement over the best baselines for predicting absolute \textit{vs} relative comprehensibility in both snippet-wise and developer-wise settings.

\subsubsection{Snippet-wise results.}
\Cref{tab:ac_rc_results_snippet} shows that 
the $\Delta_{RI}$ values are positive for all the metrics (except \BD for both the LLMs), indicating that snippet-wise RC prediction models learn more from the data than corresponding AC models.
The positive $\Delta_{RI}$ values range from 11.3\% to 138.8\%, with the highest gains observed for \BD and XGBoost.
The Mann-Whitney U test indicates statistical significance ($p < 0.05$), in favor of $\Delta_{RI} >0$ RC models, for all the metrics across all model types.
The percentage of models that outperform the best baseline is higher for RC than for AC models. Overall, 96.8\% of RC classifiers outperform the baseline, versus just 48\% of AC classifiers.
\looseness=-1

\subsubsection{Developer-wise results.}
\Cref{tab:ac_rc_results_dev} shows model-wise divergence: some RC models (RF, KNN, MLP, XGBoost and CNN) have positive $\Delta_{RI}$ for all metrics
(results are statistically significant in all cases).
However, the results are more mixed for the other models, and some AC models outperform their RC counterparts.
For the RL and AU metrics, RC models uniformly outperform their AC counterparts (all statistically significant except AU for the NB model) with only exception: \RL with GPT-5.4.

\subsubsection{Results analysis.} 
We analyze the developer-wise results in more detail to understand how effective RC prediction is compared to AC prediction. 
Of 70 model-metric combinations, we found that 48 combinations show a positive $\Delta_{RI}$ of 2.5\% to 88.7\%, while only 22 show a negative $\Delta_{RI}$ of -56.3\% to -0.4\%. Overall, this suggests that RC prediction is effective more often than AC prediction.

The most desirable case is a positive $\Delta_{RI}$ stemming from a negative AC prediction RI to a positive RC RI (compared to the baselines).
For example, AC RF classifiers for the RL metric show performance degradation (RI of -28.4\%), but RC RF classifiers outperform the baselines by 60.2\% RI. In this case, there is substantial improvement between RC and AC prediction ($\Delta_{RI}$  of 88.7\%). We found 24 model-metric combinations (out of 70) like this, with $\Delta_{RI}$ varying from 11.7\% to 88.7\%. 
Conversely, the least desirable case is negative $\Delta_{RI}$ stemming from a positive AC prediction RI to a negative RC prediction RI.
An instance of this case is LR for the \ABUFIF metric. The AC LR models improved over the baselines by 22.8\% RI but RC LR models underperformed the baselines by -9.2\% RI ($\Delta_{RI}$ of -32\%). We found only 13 such model-metric combinations (out of 70), with $\Delta_{RI}$ varying from -56.3\% to -13.7\%. Taken together, these results further support the conclusion that RC prediction is more effective than AC prediction.

\begin{tcolorbox}[boxsep=2pt, bottom=2pt]
	\vspace{0.5em}
  \textbf{\ref{rq:compare_ac_rc} Findings:} 
  At the snippet level, RC prediction is substantially more effective than AC
  prediction. Developer-wise RC results are more variable, but RC still
  outperforms AC in most model-metric comparisons. This advantage
  appears across the evaluated model architectures, suggesting that it is
  not limited to a particular model architecture. Overall, these results support
  our hypothesis that predicting relative comprehensibility is easier to learn and more effective 
  than predicting absolute comprehensibility.
  \vspace{0.5em}
\end{tcolorbox}

\section{Practitioner Perspectives on Relative Comprehensibility Prediction}
\label{sec:user_study}

While the results of \ref{rq:compare_ac_rc} show that RC models can achieve higher predictive performance than AC models, predictive performance alone does not establish practical value. An RC model that is accurate but irrelevant to real development tasks would offer limited benefit. Therefore, we conducted a survey to study whether practitioners perceive AC and RC predictions as useful, and how such predictions could fit into software development workflows. Specifically, we address the following research questions:

\begin{enumerate}[label=\textbf{RQ$_\arabic*$:}, ref=\textbf{RQ$_\arabic*$}, itemindent=0.5cm,leftmargin=0.5cm,resume]
	\item \label{rq:survey_problem}{\textit{Do practitioners consider code comprehension difficulty a relevant problem in software development?}}
	\item \label{rq:survey_ac_rc}{\textit{How useful do practitioners perceive AC and RC models for common software engineering tasks?}}
	\item \label{rq:survey_workflow}{\textit{How should code comprehensibility models be integrated into development workflows?}}
\end{enumerate}
\vspace{0.1cm}

\ref{rq:survey_problem} examines whether comprehensibility prediction addresses a problem that developers actually face. \ref{rq:survey_ac_rc} studies whether practitioners see value in AC and RC predictions for concrete development tasks. \ref{rq:survey_workflow} investigates how these models should be exposed to developers if they are used in practice.

\subsection{Methodology}

To answer the RQs, we designed an online survey (hosted on Qualtrics~\cite{Qualtrics}) about code comprehension challenges and code comprehensibility prediction models, focusing on four software engineering tasks: code review, refactoring, bug fixing, and code understanding. These tasks are common maintenance and evolution activities where developers must understand existing code before making decisions. They also help contrast AC and RC models. Code review and refactoring often involve comparing an original snippet with a changed or alternative version, which is a natural setting for RC. Bug fixing and code understanding may involve only one snippet or code region, which is a natural setting for AC.
\looseness=-1
The study methodology, including questionnaire and participant recruitment procedures, was approved by the William \& Mary Institutional Review Board under Protocol IRB-2025-582. 

\subsubsection{Survey questionnaire.}
The survey was designed following general guidelines~\cite{Groves2011} and SE-specific best practices~\cite{Pfleeger2001SENotes1, Pfleeger2001SENotes2, Pfleeger2001SENotes3, Pfleeger2001SENotes4, Pfleeger2001SENotes5, Pfleeger2001SENotes6}, through multiple collaborative sessions by the authors.
The survey first introduced code comprehensibility as the difficulty a developer has when understanding a code snippet. It then described Absolute Comprehensibility (AC) models, which predict the comprehensibility of one snippet, and Relative Comprehensibility (RC) models, which compare two snippets and predict which one is easier to understand. Examples were given about how these models operate (\ie inputs and outputs).

The survey included multiple-choice/selection questions, Likert-scale questions, and optional open-ended questions. Participants answered questions about code comprehension difficulties, prior use of related tools or metrics, perceived usefulness of AC and RC models, preferred workflow integrations, preferred output formats, and demographic background.
The full survey questionnaire is found in our replication package~\cite{repl_pack}.

\subsubsection{Participants.}
We targeted software engineers and related software practitioners. Participants were recruited through the authors' professional networks using email and social media posts, including LinkedIn. Participation was voluntary, anonymous, and uncompensated. We obtained $n=38$ valid responses to the survey, which correspond to 38 software practitioners currently working in the industry.

\Cref{tab:survey-demographics} summarizes participants' current professional role, experience, industries, and projects they typically develop. The sample included mostly software developers, but also roles such as DevOps engineers, researchers, a QA engineer, and an engineering manager. Participants also reported varied experience levels, ranging from less than one year to more than ten years, and having developed a variety of systems, notably web applications and enterprise software, for diverse industry sectors such as information technology, education, finance, and healthcare.

\begin{table}[t]
	\centering
	\small
	\caption{Participant demographics ($n=38$). Multi-select categories (*) can sum to more than $n$.}
	\label{tab:survey-demographics}
	\begin{tabular}{p{0.25\linewidth}p{0.65\linewidth}}
		\toprule
		\textbf{Category} & \textbf{Summary} \\
		\midrule
		Professional role & Software engineer/developer: 29; Researcher: 3; DevOps Engineer: 3; QA engineer/tester: 1; engineering manager: 1; Coordinator/data scientist: 1 \\
		Professional SE experience & $<1$ year: 8; 1 year: 2; 1--3 years: 9; 4--6 years: 8; 7--10 years: 6; 10+ years: 5 \\
		Project types* & Web applications: 31; enterprise software: 17; data science/ML systems: 9; mobile apps: 7; open source: 6; embedded/IoT systems: 4; others (\eg desktop apps): 4 \\
		Common industries* & Information/computing/telecommunications: 12; education: 8; finance: 4; health care: 4; professional/scientific/technical services: 4 \\
		\bottomrule
	\end{tabular}
\end{table}

\subsubsection{Response Analysis.}
We analyzed closed-ended questions using counts and percentages. For Likert-scale questions, we report the full response distribution. We used open-ended responses to help interpret the quantitative results.

\begin{figure*}[t]
	\centering
	\begin{subfigure}[t]{0.8\textwidth}
		\centering
		\includegraphics[width=\linewidth]{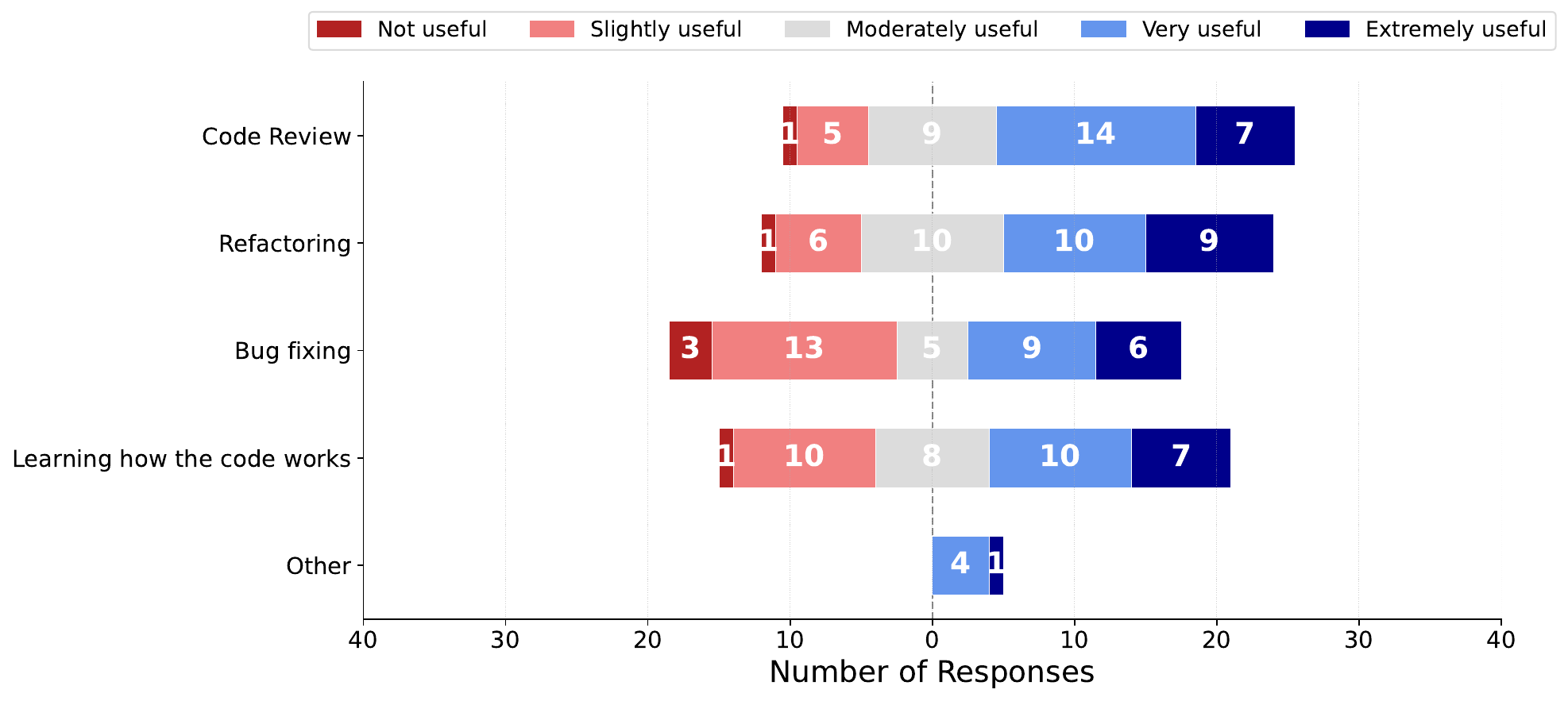}
		\caption{AC models.}
		\label{fig:ac-usefulness}
	\end{subfigure}
	\hfill
	\begin{subfigure}[t]{0.8\textwidth}
		\centering
		\includegraphics[width=\linewidth]{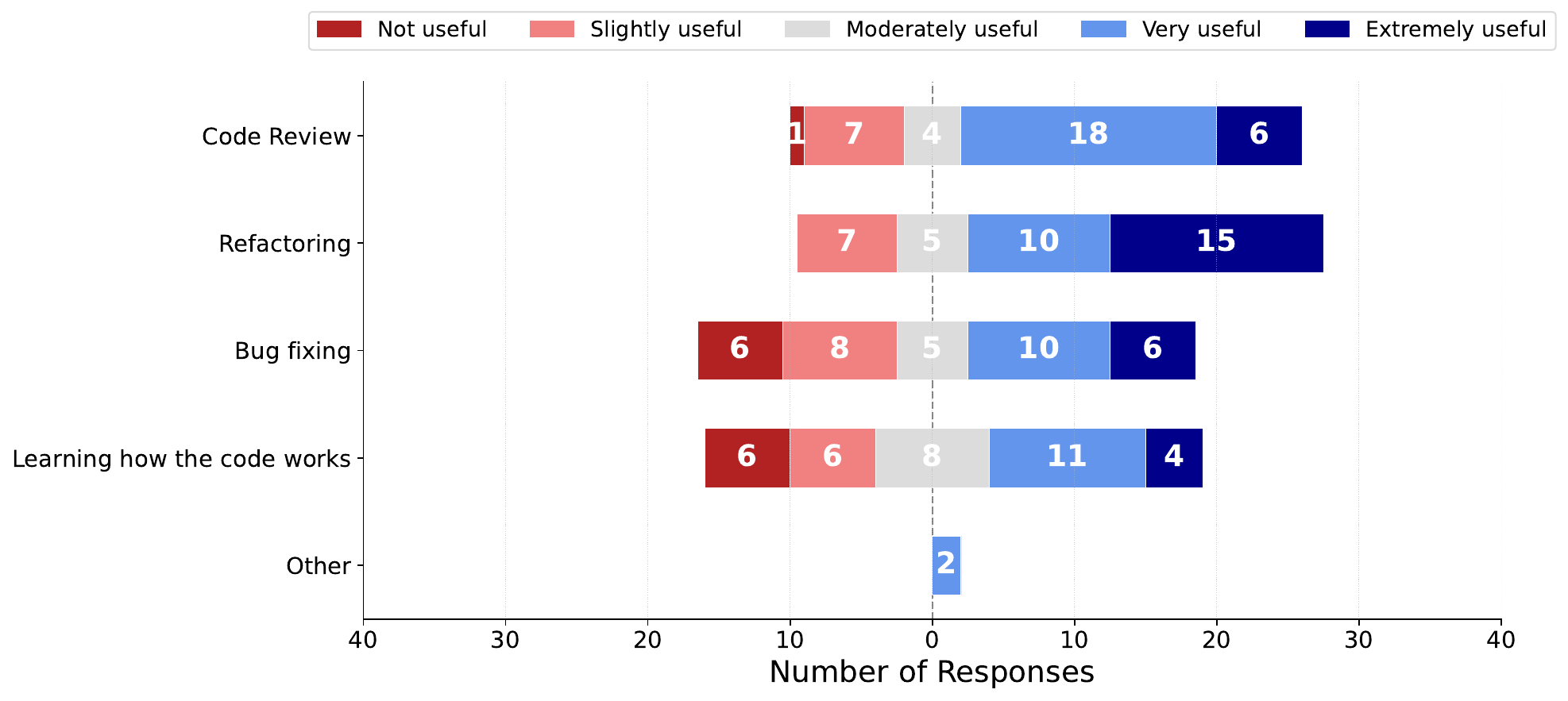}
		\caption{RC models.}
		\label{fig:rc-usefulness}
	\end{subfigure}
	\caption{Perceived usefulness of AC and RC models for code review, refactoring, bug fixing, and code understanding.}
	\label{fig:ac-rc-usefulness}
\end{figure*}

\subsection{Results}
\label{sec:practitioner-study-results}

\subsubsection{\ref{rq:survey_problem}: Relevance of Code Comprehension Difficulty.}

Most participants (35/38) reported encountering difficulty understanding existing code at least occasionally: 22 participants selected occasionally (a few times per month), 11 frequently (a few times per week), 2 very frequently (daily or almost daily), and 3 rarely (less than once a month). No participant selected never. The most common sources of difficulty were large codebases (34/38), poor identifier names (24/38), limited comments or documentation (23/38), and complex logic or algorithms (22/38).

The open-ended responses help explain these results. A software developer with 1--3 years noted that, in a large legacy system, ``the size was the largest limiting factor for understanding''~($P_{13}$). Another developer with 10+ years of experience emphasized naming, stating that ``if variables don't have descriptive names it will be difficult to follow the code''~($P_{20}$). These comments suggest that comprehensibility is not only about local code structure, but also about scale, naming, documentation, and context.

Participants were more mixed on whether measuring code comprehension difficulty is important for daily development work: 4 selected not important, 9 slightly important, 9 moderately important, 13 very important, and 3 extremely important. Thus, the results suggest that comprehension difficulty is common in practice, but explicit measurement is not equally important for every developer or task. As a developer with 4--6 years of experience explained, knowing comprehensibility ``mostly assists in the time/effort estimation, not with the actual task itself''~($P_{11}$).

\subsubsection{\ref{rq:survey_ac_rc}: Perceived Usefulness of AC and RC Models.}

Before asking about perceived model usefulness, we asked participants about prior exposure to related tools. For AC, 13/38 participants reported using tools or metrics for assessing a single snippet, most commonly SonarQube/SonarLint~\cite{SonarCube}, cyclomatic complexity, LOC, Halstead metrics, or Radon~\cite{Radon}. Only 4/38 reported using predictive AC models (\eg XGBoost or Codex). Prior exposure to RC tools was lower: 2/38 participants reported using comparative tools or metrics (\eg comparing cyclomatic complexity values), and 4/38 reported using predictive RC models, mostly LLM-based systems (\eg ChatGPT or Copilot). Thus, most responses reflect practitioners' expectations about AC and RC rather than extensive prior use of dedicated comprehensibility prediction tools.

Figure~\ref{fig:ac-rc-usefulness} shows participants' ratings of AC and RC model usefulness. Participants saw potential value in both model types, but the pattern depends on the task. RC received its strongest support for refactoring, where developers often need to decide whether a change made code easier to understand. RC was also viewed positively for code review.  A researcher with 7--10 years of experience  captured this use case directly: ``RC is very useful for refactoring tasks when we want to confirm that we are keeping the code understandable. For code review, it would be good to observe the impact of the code change''~($P_{33}$). AC remained useful for tasks where a single snippet may be assessed on its own, especially code understanding.

We also asked participants which model type, AC or RC, would be more useful for each task. The results suggest the models are complementary. Overall, 13 participants selected both models as equally useful, 12 selected RC, 11 selected AC, and 2 were unsure. For refactoring, however, RC was preferred by 21 participants, compared with 6 for AC and 9 for both. For code understanding, AC was preferred by 16 participants, compared with 6 for RC and 10 for both. 
A developer with 10+ years of experience described this task distinction as follows: ``AC is more useful when the codebase is small in early phases of development. RC is more useful during the maintenance phase when changes carry more risk''~($P_{2}$). 
A developer with 7--10 years of experience summarized the distinction more succinctly: ``RC guides improvement, AC ensures minimum standards''~($P_{18}$). 
Another developer (1--3 years) stated it more generally: ``Absolute understandability helps evaluate the quality of a single code fragment, while relative understandability is useful for comparing alternatives and choosing the clearest implementation''~($P_{30}$). 
These results suggest that RC is most compelling when a comparison target exists, while AC remains useful for standalone assessment. 

\subsubsection{\ref{rq:survey_workflow}: Workflow Integration of Comprehensibility Models.}

Participants preferred integrations that fit into existing development environments. The most common preference was an IDE extension or plugin (29/38), followed by code review annotations (21/38), inline IDE feedback (14/38), and standalone dashboards or visualizations (12/38).

Participants also preferred outputs that go beyond a raw prediction. The most common output preferences were actionable suggestions on how to make code more comprehensible (28/38), explanations of why the code is difficult to understand (27/38), numeric comprehensibility scores (21/38), and high-level summaries (11/38). 

The open-ended responses also point to actionability. An engineering manager with 10+ years of experience noted that readability information could support review and planning: ``It could be useful in code review to ensure code check in is readable''~($P_{31}$). A developer with 4--6 years of experience described RC as a way to act on comparisons: ``The RC is a tool that compares and allows me to proactively modify my code against other code sources''~($P_{24}$). These responses suggest that useful comprehensibility tools should not only identify difficult code, but also help developers decide what to improve and evaluate whether a change made the code easier to understand.

\begin{tcolorbox}[boxsep=2pt, bottom=2pt]
	\vspace{0.5em}
	\textbf{\ref{rq:survey_problem}--\ref{rq:survey_workflow} Findings:}
	Code comprehension difficulty (\ie comprehensibility) is a common problem for the surveyed practitioners, but the need to measure it varied. 
	Practitioners saw value in both AC and RC models. RC was perceived as particularly useful for refactoring and code review, where developers compare versions or alternatives. AC remained useful for standalone code understanding. Overall, the results suggest that predictive performance gains of RC models over AC models may translate into practical value. 
	Practitioners preferred model integrations in IDEs and code review tools, and they preferred outputs with explanations and actionable suggestions. These results motivate the development of user-focused comprehensibility models and tools.
	\vspace{0.5em}
\end{tcolorbox}

\section{Threats to Validity}
\label{sec:threats}

We discuss the factors that could affect the validity of our findings.

\subsection{Construct Validity.}

Our study uses several code comprehensibility proxies, including
understandability, readability, perceived understandability, and answers to
comprehension questions. These measures capture different aspects of code
comprehension difficulty and may not fully show whether a developer has correctly
understood the code. To reduce the effect of any single proxy, we evaluated AC
and RC models across all available metrics and corresponding data from prior studies~\cite{Scalabrino:TSE19,Raymond:TSE10}, and based our conclusions on the overall
results.

For snippet-wise prediction, we combined individual proxy measurements
into one value for each snippet. We used the mean and rounded the result when a
discrete class was needed. Other approaches, such as using the median, mode, or
majority vote, could produce different labels. The threshold used to decide
whether two snippets are similarly comprehensible also affects the RC labels.
We therefore evaluated snippet-wise RC using three threshold values. The main
trends were similar across them, which suggests that the results are not tied
to one threshold. However, other aggregation and labeling choices could still
lead to different results.

We compared AC and RC prediction using relative improvement over the strongest na\"ive
baseline for each task. This was needed because AC and RC have different output
spaces and class distributions, so their absolute performance scores are not
directly comparable. However, RI depends on the selected baselines. We
therefore also report weighted F1, MCC, and Cohen's kappa for each model.

The practitioner survey measured perceived usefulness, not the actual value of
AC or RC tools used in practice. Participants evaluated descriptions and
examples of possible tools rather than complete tools used in real development
tasks. The survey results should therefore be interpreted as evidence of
expected usefulness and preferred ways of model integration.

\subsection{Internal Validity.}

The results may be affected by the selected models, hyperparameters, features,
data processing steps, and evaluation procedure. For the classical
machine learning models, we selected hyperparameters using nested
cross-validation. We also performed feature selection, normalization, and
class balancing using only the training data. This reduces the risk of test
data affecting model development. We used fixed random seeds where possible
and provided the model configurations and scripts in our replication package~\cite{repl_pack}.

We implemented the code features using the definitions reported by Scalabrino
\etal~\cite{Scalabrino:TSE19}. Some definitions were unclear, and we found
differences between the values produced by the original code and those
included in their replication package. We removed features that we could not
reproduce reliably and implemented the remaining features using the Java 8
Language Specification~\cite{javaspec}. Other implementations or feature sets
could produce different results.

Different prompt designs may also affect LLM performance. Prior work has shown
that small prompt changes can lead to different results~\cite{Zhao:ICML21}.
We used fixed prompt templates and the same task description across comparable
settings. However, we did not optimize the prompts, use few-shot examples, or
fine-tune the LLMs. The LLM results therefore describe the evaluated zero-shot
settings, not the best performance the models could achieve.
LLM outputs may also vary across repeated runs. We ran each prompt three times and checked the consistency of
the predictions across runs. 

The survey results may be affected by how AC and RC were explained and by the
examples included in the questionnaire. Some participants may also have had
little prior knowledge of predictive comprehensibility models, especially RC
models. We gave all participants the same definitions and examples, but they
may still have interpreted the proposed tools differently.

\subsection{External Validity.}

The two datasets contain 150 Java snippets from open-source projects. Although
the snippets come from several projects and include non-trivial methods, the
results may not generalize to other programming languages, proprietary
systems, other application domains, or code with different structure and
documentation practices.
The snippets are also limited in size. The results may differ for larger units
such as classes or files  where understanding depends
more on broader context and architecture.

Most participants in the original human studies were students. Their
comprehension measurements may not represent those of experienced professional
developers. The developer-wise results are also limited by the developer
features available in the original datasets, which capture only part of the
differences among developers.

We evaluated ten model types, including two LLMs. The results may not
generalize to other model families, newer model versions, smaller models, or
fine-tuned models. In particular, our LLM
results apply only to the evaluated models under zero-shot prompting.

The practitioner survey included 38 participants recruited mainly through the
authors' professional networks. The sample is relatively small and may not represent the
broader population of software developers. Participants may differ from other
developers in experience, role, domain, or interest in code quality tools.
The survey results should therefore be viewed as exploratory.

\section{Discussion and Implications}
\label{sec:discussion}

This work found that predicting the comprehensibility of a code snippet relative to another (RC: relative comprehensibility) is generally more effetive than predicting the comprehensibility of a snippet in isolation (AC: absolute comprehensibility). Prior work has focused almost
exclusively on AC prediction. Our results suggest that RC is a promising
alternative, particularly for snippet-wise prediction, where RC consistently
outperforms AC across different comprehensibility proxies, datasets, and
 model architectures (state-of-the art machine and deep learning models, including LLMs). Developer-wise RC prediction is more challenging,
but it still outperforms AC in most model-metric comparisons.

These findings have implications for both future research and developer tools.
Our practitioner survey complements the predictive evaluation. Participants
reported that understanding code is a common challenge in their daily work and
perceived RC predictions as particularly useful for tasks that naturally
involve comparing code alternatives, such as refactoring and code review.
Participants also preferred predictions that include explanations and
actionable suggestions rather than only a label or score.

\textbf{Importance of Code Comprehensibility.} 
This study focuses on the fundamental problem of automatically measuring and predicting human code comprehension difficulty (\ie code comprehensibility prediction), which is essential for supporting developers in SE tasks like refactoring, code review, and debugging.
As prior work indicates \cite{Borstler:2023developers,Maalej:TOSEM14,Scalabrino:TSE19,Raymond:TSE10}, code comprehensibility is fundamental to
developing and maintaining high-quality software.
Our survey provides additional evidence that this is a common practical
problem: 92.1\% of participants reported facing difficulty understanding code
at least a few times per month.

\textbf{RC For Developer Tools.}
The best RC models, such as the snippet-wise XGBoost models (row ``XGBoost'' in \Cref{tab:rc_results_snippet}), achieve $wF1$ scores that are acceptable in an absolute sense: between
about 0.811 and 0.914. This generally implies that the models can correctly predict the relative comprehensibility of a snippet pair in \textasciitilde 8-9/10 cases. 
These
results indicate strong overall performance across the three RC classes and
suggest that RC models could support developer-facing tools. In contrast, AC
models rarely achieve similar performance.

Developer-wise RC prediction remains more difficult. Except for \AU, the best
models achieve more modest performance (between 0.71 and 0.761 $wF1$), suggesting that the available
code and developer features are not sufficient to accurately predict how an individual
developer will judge a snippet pair. Future research is needed to develop more accurate models and assess how useful this accuracy can be for developers. Overall, the current findings are
more promising for tools that estimate aggregate judgments than for tools that
personalize predictions for individual developers.

Although these results are encouraging, prediction performance alone does not
demonstrate practical usefulness. Field studies are needed to evaluate whether
RC-based tools help practitioners complete SE tasks more
effectively. Our survey suggests that explanations and actionable suggestions
will be important for the adoption of such tools.

\textbf{RC for Code Comprehensibility Studies.} 
Our results also have implications for future research on code
comprehensibility. Existing studies typically collect absolute judgments for
individual snippets. In contrast, RC focuses on comparing two snippets. Since
our results show that RC is easier to predict than AC, future studies should
consider collecting pairwise judgments directly and investigate whether they
produce more consistent and useful datasets for developing predictive models.

\textbf{Validating Refactored Code.}
Refactoring is a common software development practice aimed at improving code quality, focusing on factors such as reducing complexity, eliminating code smells, and enhancing code optimization.
A key aspect of refactoring is ensuring that the code becomes more comprehensible~\cite{Sellitto:SANER2022}, as this directly impacts maintainability.
Because refactoring naturally compares an original
version with a refactored version, RC is a good match for this task. An RC
model could predict whether a refactoring improves, reduces, or preserves
comprehensibility, providing developers with feedback before the code is
submitted for review.

This application is also consistent with our practitioner survey. Participants
identified refactoring as one of the tasks that could
benefit most from RC prediction. Since our study did not evaluate RC on
pairs of semantically equivalent code, future work should investigate whether
RC models generalize to refactoring scenarios and whether the provided
feedback improves developer decisions.

\textbf{Identifying Where to Refactor.}
Identifying which parts of the code need refactoring can be challenging. An effective AC model would be ideal for this task, as it would assign a comprehensibility score to each candidate location, allowing for a simple ranking to guide refactoring. This approach scales linearly with the program size.
\looseness=-1
However, as we have shown, AC ML models are often ineffective. While an RC model could also be used to identify refactoring candidates, it would be computationally expensive. This is because it would require comparing each candidate location to all others, selecting the one considered ``less comprehensible'' than the most others. This method results in a quadratic number of model invocations based on the number of candidate locations. Fortunately, this complexity is reduced if developers already have refactoring candidates, allowing the RC model to assess only those locations.

\textbf{Improving Code Review.}
Code reviews are an essential part of software development for maintaining high code quality. Since code review inherently involves comparing two versions of code (\ie the original and the proposed changes), RC models are particularly well-suited for this process. An RC model could enhance code reviews by identifying changes that significantly reduce code comprehensibility, allowing reviewers to focus on areas that may impact maintainability and clarity.
\looseness=-1

This application was also strongly supported by our practitioner survey.
Participants viewed code review as one of the most promising uses of RC
prediction, but they also indicated that predictions should be accompanied by
clear explanations and actionable suggestions. Future work should therefore
evaluate RC models on real code changes and study whether such feedback
improves review quality or efficiency.

\section{Conclusion}
\label{sec:conclusions}

This empirical study investigated relative 
code comprehensibility (RC) as an alternative to predicting absolute 
code comprehensibility (AC) using learning-based models. We trained 
and evaluated ten model architectures spanning classical machine 
learning, a convolutional
neural network, and two large language models, across seven 
comprehensibility metrics from two distinct human-subject datasets. 
We compared the models with na\"ive
baselines in both snippet-wise and developer-wise prediction settings.

Our results show that AC models often fail to outperform the baselines and
achieve a maximum average relative improvement of 33.4\%, which is consistent
with prior work~\cite{Scalabrino:TSE19}. In contrast, RC models perform much
better, especially in the snippet-wise setting. They achieve average relative
improvements of up to 159.8\% for snippet-wise prediction and 74.7\% for
developer-wise prediction. Overall, 96.8\% of the evaluated RC configurations
outperform the strongest baseline, compared with 48\% of the AC configurations.
The snippet-wise advantage also appears across the evaluated model
architectures, suggesting that it is not limited to a particular one.
Developer-wise results are more variable, although RC still outperforms AC in
most model-metric comparisons.

To complement the predictive evaluation, we surveyed 38 software
practitioners about the perceived usefulness of AC and RC models and how
such models could be integrated into their development workflows. Participants
perceived both forms of prediction as potentially useful, but showed stronger
support for RC in comparison-oriented tasks, particularly code review  and refactoring. They also preferred
model outputs that explain the prediction and provide actionable suggestions.
These findings identify promising applications and design requirements for
future RC-based developer tools.
\looseness=-1

Overall, our findings suggest that relative prediction is a promising
direction code comprehensibility modeling.
Future work should develop more accurate RC models, evaluate these models in real development settings, and study
whether they can help practitioners write, review, and maintain more
comprehensible software.

\balance
\bibliographystyle{ACM-Reference-Format}
\bibliography{references}

\end{document}